\documentclass[structabstract]{aa}  
\usepackage{graphicx}
\usepackage{float}
\usepackage{indentfirst}
\usepackage{lscape}
\usepackage{longtable}
\usepackage{txfonts}
\begin{document}
   \title{A combined IRAM and Herschel/HIFI study of cyano(di)acetylene in Orion KL: tentative detection of DC$_3$N\thanks{Based on observations carried out with the IRAM 30m telescope and with the Herschel telescope (HIFI instrument). IRAM is supported by INSU/CNRS (France), MPG (Germany), and IGN (Spain). Herschel is an ESA space observatory with science instruments provided by European-led Principal Investigator consortia and with important participation from NASA.}}


   \author{G. B. Esplugues
          \inst{1},
          J. Cernicharo
          \inst{1},
          S. Viti  
          \inst{2},    
          J. R. Goicoechea
          \inst{1},
          B. Tercero
          \inst{1},
          N. Marcelino
          \inst{3},
          Aina Palau
          \inst{4}, 
          T. A. Bell
          \inst{1},
          E.A. Bergin
          \inst{5},
          N. R. Crockett 
          \inst{5},
          \and
          S. Wang
          \inst{5}.        
         }

   \institute{Centro de Astrobiolog\'ia (CSIC-INTA), Ctra. de Torrej\'on-Ajalvir, km. 4, E-28850 Torrej\'on de Ardoz, Madrid, Spain\\
              \email{espluguesbg@cab.inta-csic.es}
        \and
            Department of Physics \& Astronomy, University College London, Gower St. London WC1E 6BT.
         \and
                      National Radio Astronomy Observatory, 520 Edgemont Road, Charlottesville, VA 22903, USA. 
        \and
Institut de Ci$\grave{e}$ncies de l'Espai (CSIC-IEEC), Campus UAB-Facultat de Ci$\grave{e}$ncies, Torre C5-parell 2, E-08193 Bellaterra, Barcelona, Spain.
        \and
Department of Astronomy, University of Michigan, 500 Church St, Ann Arbor, MI 48109, USA.
             }

   \date{Received ; accepted }

 
  \abstract   
   {We present a study of cyanoacetylene (HC$_3$N) and cyanodiacetylene (HC$_5$N) in Orion KL, through observations from two lines surveys performed with the IRAM 30-m telescope and the HIFI instrument on board the Herschel telescope. The frequency ranges covered are 80-280 GHz and 480-1906 GHz.} 
   {This study (divided by families of molecules) is part of a global analysis of the physical conditions of Orion KL, as well as molecular abundances in the different components of this cloud.}
   {We modeled the observed lines of HC$_3$N, HC$_5$N, their isotopologues (including DC$_3$N), and vibrational modes, using a non-LTE radiative transfer code. 
In addition, to investigate the chemical origin of HC$_3$N and DC$_3$N in Orion KL, we have used a time-dependent chemical model.}
   {We detect 40 lines of the ground vibrational state of HC$_3$N and 68 lines of its $^{13}$C isotopologues. We also detect 297 lines of six vibrational modes of this molecule ($\nu$$_7$, 2$\nu$$_7$, 3$\nu$$_7$, $\nu$$_6$, $\nu$$_5$, and $\nu$$_6$+$\nu$$_7$) and 35 rotational lines of the ground vibrational state of HC$_5$N.
We report the first tentative detection of DC$_3$N in a giant molecular cloud. We have obtained a DC$_3$N/HC$_3$N abundance ratio of 0.015$\pm$0.009, comparable to typical D/H ratios of cold dark clouds.
We provide column densities for all species and derived isotopic and molecular abundances. 
We have also performed a 2$\arcmin$$\times$2$\arcmin$ map around Orion IRc2 and we present maps of HC$_3$N lines with energies from 34 to 154 K and maps of lines of the HC$_3$N vibrational modes $\nu$$_6$ and $\nu$$_7$ with energies between 354 and 872 K. 
In addition, a comparison of our results for HC$_3$N with those in other clouds has allowed us to derive correlations between the column density, the FWHM, the mass, and the luminosity of the clouds.
}
   {The high column densities of HC$_3$N, in particular of the ground vibrational state and the vibrational mode $\nu$$_7$, obtained in the hot core, make this molecule an excellent tracer of hot and dense gas. In addition, the large frequency range covered reveals the need to consider a temperature and density gradient in the hot core in order to obtain better line fits.
The high $D$/$H$ ratio (comparable to that obtained in cold clouds) that we derive suggests a deuterium enrichment. Our chemical models indicate that the possible deuterated HC$_3$N present in Orion KL is formed during the gas-phase. This fact provides new hints concerning the processes leading to deuteration.}

   \keywords{survey-stars: formation - ISM: abundances - ISM: clouds - ISM: molecules - Radio lines: ISM}
   \titlerunning{Study of cyano(di)acetylene in Orion KL: detection of DC$_3$N}
   \authorrunning{G. B. Esplugues et al.}
   \maketitle
%


\section{\textbf{Introduction}}

The Orion region is one of the best places to study the formation of high-mass stars due to its proximity ($\simeq$414 pc, Menten et al. 2007) and the presence of compact radio sources, molecular outflows (Wright et al. 1996), a molecular hot core, and an IR luminosity of $\sim$10$^{5}$L$_\odot$ in the central part of the region, indicating the presence of embedded massive protostars (Nissen et al. 2007). 
The Becklin-Neugebauer ($BN$) object was the first source of infrared radiation discovered in the Orion nebula (Becklin $\&$ Neugebauer et al. 1967). The luminosity of the $BN$ object is $L$$\simeq$1500L$_\odot$ and it lies about 12$\arcsec$ north of the Kleinmann-Low (KL) nebula.
Menten $\&$ Reid (1995) detected the radio continuum emission of IR source $n$ and the very embedded radio source $I$ (located a few arcseconds to the south of the young and massive star IRc2), which could be a binary system with a total mass of $\sim$20M$_\odot$, according to measurements of the proper motions of this source and the object $BN$ (Goddi et al. 2011). Source $I$ has been proposed as a possible driver of the two outflows observed in Orion KL: a high-velocity (30-100 km s$^{-1}$), wide-angle ($\sim$1 rad) outflow that extends northwest-southeast over 0.3 pc, and a low-velocity ($\sim$18 km s$^{-1}$) elongated northeast-southwest (Genzel \& Stutzki 1989, Greenhill et al. 1998, Zapata et al. 2009). 

Orion BN/KL contains a large amount of dense gas, evidenced by the presence of H$_{2}$O, OH, and SiO masers. In fact, source $I$ coincides with the centroid of the SiO maser distribution (Plambeck et al. 2009). These high densities and the large dust extinction (A$_V$$\sim$1000) obscure the innermost regions in the near and mid-IR. Therefore, to study the innermost regions of the hot core, it is necessary to observe at radio wavelengths rotational transitions excited by collisions in these high density zones. Moreover, the observation of highly vibrationally excited states can provide information on the infrared radiation field in the same regions.

Orion KL also constitutes an excellent astrochemical laboratory due to its rich and complex chemistry. Many molecules (tracers of shocks, cold gas, hot and dense gas, etc.) have been observed in this cloud, in particular, organic molecules associated mainly with the hot core (richer in N-bearing organics) and the compact ridge (richer in O-bearing organics) component (Wang et al. 2009).
One very abundant molecule in the hot core of Orion KL is HC$_{3}$N. Interstellar cyanoacetylene, HC$_{3}$N, was discovered by Turner (1971), who detected emission in the $J$=1-0 line toward Sgr B2. This molecule is an excellent tracer of hot and dense regions affected by high extinction; its vibrational levels are mainly excited by mid-IR radiation (de Vicente et al. 2000) and the relatively low energy of its bending modes makes them easy to detect in many vibrationally excited states. Another advantage of studying HC$_{3}$N is that, although it is abundant in warm clouds, the radio lines are usually optically thin (Morris et al. 1976). In addition, these small optical depths allow an accurate determination of the $^{12}$C/$^{13}$C abundance ratio from the observation of the $^{13}$C isotopologues.
The deuterated counterpart of cyanoacetylene (DC$_{3}$N) was detected for the first time in the cold interstellar cloud Taurus Molecular Cloud 1 by Langer et al. (1980). Later it has been observed in L1498, L1544, L1521B, L1400K, and L1400G by Howe et al. (1994), in L1527 by Sakai et al. (2009) and in Cha-MMS1 by Cordiner et al. (2012). However, DC$_{3}$N has never been observed so far in giant molecular clouds.

Most deuterated species are produced in molecular environments characterised by low temperatures ($T$$\leq$20 K) (Millar et al. 1989). Although these temperatures are typical of low-mass pre-stellar cores, many deuterated molecules such as DCN (Jacq et al. 1999), DNC (Tatematsu et al. 2010), CHD$_{2}$OH (Parise et al. 2002), or D$_{2}$CO (Fuente et al. 2005) have been detected in hot cores, where the gas kinetic temperature is $\textgreater$100 K. In the case of Orion, some of the first detected deuterated molecules were CH$_{3}$OD (Mauersberger et al. 1988) and CH$_{2}$DOH (Jacq et al. 1993).  
The formation of deuterated molecules occurs through exchange reactions (Howe $\&$ Millar et al. 1993) where deuterated ions transfer the deuterium to neutral species in ion-molecule reactions. These reactions normally dominate at low temperatures, but there are some reactions such as

\begin{equation}
\mathrm{CH^{+}_{3} + HD \leftrightarrow CH_{2}D^{+} + H_{2} + 370 K} 
\end{equation}

\noindent which are energetically favorable at high temperatures that can lead to a variety of deuterated molecules in regions with temperatures up to 100 K (Howe $\&$ Millar et al. 1993).

In this paper we focus on the IRAM line survey, first presented by Tercero et al. (2010), and on Herschel/HIFI observations (from the HEXOS program, Bergin et al. 2010, Crockett et al. 2010), covering the frequency ranges 80-280 and 480-1910 GHz, respectively. Observations are described in Sect. \ref{section:observations}.
We study the molecules HC$_{3}$N, its isotopologues, and HC$_{5}$N, and report the first detection of DC$_{3}$N in Orion KL.
Here we present (Sect. \ref{section:results}) more than 400 lines from HC$_3$N for the ground state, different vibrational states ($\nu$$_5$, $\nu$$_6$, $\nu$$_7$, 2$\nu$$_7$, 3$\nu$$_7$, and $\nu$$_6$+$\nu$$_7$), and isotopologues, and 35 lines of HC$_5$N. We also present (Sect. \ref{subsection:maps}) maps over a region 2$\arcmin$$\times$2$\arcmin$ around Orion IRc2 of two transitions of HC$_3$N, two of HC$_3$N $\nu$$_7$, and one of HC$_3$N $\nu$$_6$.
Unlike previous works, we use a non-LTE radiative transfer code (LVG) to derive physical and chemical parameters, such as column densities or abundances (Sect. \ref{section:column_densities}).
In addition, we compare some of our results for HC$_3$N with results obtained by other authors in a sample of 22 molecular cloud cores (Sect. \ref{section:HC3N_other_sources}). We study possible correlations between the line widths, the mass of the cloud cores, the infrared luminosity, the galactocentric distance, and the column density of HC$_3$N.
In Sects. \ref{section:discussion} we present chemical models of the hot core of Orion KL in order to study the origin of HC$_3$N and DC$_3$N, and we discuss the results. Finally, in Sect. \ref{section:summary} we summarize our conclusions.

\section{Observations}
\label{section:observations}
\subsection{IRAM 30m}

  The observations were carried out using the IRAM 30m radiotelescope during September 2004 and 
March 2005 
Four SiS receivers operating at 1.3, 2 and 3 mm were used simultaneously with image sideband rejections. 
The intensity scale was calibrated using two absorbers at different temperatures and using the Atmospheric Transmission Model (ATM, Cernicharo 1985; Pardo et al. 2001). 
Observations were made in the balanced wobbler-switching mode, with a wobbling frequency of 0.5 Hz and a beam throw in azimuth of $\pm$240$\arcsec$. As backends we used two filter banks with 512x1 MHz channels and a correlator providing two 512 MHz bandwidths and 1.25 MHz resolution (see Esplugues et al. 2013 for more details). We pointed the telescope towards the IRc2 source at $\alpha$$_{J2000}$=5$^{h}$ 35$^{m}$ 14.5$^{s}$, $\delta$$_{J2000}$=-5$\degr$ 22$\arcmin$ 30.0$\arcsec$. 


To analyze the survey we considered only the lines with antenna temperatures higher than 0.02 K (i.e. $\textgreater$3$\sigma$). The data were processed using the IRAM GILDAS software\footnote{http://www.iram.fr/IRAMFR/GILDAS} (developed by the Institut de Radioastronomie Millim\'etrique). We present the spectra in units of main beam temperature $T$$_{\mathrm{MB}}$, which is defined as

\begin{equation}
T_{\mathrm{MB}}=\left(T^{\star}_{\mathrm{A}}/\eta_{\mathrm{MB}}\right),
\end{equation}

\begin{table}
\caption{IRAM 30m telescope efficiency data along the covered frequency range.}               
\centering          
\label{table:tablaeficiencias}
\begin{tabular}{c c c}     
\hline\hline                    
Frequency &  $\eta$$_{\mathrm{MB}}$ & HPBW \\
(GHz) & & ($\arcsec$)\\
\hline                    
   86  & 0.82 & 29.0  \\
   100 & 0.79 & 22.0  \\
   145 & 0.74 & 17.0  \\
   170 & 0.70 & 14.5  \\
   210 & 0.62 & 12.0  \\
   235 & 0.57 & 10.5  \\
   260 & 0.52 & 9.5   \\
   279 & 0.48 & 9.0   \\
\hline                  
\end{tabular}
\end{table}

\noindent where $\eta$$_{\mathrm{MB}}$ is the main beam efficiency and $T$$^{\star}_{\mathrm{A}}$ the antenna temperature. Table \ref{table:tablaeficiencias} shows the half power beam width (HPBW) and the mean beam efficiencies over the covered frequency range.


We also used the 30m telescope to map a region around Orion IRc2 at 1.3 mm. We covered the whole frequency range (216-250 GHz) using the HERA heterodyne receiver, and the ranges 200-216 GHz and 250-285 GHz using EMIR E230 and E330 receivers, over a region of 2$\arcmin$$\times$2$\arcmin$ with a 4$\arcsec$ spacing. The survey was performed in on-the-fly mode using position switching with a emission-free reference position at an offset (-600$\arcsec$,0) from IRc2. As backends we used WILMA with 2 MHz spectral resolution.
The maps presented in Sect \ref{subsection:maps} were observed using HERA (tuned at frequencies of 227.9 and 236.9 GHz), and the E090 receivers (at 109.983 GHz).  Observations at 1.3 mm were performed in December 2008, with opacities $\sim0.1$ at 1.3 mm and 1-1.5 mm of precipitable water vapor. 
The 3 mm data was observed in February 2010, under poorer weather conditions (opacities of 0.3-0.4 and 5 mm of pwv). 
Pointing was checked every 1-1.5 hrs on nearby and strong quasars, with errors better than 4-5$\arcsec$. 
Data reduction was also performed using the IRAM GILDAS software.

\subsection{Herschel/HIFI}

The HIFI line survey was observed as part of the HEXOS Guaranteed Time Key Program.
The HIFI instrument (de Graauw et al. 2010), on board $Herschel$, observed a bandwidth of approximately 1.2 THz in the frequency range 488-1902 GHz, with gaps between 1280-1430 GHz and 1540-1570 GHz. Most of observations were carried out between March 2010 and March 2011. The spectral resolution is 1.1 MHz corresponding to 0.2-0.7 km s$^{-1}$. HIFI is a double sideband system where, in the conversion to frequencies detectable by the spectrometers, spectral features in the opposite sideband appear superposed at a single frequency. As a part of the spectral scan observation, different settings of the local oscillator (LO) are observed and the double sideband is deconvolved to isolate the observed sideband. It was applied the standard HIFI deconvolution using the \textit{doDeconvolution} task within HIPE (see Bergin et al. (2010) and Crockett et al. (2010) for more details).
The spectral scans for each band were taken in dual beam switch (DBS) mode, using the wide band spectrometer (WBS). For bands 1-5, the telescope was pointed toward coordinates $\alpha$$_{J2000}$=5$^{h}$ 35$^{m}$ 14.3$^{s}$, $\delta$$_{J2000}$=-5$\degr$ 22$\arcmin$ 33.7$\arcsec$, midway between the Orion hot core and the compact ridge. For bands 6-7, where the beam size is smaller (see Table \ref{table:tablaeficienciasHIFI}), the telescope was pointed directly toward the hot core at coordinates $\alpha$$_{J2000}$=5$^{h}$ 35$^{m}$ 14.5$^{s}$, $\delta$$_{J2000}$=-5$\degr$ 22$\arcmin$ 30.9$\arcsec$. We assume the nominal absolute poiting error (APE) for Herschel of 2.0$\arcsec$ (Pilbratt et al. 2010). The data were reduced using the standard HIPE (Ott 2010) pipeline version 5.0. We also present the spectra in units of main beam temperature $T$$_{\mathrm{MB}}$.

\begin{table}
\caption{HIFI efficiency data along the covered frequency range.}               
\centering          
\label{table:tablaeficienciasHIFI}
\begin{tabular}{c c c c}     
\hline\hline                    
Bands & Frequency range &  $\eta$$_{\mathrm{MB}}$ & HPBW \\
      & (GHz) &                       & ($\arcsec$)\\
\hline                    
1     &  488.1-628.1   & 0.685 & 39  \\
2     &  642.3-792.9   & 0.681 & 30  \\
3     &  807.1-952.9   & 0.677 & 25  \\
4     &  967.1-1112.8  & 0.670 & 21  \\
5     &  1116.2-1241.8 & 0.662 & 19  \\
6     &  1429.2-1699.8 & 0.645 & 13  \\
7     &  1699.2-1902.8 & 0.632 & 13  \\
\hline                  
\end{tabular}
\end{table}


\section{Results}
\label{section:results}
 
\subsection{Lines profiles}

Cyanoacetylene is a linear molecule and therefore its rotational spectrum is very simple. Owing to its large moment of inertia, it has a small rotational constant $B$ and its rotational transitions $J$$\rightarrow$$J$-1 are relatively close in frequency. Like other species containing $^{14}$N, cyanoacetylene has a hyperfine structure due to the interaction of the nitrogen nuclear spin with the rotation of the molecule. 

\begin{figure*}
   \centering 
   \includegraphics[angle=0,width=15.3cm]{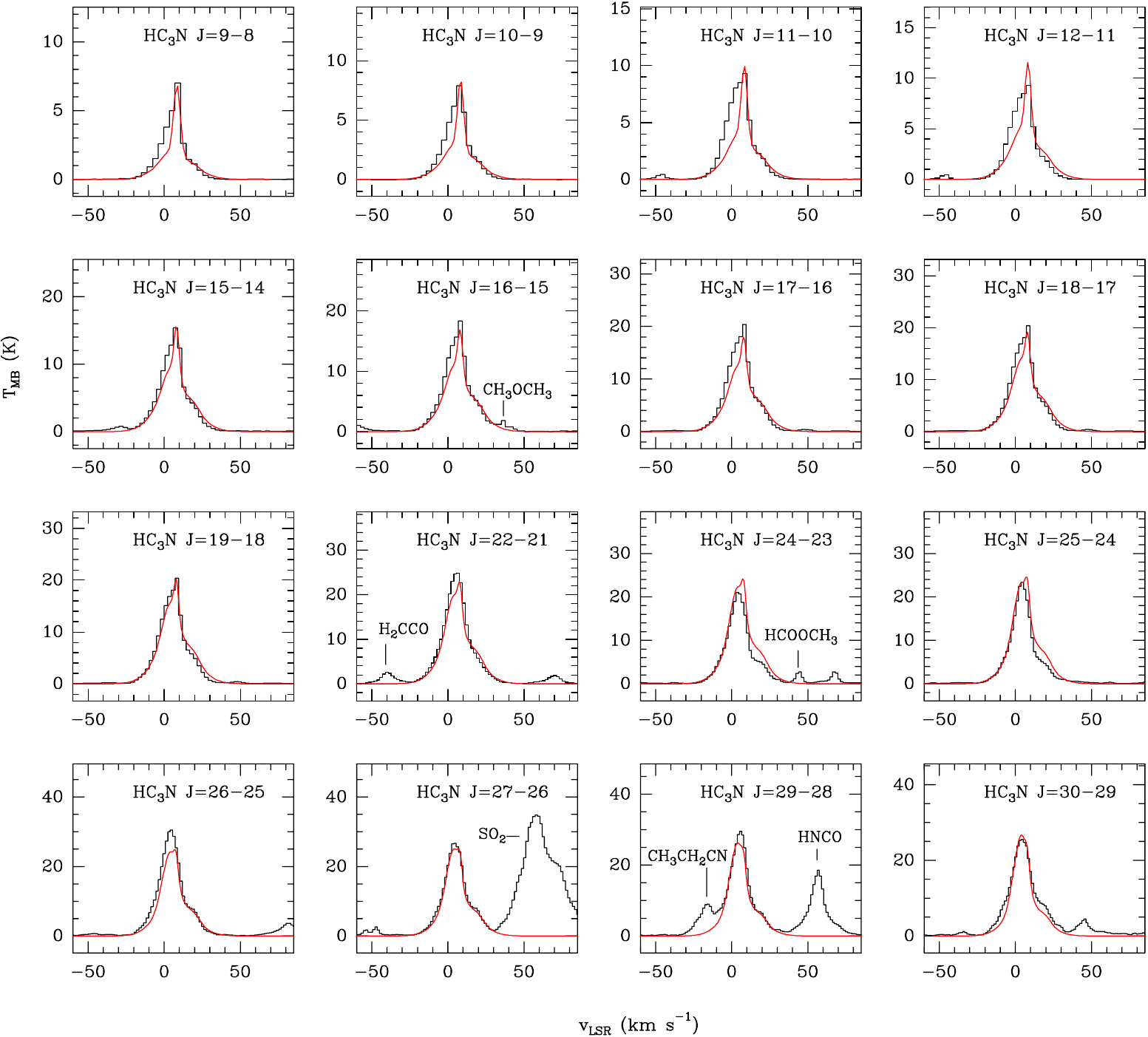}
   \caption{Observed pure rotational lines of HC$_{3}$N (black histogram) in the IRAM line survey. Best fit LVG model results are shown in red.}
    \label{figure:HC3N_mix_30m}
   \end{figure*}

\begin{figure*}
   \centering 
   \includegraphics[angle=0,width=18cm]{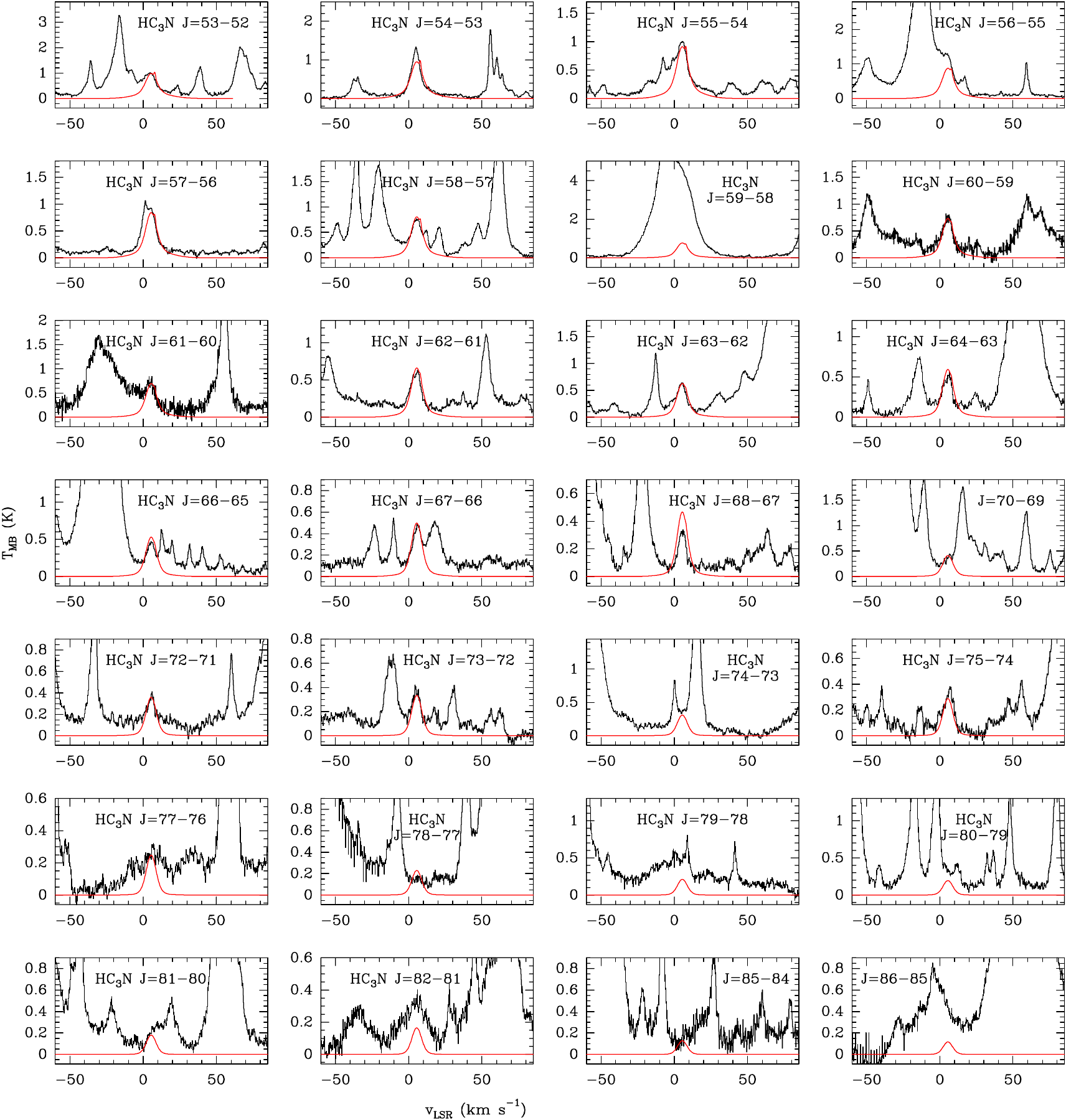}
   \caption{Observed pure rotational lines of HC$_{3}$N (black histogram spectra) with the HIFI survey. Best fit LVG model results are shown in red.}
    \label{figure:HC3N_mix_hifi}
   \end{figure*}

 \subsubsection{HC$_{3}$N}

Figures \ref{figure:HC3N_mix_30m} and \ref{figure:HC3N_mix_hifi} show the lines of HC$_{3}$N (together with our best fit LVG model, see Sect. \ref{section:column_densities}) observed with the IRAM and the Herschel-HIFI survey, respectively. 
The 18 observed pure rotational lines between 3-1.3 mm belong to transitions between $J$=9-8 and $J$=30-29 with a range of energy of $E$$_{\mathrm{up}}$=19-230 K (see Table \ref{table:tab_HC3N_vib}). All these lines are strong, with 8$\textless$$T$$_{\mathrm{MB}}$$\textless$31 K, and are barely blended with other species. The observed lines with HIFI are much weaker than those obtained with IRAM-30m, $T$$_{\mathrm{MB}}$$\textless$1.5 K, due to their high energies $E$$_{\mathrm{up}}$=624-1826 K.
Assuming that the line emission is optically thin and the level populations are thermalized (Goldsmith \& Langer et al. 1999), we fitted the observed lines with Gaussian profiles using the CLASS software
to derive the contribution of each cloud spectral component (see Table \ref{table:sHC3N_parameters_gaussian} in Appendix). We treated the radial velocity ($V$$_{\mathrm{LSR}}$), the line-width and the antenna temperature as free parameters in the fits. We took into account the typical velocity ranges of each component of Orion KL to consider the best fit and to avoid possible degeneracy. The contribution to the intensity arises from different velocity components: a wide component present in all lines, a very narrow component centered at $\simeq$10 km s$^{-1}$ which affects mainly the lowest transitions, and another component wider ($\simeq$10 km s$^{-1}$) than the previous which affects mainly the transitions with $J$$\textgreater$20 and that is centered at about 5 km s$^{-1}$. 
For the HIFI case, the contribution to the emission from a wide ($\sim$25 km s$^{-1}$) component dominates the observed line profiles.

\begin{figure*}[ht!]
   \centering 
   \includegraphics[angle=0,width=16.5cm]{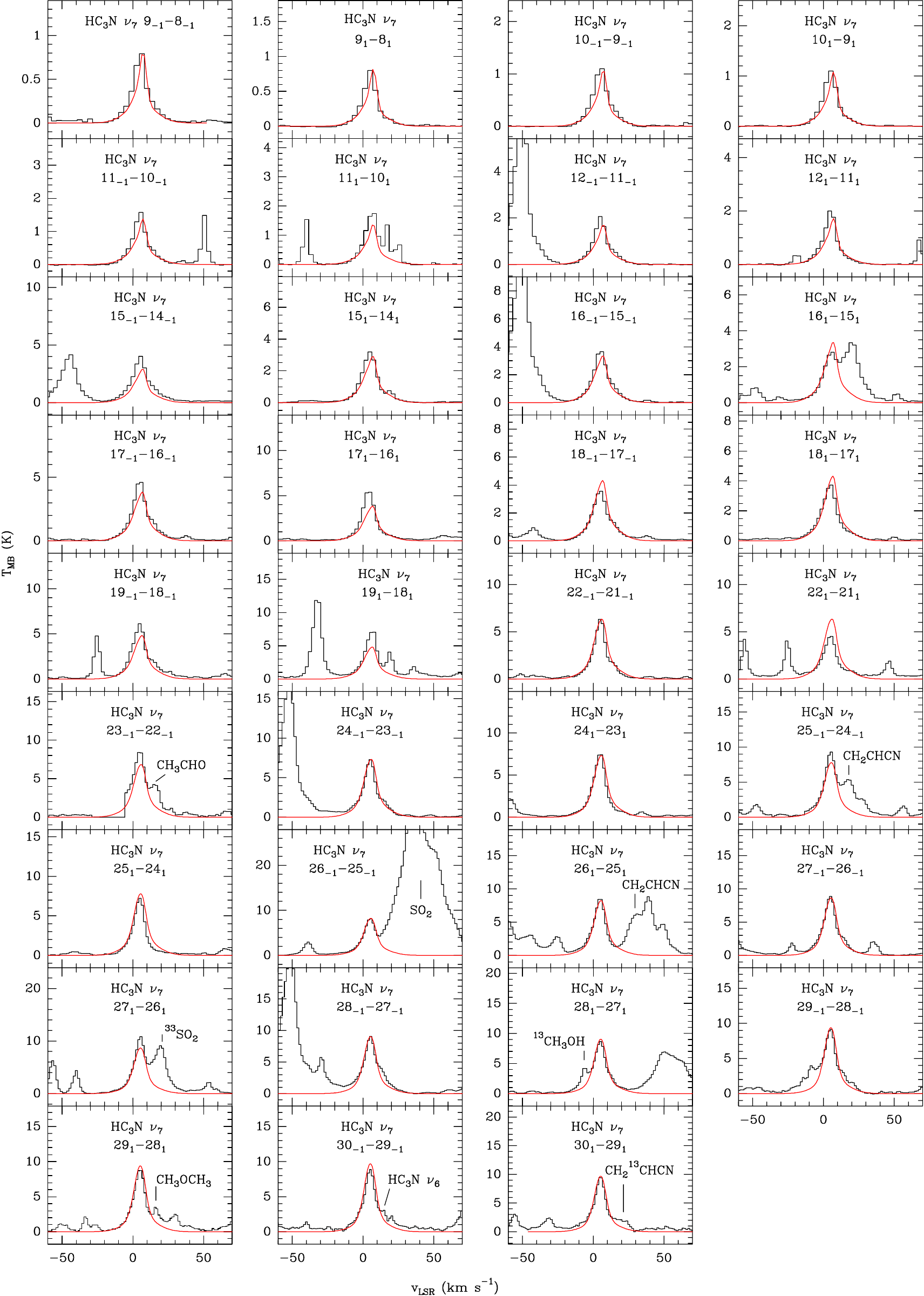}
   \caption{Observed pure rotational lines of HC$_{3}$N $\nu$$_{7}$ (black histogram) in the IRAM survey. Best fit LTE model results are shown in red.}
    \label{figure:HC3N_V7_mix_30m}
   \end{figure*}

We have also detected rotational lines from six vibrational states of HC$_{3}$N: $\nu$$_{7}$, 2$\nu$$_{7}$, 3$\nu$$_{7}$, $\nu$$_{6}$, $\nu$$_{6}$+$\nu$$_{7}$, and $\nu$$_{5}$. The modes $\nu$$_{5}$, $\nu$$_{6}$, and $\nu$$_{7}$ are doubly degenerate in order of decreasing energy (954, 718, and 320 K, respectively). With rotation, their degeneracy is broken, which is known as $l$-doubling. The modes $\nu$$_{1}$, $\nu$$_{2}$, and $\nu$$_{3}$ have energies higher than 2000 K and they have not been detected in this study, nor has the $\nu$$_{4}$ mode.
We observed 36 rotational transitions of HC$_{3}$N $\nu$$_{7}$ in our IRAM line survey between $J$=9-8 and $J$=30-29 (see Fig. \ref{figure:HC3N_V7_mix_30m} and Sect. \ref{section:column_densities} for the best fit LTE model). In the HIFI data, 30 lines from $J$=53-52 to $J$=77-76 are detected in the range of the survey (see Fig. \ref{figure:HC3N_V7_mix_hifi}). From line profiles at mm frequencies, we see that the line intensity increases with the rotational quantum number $J$, whereas in the HIFI lines the opposite occurs (see Table \ref{table:tab_HC3N_vib} where spectroscopic parameters for HC$_{3}$N and its vibrational modes\footnote{Spectroscopic parameters for HC$_{3}$N (dipole moment $\mu$=3.73 D) have been obtained from  DeLeon et al. (1985), de Zafra (1971), Mbosei (2000), Creswell et al. (1977), Chen et al. (1991), Yamada (1995), and Thorwirth et al. (2000). For $^{13}$C isotopologues they have been obtained from Creswell et al. (1977), Mallinson \& de Zafra (1978), and Thorwirth et al. (2001). For the vibratonal mode $\nu$$_{7}$ and its $^{13}$C isotopologues, they were obtained from Thorwirth et al. (2000) and Thorwirth et al. (2001), while the spectroscopic parameters for the rest of the vibrational modes were obtained from Yamada \& Creswell (1986), Thorwirth et al. (2000), and Mbosei (2000).} are shown). The intensity of each line depends (among other paremeters) on the energy of the level and mainly on the frequency, but the covered frequency range for a given transition range is much larger in the case of HIFI than in IRAM-30m. This behaviour produces different trends in the lines observed with both telescopes. 
On the other hand, it is evident that a wide component, as well as a narrow compontent centered at $\simeq$5 km s$^{-1}$, play an important role in the contribution to the emission of all these lines. 

With respect to the mode 2$\nu$$_{7}$ of HC$_{3}$N, there are 46 detected lines in the range covered by the IRAM survey. These transitions cover the range $J$=9-8 to $J$=30-29 and are shown in Fig. \ref{figure:HC3N_2V7_mix_30m} (see Appendix and Sect. \ref{section:column_densities} for the best fit LTE model). Within the Herschel line survey, there are 19 detected lines (see Fig. \ref{figure:HC3N_2V7_mix_hifi}) covering the frequency range of 484-660 GHz (higher transitions are too weak and/or blended by stronger emission from other species). From all line profiles and the range of energy ($E$$_{\mathrm{up}}$=665-1800 K), we deduce that the main contribution must arise from a narrow and hot component.

27 transitions of HC$_{3}$N 3$\nu$$_{7}$ are in the frequency range of the survey at 3, 2 and 1.3 mm (Table \ref{table:tab_HC3N_vib}). Figure \ref{figure:HC3N_3V7_mix_30m} shows some of these lines. In this vibrational mode there are four $l$-transitions, although our spectral resolution can not always can resolve all of them. The lines observed with Herschel are very weak and mostly hidden by stronger species, therefore we do not consider detection of this vibrational mode of HC$_{3}$N with the Herschel telescope.

We have detected 15 transitions of HC$_{3}$N $\nu$$_{6}$ in the IRAM line survey and 13 in the HIFI survey. Lines at 3 mm are weak, with $T$$_{\mathrm{MB}}$$\sim$0.3 K, while lines at 2 and 1.3 mm present intensities between 0.5 and 2.5 K (see Fig. \ref{figure:HC3N_V6_mix_30m} in Appendix). Similar to the two previous cases, lines from the Herschel survey are weak and/or blended with stronger species (see Fig. \ref{figure:HC3N_V6_mix_hifi}).
We also detect 42 transitions of $\nu$$_{6}$+$\nu$$_{7}$ (Fig. \ref{figure:HC3N_V6+V7_mix_30m}), dominated by contributions coming from narrow components. We do not detect lines from these vibrational modes with HIFI, due to the high energies of the levels (up to 1500 K) and to the presence of stronger lines.

We detect in total 22 lines of HC$_{3}$N $\nu$$_{5}$ (see Table \ref{table:tab_HC3N_vib}, and a selection of these lines in Fig. \ref{figure:HC3N_V5_mix_30m}). The rest of them are hidden by species with higher intensity and we do not consider them in this work. Given the high energies ($>$970 K) of these transitions, we deduce that the emission should arise from a very hot component. In addition, from the line profiles this component must be narrow ($\leq$10-11 km s$^{-1}$).

We have also observed several lines, both of the ground and the vibrational state $\nu$$_{7}$, arising from the 13C isotopic substitutions of HC$_{3}$N. In Table \ref{table:tab_HC3N_vib} (see Appendix) we list all detected transitions for the $^{13}$C isotopologues of the ground state and their energies. Within the IRAM survey, we detected 13 lines of H$^{13}$CCCN (Fig. \ref{figure:H13CCCN_mix_30m}), 18 lines of HC$^{13}$CCN (Fig. \ref{figure:HC13CCN_mix_30m}), and 16 lines of HCC$^{13}$CN (Fig. \ref{figure:HCC13CN_mix_30m}). From the Herschel line survey, we detected 21 lines from isotopologues in total. 
From their line profiles, we infer a contribution from a very narrow component which mainly affects the lowest frequencies, and a wide component important for transitions with $J$$\textgreater$20. 
For the mode $\nu$$_{7}$ we detected, (only in the IRAM survey), 47 isotopic lines (see Appendix, Figs. \ref{figure:H13CCCN_V7_mix_30m}, \ref{figure:HC13CCN_V7_mix_30m}, \ref{figure:HCC13CN_V7_mix_30m}, and Table \ref{table:tab_HC3N_vib}). The temperatures of these lines are $T$$_{\mathrm{MB}}$$\textless$0.7 K and some of them are blended with other species. We have not detected any transition arising from the $^{15}$N isotopologue.

 \subsubsection{DC$_{3}$N}

\begin{figure}
   \centering 
   \includegraphics[angle=0,width=8.8cm]{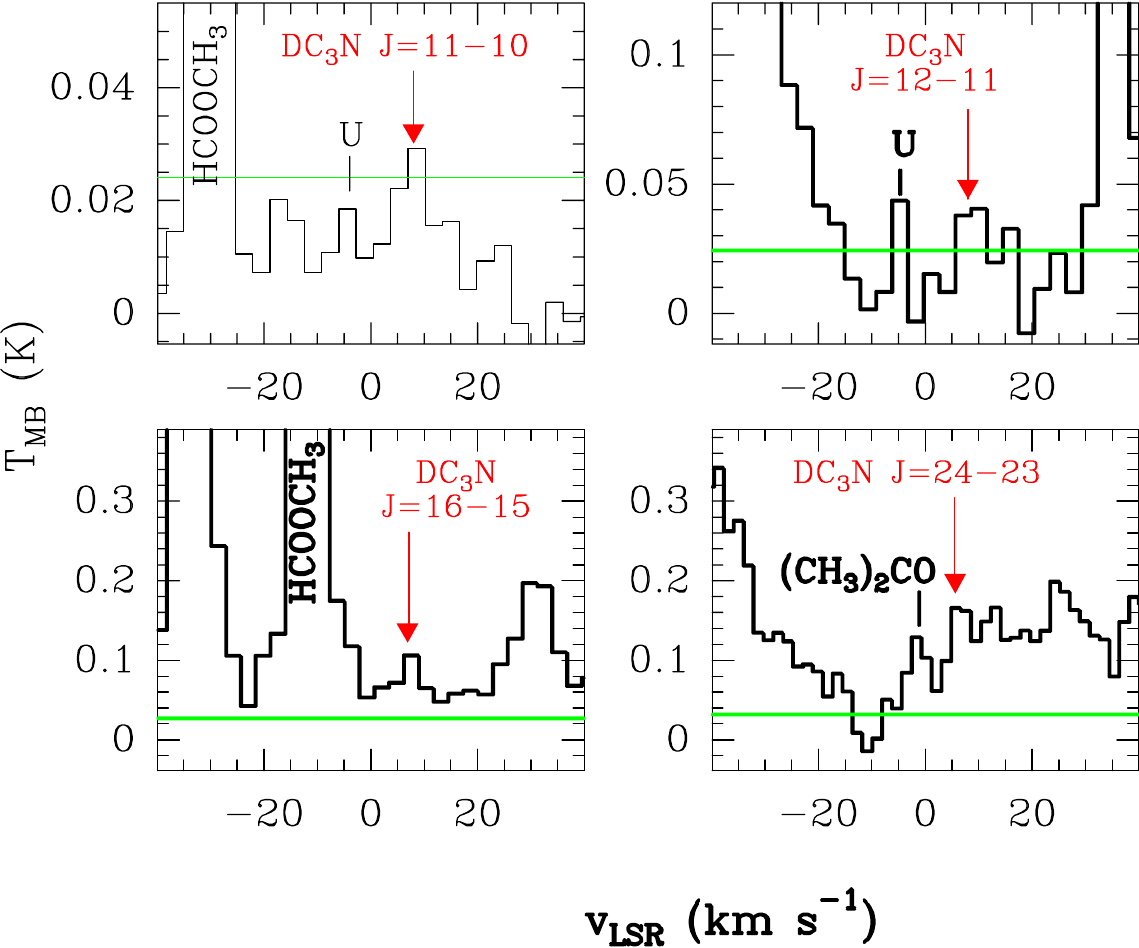}
   \caption{Tentative detection of four lines of DC$_{3}$N. The horizontal green line indicates the 3$\sigma$ threshold used for line identification.}
    \label{figure:DC3N_solo_detectadas}
   \end{figure}

\begin{figure}
   \centering 
   \includegraphics[angle=-90,width=6.0cm]{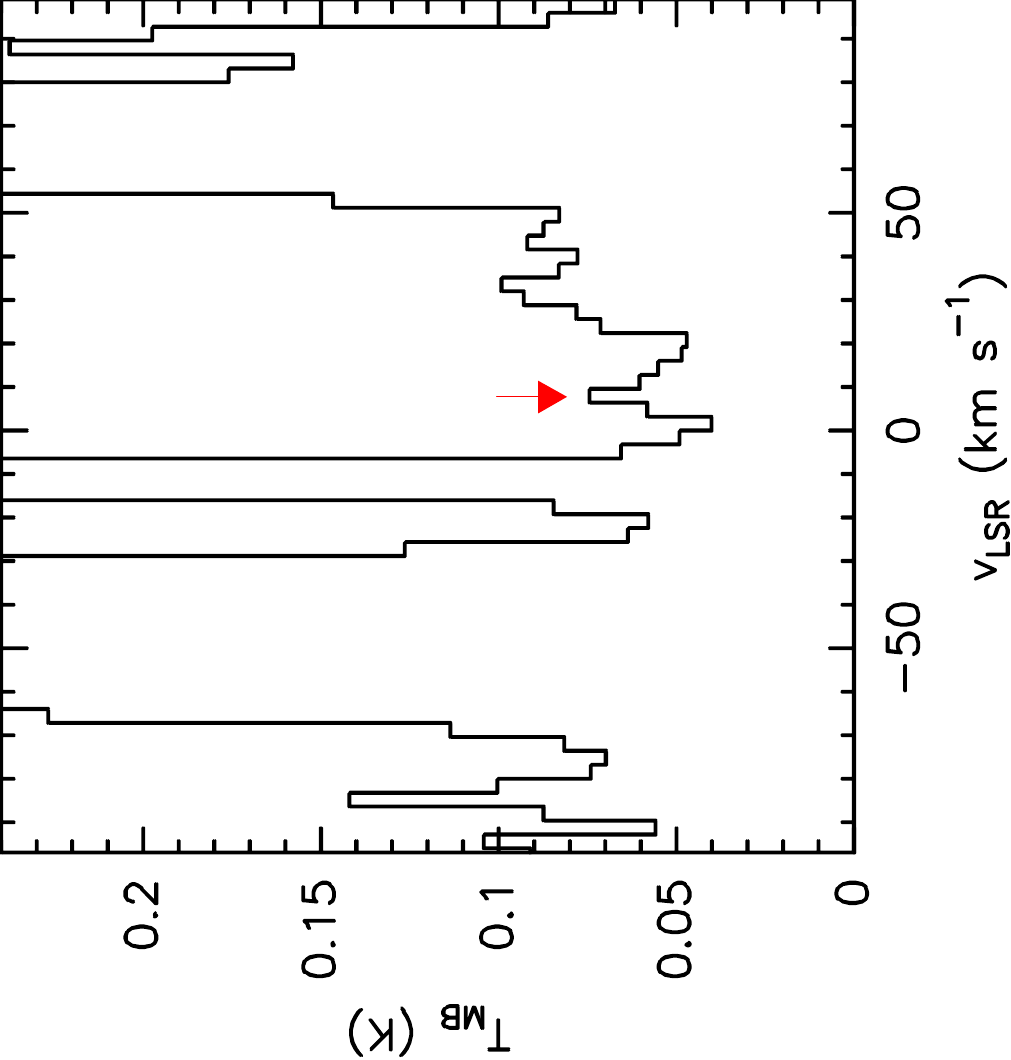}
   \caption{Average of the spectra shown in Fig. \ref{figure:DC3N_solo_detectadas}. The red arrow indicates the resulting DC$_{3}$N emission.}
    \label{figure:average_DC3N}
   \end{figure}

\begin{figure*}
   \centering 
   \includegraphics[angle=0,width=16cm]{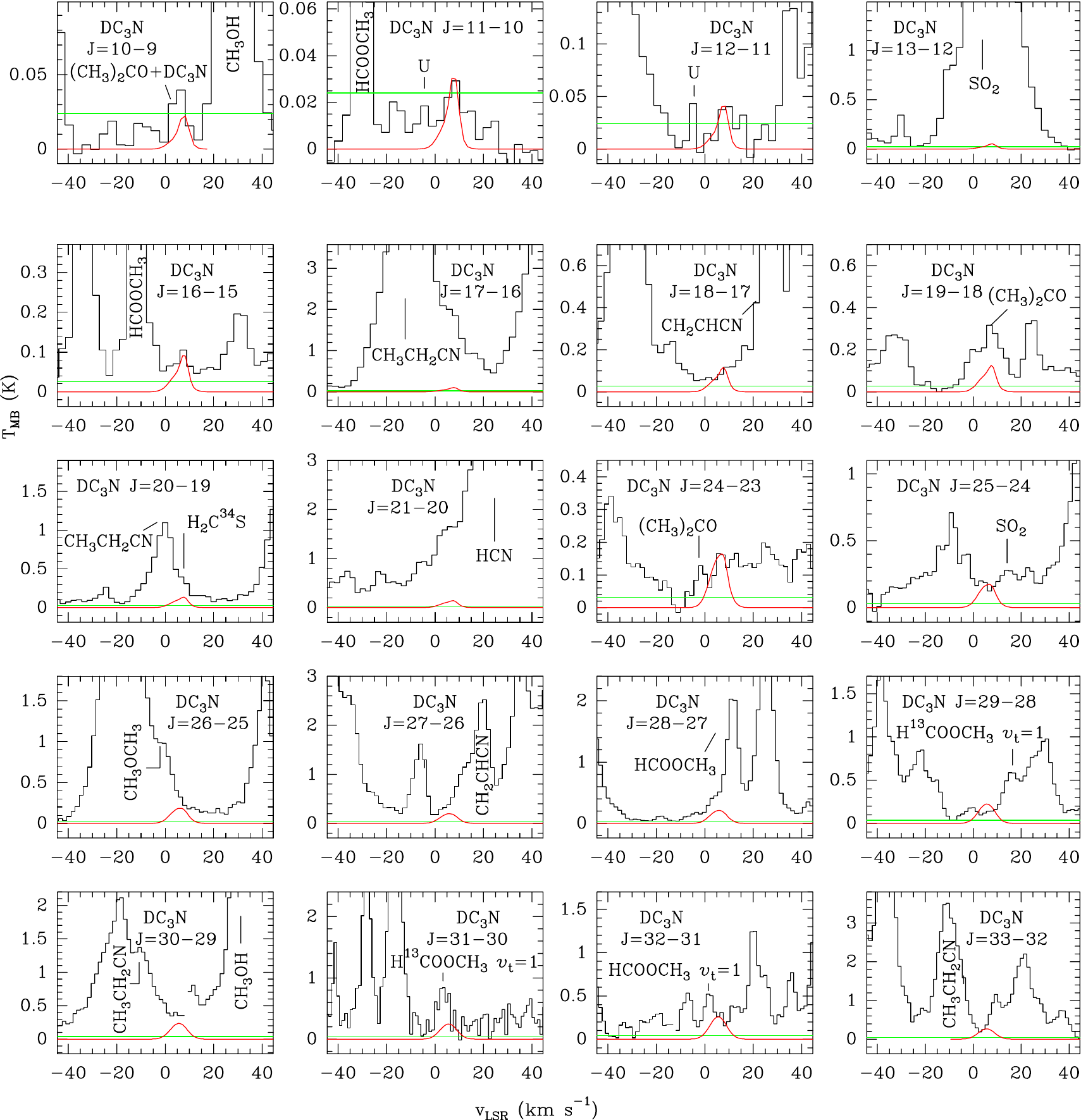}
   \caption{Observed spectra of DC$_{3}$N (black histogram) in the IRAM survey. Best fit LVG model results are shown in red. The horizontal green line indicates the 3$\sigma$ threshold used for line identification.}
    \label{figure:DC3N}
   \end{figure*}

There are 20 rotational transitions (listed in Table \ref{table:DC3N} together with their spectroscopic parameters\footnote{Spectroscopic parameters for DC$_{3}$N (dipole moment $\mu$=3.73 D) have been obtained from Mallinson \& de Zafra (1978), Tack \& Kukolich (1983), and DeLeon et al. (1985).}, see Appendix) of deuterated cyanoacetylene, that fall in the range of the IRAM survey. Four of these lines are not overlapped with other species and they are identified as DC$_{3}$N (see Fig. \ref{figure:DC3N_solo_detectadas}). 
However, in this figure we also observe that two of the four lines are very close to the detection limit, 3$\sigma$, (green horizontal line). 
In order to increase the signal-to-noise and therefore be more confident of the likely detection of DC$_{3}$N, we have averaged the four spectra (accounting for the width of the channels in each spectrum). The result is shown in Fig. \ref{figure:average_DC3N}, where the red arrow indicates the resulting line profile of the average of the four DC$_{3}$N lines. This profile presents a Gaussian shape (centered at 7.76 km s$^{-1}$), is not blended with other species, and its main beam temperature is $\textgreater$0.07 K ($>$5$\sigma$). 
This suggests that there is a molecule which has emission lines at the frequencies listed in Table \ref{table:DC3N}, producing emission consistent with the expected velocity (5-10 km s$^{-1}$, corresponding to the typical velocity range of the Orion KL components). We have also checked, using the MADEX catalog (see Sect. \ref{section:column_densities}), that no lines from other species are expected at the frequencies of DC$_{3}$N. Hence, altogether this suggests that the molecule responsible for this emission is likely DC$_{3}$N.
The four transitions, their frequencies, energies (whose range is $E$$_{\mathrm{up}}$=26-122 K), and their observed velocities $V$$_{\mathrm{LSR}}$ are listed in Table \ref{table:only_detected_DC3N}. 
For the ground state of DC$_{3}$N, we show in Fig. \ref{figure:DC3N} all the DC$_{3}$N lines within the IRAM survey, together with our best fit LVG model (see Sect. \ref{section:column_densities}). 
The line profiles of the four features, which are not overlapped with other species, indicate that the contribution comes from narrow ($<$11 km s$^{-1}$) velocity components (typical line width range of the compact ridge and the hot core). 
We have not detected lines of vibrational states nor isotopologues of DC$_{3}$N. We have also not detected lines of this species in the frequency range covered by Herschel due to their weakness and/or the presence of stronger species.

\begin{table}
\caption{Tentative lines of DC$_{3}$N.}   
\centering   
\begin{tabular}{c c c c c }  
\hline\hline       
Species & Transition & Frequency  & $V$$_{\mathrm{LSR}}$ & $E$$_{\mathrm{up}}$  \\
         &            &   (MHz)   &   (km s$^{-1}$) &   (K) \\
\hline                    
   DC$_{3}$N & 11-10 & 92872.37  & 8.6 & 26.7  \\
   DC$_{3}$N & 12-11 & 101314.82 & 9.9 & 31.6 \\
   DC$_{3}$N & 16-15 & 135083.18 & 7.4 & 55.1 \\
   DC$_{3}$N & 24-23 & 202610.90 & 6.6 & 121.6 \\ 
\hline
\label{table:only_detected_DC3N}                  
\end{tabular}
\end{table}

\subsubsection{HC$_{5}$N}

Like HC$_{3}$N, cyanodiacetylene (HC$_{5}$N) is a linear molecule, H-C$\equiv$C-C$\equiv$C-C$\equiv$N, with a large dipole moment. It was detected for the first time in the interstellar medium by Avery et al. (1976) in Sgr B2. Later it was also detected in several dark clouds and in Orion KL by Bujarrabal et al. (1981) at wavelengths near 3 mm.

\begin{figure*}[ht!]
   \centering 
   \includegraphics[angle=0,width=15.3cm]{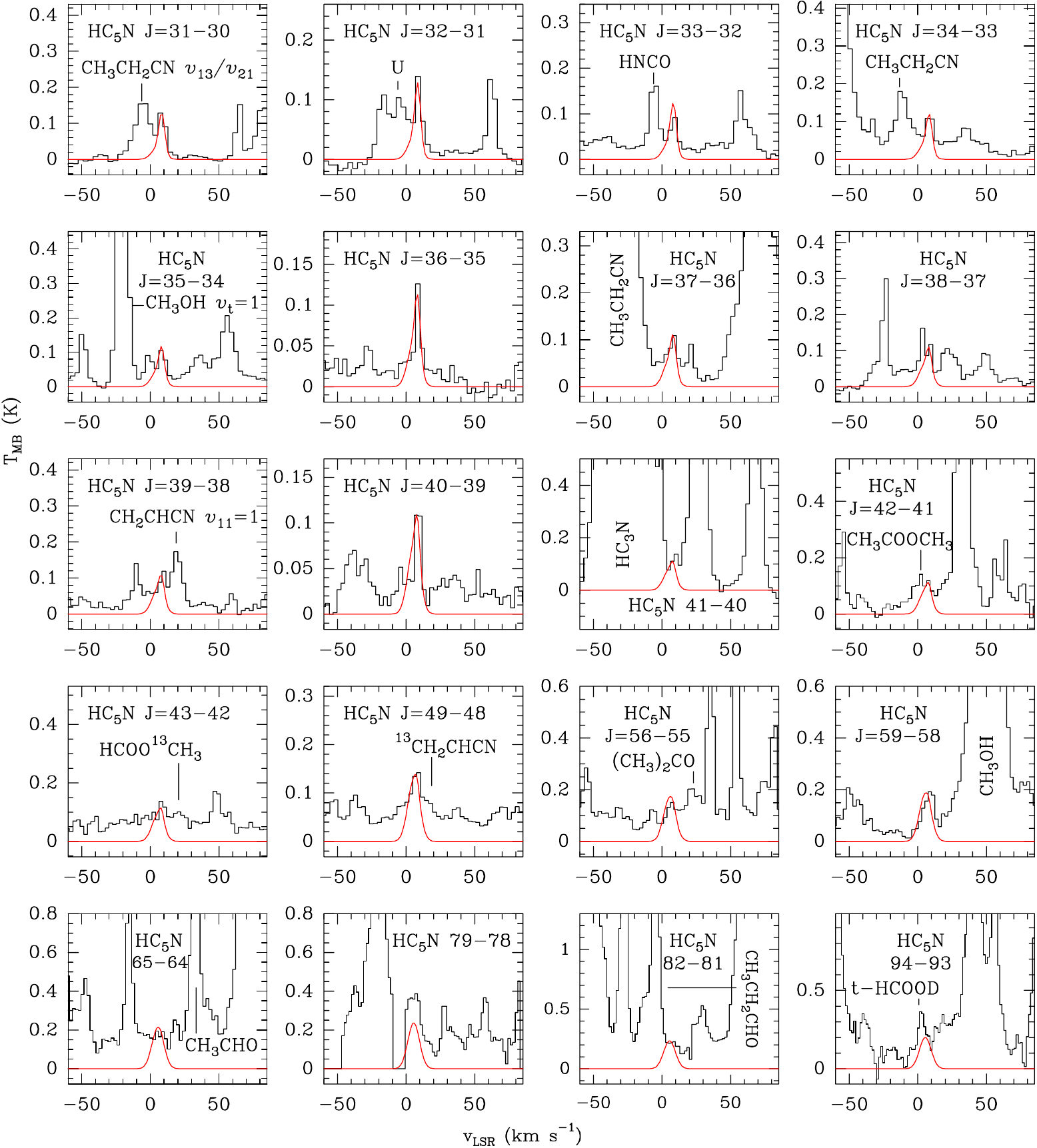}
   \caption{Sample of observed lines of HC$_{5}$N (black histogram) in the IRAM survey. Best fit LVG model results are shown in red.}
    \label{figure:HC5N_30m}
   \end{figure*}

We have detected 35 rotational lines of HC$_{5}$N in Orion KL in the IRAM line survey. Figure \ref{figure:HC5N_30m} shows a sample of transitions, together with our best fit LVG model lines (see Sect. \ref{section:column_densities}). The transitions of HC$_{5}$N cover the range $J$=30 to $J$=103 and their energies (listed in Table \ref{table:tab_HC5N_vib}, together with other spectroscopic parameters\footnote{Spectroscopic parameters for HC$_{5}$N (dipole moment $\mu$=4.33 D) have been obtained from Bizzocchi et al. (2004) and Kroto et al. (1976).}) range between $E$$_{\mathrm{up}}$=63-685 K. All lines present a small line width ($\lesssim$10 km s$^{-1}$), especially lines at 3 mm, therefore the emission must arise from narrow components, such as the ridge or the hot core. We have not detected lines in the HIFI data because of the high energies, $E$$_{\mathrm{up}}$$\textgreater$2100 K, involved in these transitions. No lines arising from the $^{13}$C isotopologues of HC$_{5}$N have been detected.

\subsection{2$\arcmin$$\times$2$\arcmin$ maps around IRc2}
\label{subsection:maps}

\begin{figure*}
   \centering 
   \includegraphics[angle=-90,width=16.8cm]{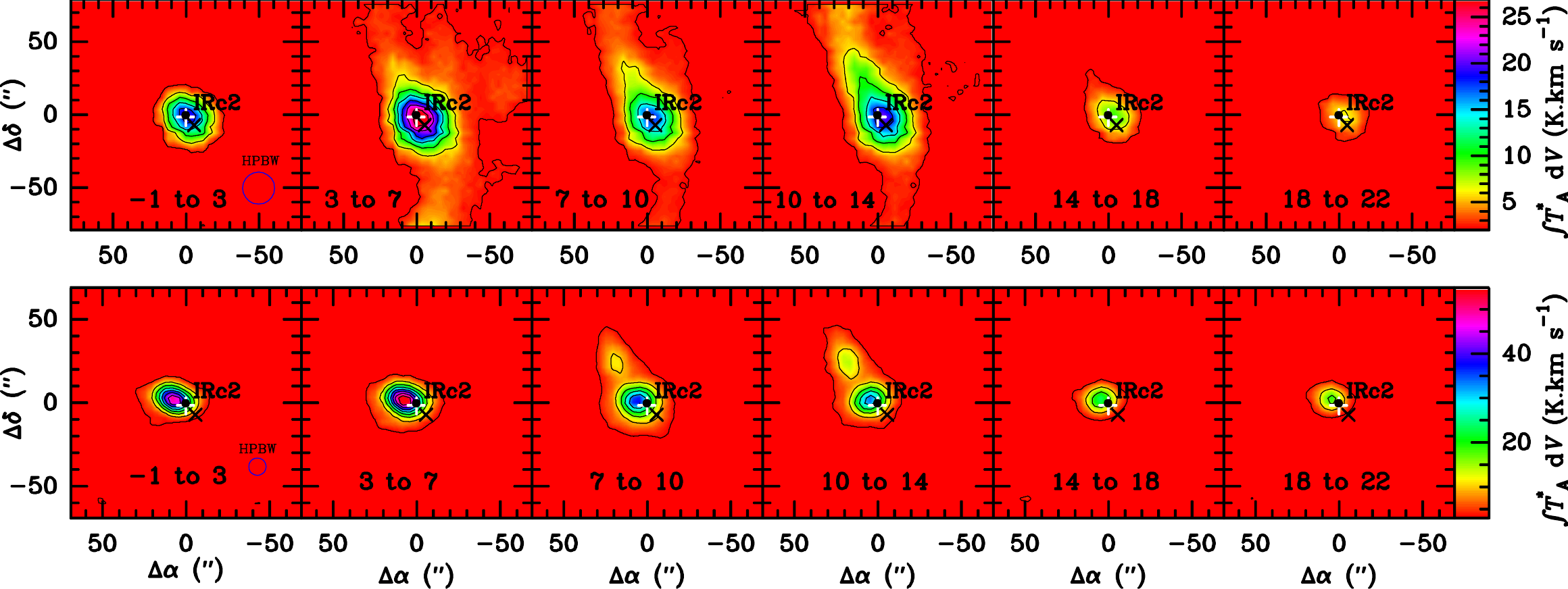}
   \caption{HC$_{3}$N line integrated intensity maps for different velocity ranges (indicated at the bottom of each panel). Row 1 shows the transition $J$=12-11 with $E$$_{\mathrm{up}}$=34.1 K. The interval between contours is 4 K km s$^{-1}$ and the minimum contour is 2 K km s$^{-1}$. Row 2 shows the transition $J$=26-25 with $E$$_{\mathrm{up}}$=153.3 K. The interval of contours is 7 K km s$^{-1}$ and the minimum contour is 3 K km s$^{-1}$. The white plus indicates the position of the hot core and the black cross the position of the compact ridge. The beam size is also shown with a blue circle.}
    \label{figure:HC3N_maps}
   \end{figure*}

From the 2D survey data of Orion KL (Marcelino et al. in preparation) taken with the IRAM-30m telescope, we present maps of the integrated intensity of HC$_3$N over different velocity ranges. Figure \ref{figure:HC3N_maps} shows the mapped transitions $J$=12-11 and $J$=26-25, which correspond to energy levels $E$$_{\mathrm{up}}$=34.1 and 153.3 K, respectively. In both transitions, the maximum integrated intensity is found in the velocity range 3-7 km s$^{-1}$ (typical velocities of the hot core). However, whereas at low energies the peak is located around the IRc2 position (in agreement with the HC$_3$N interferometric observations from Wright et al. 1996), at higher energies (lower panel) it is slightly shifted to the east for all velocity ranges. 
In the upper panel we can see extended emission along the molecular ridge in the N-S direction. In the velocity range 7-14 km s$^{-1}$ there is a clear elongation of the emission towards the NE, corresponding to the extended ridge. All this extended emission is the result of the low energy of this transition ($\sim$34 K). Maps of the $J$=26-25 transition represented in the lower panels, with an energy level four times higher, show the same elongation but the emission is less extended. 
In addition, we observe a second emission peak centered at $\sim$20$\arcsec$ east and $\sim$25$\arcsec$ north of IRc2. In Sect. \ref{section:discussion} we explain its possible origin.
At velocities larger than 14 km s$^{-1}$, the integrated intensity decreases, and becomes more compact. 



\begin{figure*}
   \centering 
   \includegraphics[angle=-90,width=16.7cm]{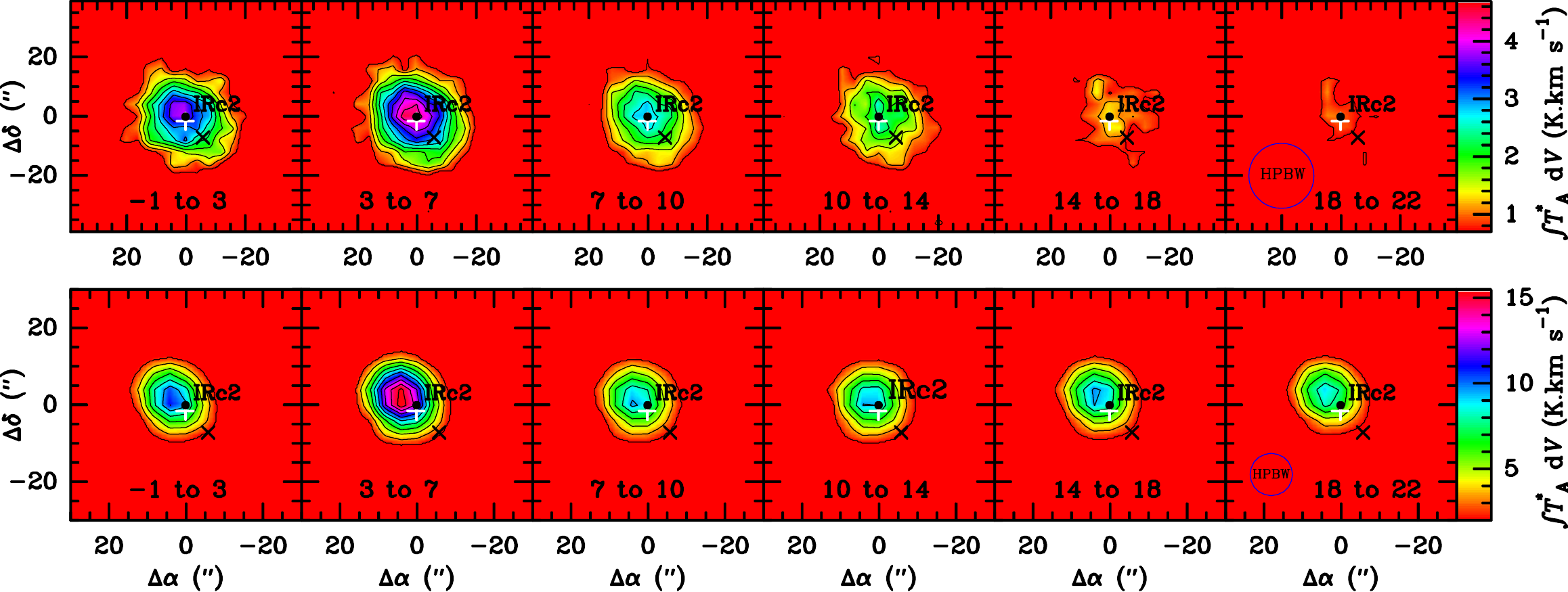}
   \caption{HC$_{3}$N $\nu$$_{7}$ line integrated intensity maps for different velocity ranges (indicated at the bottom of each panel). Row 1 shows the transition $J$=12$_{-1}$-11$_{-1}$ with $E$$_{\mathrm{up}}$=354.2 K. The interval between contours is 0.4 K km s$^{-1}$ and the minimum contour is 0.7 K km s$^{-1}$. Row 2 shows the transition $J$=25$_{-1}$-24$_{-1}$ with $E$$_{\mathrm{up}}$=462.3 K. The interval of contours is 1.5 K km s$^{-1}$ and the minimum contour is 2 K km s$^{-1}$. The white plus indicates the position of the hot core and the black cross the position of the compact ridge. The beam size is also shown with a blue circle. The velocity ranges 14-18 and 18-22 km s$^{-1}$ show integrated intensity contaminated by CH$_{2}$CHCN since there is an overlap of this species with the line of HC$_{3}$N $\nu$$_{7}$ $J$=25$_{-1}$-24$_{-1}$.}
    \label{figure:HC3N_V7_maps}
   \end{figure*}

\begin{figure*}
   \centering 
   \includegraphics[angle=-90,width=16.7cm]{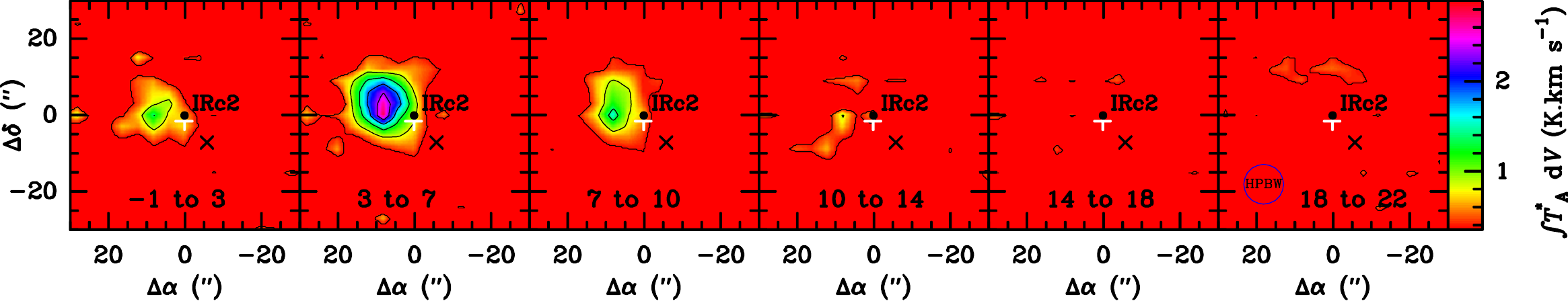}
   \caption{HC$_{3}$N v$_{6}$ line integrated intensity maps for different velocity ranges (indicated at the bottom of each panel) for the transition $J$=26$_{-1}$-25$_{-1}$ with energy $E$$_{\mathrm{up}}$=871.2 K. The interval between contours is 0.5 K km s$^{-1}$ and the minimum contour is 0.35 K km s$^{-1}$. The white plus indicates the position of the hot core and the black cross the position of the compact ridge. The beam size is also shown with a blue circle.}
   \label{figure:HC3N_V6_maps}
   \end{figure*}


Figure \ref{figure:HC3N_V7_maps} shows integrated intensity maps for two transitions of the vibrational mode HC$_3$N $\nu$$_7$ ($J$=12$_{-1}$-11$_{-1}$ and $J$=25$_{-1}$-24$_{-1}$). Similar to the ground state of HC$_3$N, the peak of emission is observed in the range 3-7 km s$^{-1}$. In this case, however, the emission presents a circular shape centered around IRc2, with the peak shifted a few arcseconds ($\sim$3$\arcsec$) towards the NE of IRc2 for the high energy transition. The same behavior was found for the mapped transition at high energy of the ground state of HC$_3$N, indicating the possible presence of a warmer region in this direction.
Another feature is that for high velocities ($V$$_{\mathrm{LSR}}$$\textgreater$14.5 km s$^{-1}$), we barely find emission of HC$_3$N $\nu$$_7$ from the low frequency transition (upper panel), while for the high frequency transition we observe an important contribution from this species. This line is overlapped with CH$_2$CHCN (24$_{(5,19)}$-23$_{(5,18)}$), see Fig. \ref{figure:HC3N_V7_mix_30m}, at high velocities (15-25 km s$^{-1}$). Therefore, the emission that we obtain in the maps in these ranges is emission from CH$_2$CHCN instead of HC$_3 $N $\nu$$_7$.

We also present an integrated emission map for the transition $J$=26$_{-1}$-25$_{-1}$ of HC$_3$N $\nu$$_6$ (see Fig. \ref{figure:HC3N_V6_maps}). The largest contribution to the emission also comes from the velocity range 3-7 km s$^{-1}$. All the emission is located to the NE of IRc2; unlike the ground state of HC$_3$N and its vibrational mode $\nu$$_7$, there is no emission from the west of IRc2. We do not find emission either in the range 14-22 km s$^{-1}$. 

\section{Analysis}
\label{section:column_densities}

We have analyzed the non-LTE excitation and radiative transfer of the DC$_{3}$N, HC$_{3}$N, and HC$_{5}$N lines. We used the LVG (large velocity gradient) approximation, where the width of the lines is due to the existence of large velocity gradients along the line of sight through the cloud, so the speed variation between relatively close points is greater than the local velocity dispersion. Only for the hot core do we consider local thermodynamic equilibrium (LTE) approximation, due to the higher gas density so that most of transitions are thermalized to the gas temperature ($T$$_{\mathrm{rot}}$$\simeq$$T$$_{\mathrm{K}}$).
The code (MADEX) has been developed by Cernicharo (2012), where corrections for the beam dilution of each line, depending on the different beam sizes at different frequencies, are included. In addition, this code takes into account the size of each component and its offset position with respect to IRc2. The LVG models are based on the Goldreich \& Kwan (1974) formalism.


In order to apply the LVG code, we have considered the same cloud components for Orion KL used in Esplugues et al. (2013) for the analysis of SO and SO$_{2}$. Most of these cloud components are considered as well in previous Orion KL interferometric studies such as Wright et al. (1996) and Beuther et al. (2005). Taking into account the mapped transitions and the parameters obtained from the Gaussian fits (see Sect. \ref{section:HC3N_other_sources}) of the line profiles of HC$_{3}$N, we assume uniform physical conditions of density, kinetic temperature, line-width and radial velocity for each component (see Table \ref{table:components_LVG}), so that the column density of each observed species is the only free parameter. However, in order to fit the lines profiles of HC$_{3}$N (ground and vibrational states) obtained with Herschel observations, a density and temperature stratification/gradient in the hot core is needed (as was predicted by Esplugues et al. (2013) with IRAM observations of SO$_{2}$ transitions with high energies, $>$700 K). This gradient is simulated by adding a hotter inner component to the typical hot core conditions (see Table \ref{table:components_LVG}). 
The final considered fit is the one that reproduces more line profiles better from transitions covering a ide energy range within a $\sim$20$\%$ of the uncertainty in the line intensity.
Given the different sources of uncertainty such as spatial overlap of the cloud components, modest angular resolution, and the lack of collisional rates for some vibrational modes such as $\nu$$_{7}$, 2$\nu$$_{7}$, and 3$\nu$$_{7}$ to consider LVG instead of LTE in components different to the hot core discussed in Esplugues et al. (2013), we have estimated the uncertainty in the model intensity predictions to be 25$\%$ for the HC$_{3}$N results, 30$\%$ for HC$_{5}$N (higher uncertainty for HC$_{5}$N with respect to HC$_{3}$N because of the lines are weaker and more blended with other species), and 35$\%$ for DC$_{3}$N (due to the overlap with other species) and for the vibrational modes of   HC$_{3}$N (because of considering LTE instead of LVG).
In all the fits, we have checked the contribution of each component separately from the rest. The adopted fit is that which best reproduces the majority of the observed line profiles for both surveys (IRAM and HIFI), within the $\sim$15$\%$ of uncertainty in the intensity (see Esplugues et al. 2013 for more details about the technique used to fit the line profiles).

\subsection{HC$_{3}$N}
\label{subsection:HC3N}

\begin{table*}
\caption{Physical parameters adopted in the radiative transfer code for the Orion KL cloud components.}             
\centering   
\begin{tabular}{llllllll}     
\hline\hline 
 &   Source             & \multicolumn{2}{c}{Offset$_{\mathrm{(IRc2)}}$} &  &  &  &  \\     

 Component  &  diameter & ${\mathrm{IRAM}}$ & ${\mathrm{HIFI}}$  & $n$(H$_{2}$)  &  $T$$_{\mathrm{K}}$ & $\triangle$$V$$_{\mathrm{FWHM}}$ & $V$$_{\mathrm{LSR}}$ \\
           & ($\arcsec$)     & ($\arcsec$)  & ($\arcsec$)      &  (cm$^{-3}$)   & (K) & (km s$^{-1}$) & (km s$^{-1}$) \\  
\hline 
                   
   Extended ridge (ER)              & 120  & 0  & 0    & 10$^{5}$    & 60  & 4   & 8.5 \\  
   Compact ridge (CR)               & 15   & 7  & 3    & 10$^{6}$    & 110 & 3   & 8 \\
   High velocity plateau (HP)       & 30   & 4  & 4    & 10$^{6}$    & 100 & 30  & 11\\
   Plateau (PL)                     & 20   & 0  & 0    & 5$\times$10$^{6}$  & 150 & 25  & 6 \\
   Outer hot core (HC1)             & 10   & 2  & 2    & 1.5$\times$10$^{7}$& 220 & 10  & 5.5 \\
   Inner hot core (HC2)                   & 7    & 4  & 4    & 5$\times$10$^{6}$  & 310 & 7   & 5.5\\
   20.5 km s$^{-1}$ component       & 5    & 3  & 3    & 5$\times$10$^{6}$  & 90  & 7.5 & 20.5 \\
\hline
\label{table:components_LVG}                  
\end{tabular}
\end{table*}

\begin{table*}
\caption{Column densities, $N$, for HC$_{3}$N (its isotopologues and vibrationally excited states, including DC$_{3}$N) and HC$_{5}$N.}             
\centering   
\begin{tabular}{c c c c c c c c}     
\hline\hline       
Species                   & Extended ridge & Compact ridge & High velocity   & Plateau & Outer hot  & Inner hot core & 20.5 km s$^{-1}$ \\ 
                          & (ER)           & (CR)          & plateau (HP)    & (PL)    & core (HC1)          & (HC2)    & component\\ 
                          &$N$$\times$10$^{14}$(cm$^{-2}$) &$N$$\times$10$^{14}$(cm$^{-2}$) &$N$$\times$10$^{14}$(cm$^{-2}$) &$N$$\times$10$^{14}$(cm$^{-2}$) &$N$$\times$10$^{14}$(cm$^{-2}$) &$N$$\times$10$^{14}$(cm$^{-2}$) &$N$$\times$10$^{14}$(cm$^{-2}$) \\
\hline                    
 HC$_{3}$N                & 0.8$\pm$0.2       & 2.0$\pm$0.5     & 5$\pm$1        & 7$\pm$2         & 10$\pm$3       & 7$\pm$2       & 3.4$\pm$0.6 \\  
 H$^{13}$CCCN             & 0.03$\pm$0.01     & 0.08$\pm$0.02   & 0.20$\pm$0.06  & 0.3$\pm$0.1      & 0.6$\pm$0.2    & 2.4$\pm$0.7   & ...           \\
 HC$^{13}$CCN             & 0.010$\pm$0.003   & 0.10$\pm$0.03   & 0.20$\pm$0.06  & 0.10$\pm$0.03   & 0.7$\pm$0.2    & 1.2$\pm$0.4   & ...           \\
 HCC$^{13}$CN             & 0.010$\pm$0.003   & 0.09$\pm$0.03   & 0.20$\pm$0.06  & 0.10$\pm$0.03   & 1.0$\pm$0.3    & 0.9$\pm$0.3   & ...           \\
 HC$_{3}$N $\nu$$_{5}$    & 0.03$\pm$0.01     & ...             & ...            & ...             & ...            & 3$\pm$1       & ...           \\
 HC$_{3}$N $\nu$$_{6}$    & 0.015$\pm$0.005   & 0.08$\pm$0.03   & ...            & 1.0$\pm$0.4     & 0.5$\pm$0.2    & 3$\pm$1       & ...           \\
 HC$_{3}$N $\nu$$_{7}$    & 0.25$\pm$0.09     & 0.10$\pm$0.04   & ...            & 4$\pm$1         & 8$\pm$3        & 6$\pm$2       & ...           \\
 HC$_{3}$N 2$\nu$$_{7}$   & 0.08$\pm$0.03     & 0.08$\pm$0.03   & ...            & 0.4$\pm$0.1     & 1.5$\pm$0.5    & 5$\pm$2       & ...           \\
 HC$_{3}$N 3$\nu$$_{7}$   & 0.010$\pm$0.004   & 0.05$\pm$0.02   & ...            & 0.20$\pm$0.07   & 0.9$\pm$0.3    & 1.2$\pm$0.4   & ...           \\
 HC$_{3}$N $\nu$$_{6}$+$\nu$$_{7}$ & 0.015$\pm$0.005 & 0.025$\pm$0.009 & ...     & 0.07$\pm$0.02   & ...            & 0.15$\pm$0.05 & ...          \\
 H$^{13}$CCCN $\nu$$_{7}$ & 0.013$\pm$0.005   & 0.010$\pm$0.004 & ...            & 0.10$\pm$0.04   & 0.25$\pm$0.09  & 0.3$\pm$0.1   & ...           \\
 HC$^{13}$CCN $\nu$$_{7}$ & 0.013$\pm$0.005   & 0.010$\pm$0.004 & ...            & 0.10$\pm$0.04   & 0.25$\pm$0.09  & 0.3$\pm$0.1   & ...           \\
 HCC$^{13}$CN $\nu$$_{7}$ & 0.013$\pm$0.005   & 0.010$\pm$0.004 & ...            & 0.10$\pm$0.04   & 0.25$\pm$0.09  & 0.3$\pm$0.1   & ...           \\
\hline 
 DC$_{3}$N                & ...               & 0.027$\pm$0.009 & ...           &  ...             & 0.15$\pm$0.05  & ...           & ...          \\
\hline 
 HC$_{5}$N                & 0.05$\pm$0.02     & 0.030$\pm$0.009 & ...            &  ...            & 0.7$\pm$0.2    & ...           & ...          \\
\hline 
\label{table:HC3N_column_densities}                 
\end{tabular}
\end{table*}

The results of the LVG code indicate that the HC$_{3}$N lines present typical optical depths $\tau$$<$0.5. The column density of this molecule has been derived using HC$_{3}$N-pH2 collisonal rates from Wernli (2007), and the modeled line profiles are shown in Figs. \ref{figure:HC3N_mix_30m} (IRAM) and \ref{figure:HC3N_mix_hifi} (Herschel).
The cloud component with the largest HC$_{3}$N column density is the hot core HC1, with $N$(HC$_{3}$N)=(1.0$\pm$0.2)$\times$10$^{15}$ cm$^{-2}$. This agrees with the result obtained by Wright et al. (1996), White et al. (2003), and Persson et al. (2007) who found HC$_{3}$N column densities of 2$\times$10$^{15}$, 10$^{15}$, and 1.5$\times$10$^{15}$ cm$^{-2}$, respectively. The inner hot core (HC2) and the plateau (PL), with assumed temperatures of $T$$_{\mathrm{k}}$=310 and 150 K respectively, present the same column density: $N$(HC$_{3}$N)=(7$\pm$1)$\times$10$^{14}$ cm$^{-2}$. The HC2 component is most visible in the transitions observed with HIFI (transitions with high energy), whereas the HC1 and PL components mainly affect lines at 1.3 and 2 mm. 
Another component with important contribution to the emission is the high velocity plateau, whose column density is $N$(HC$_{3}$N)=(5$\pm$1)$\times$10$^{14}$ cm$^{-2}$. This component contributes significantly to the intensity of lines at 3 mm. 
In order to obtain a good fit to the lines at low frequencies, we also needed to consider a contribution from the extended ridge, with a column density of $N$(HC$_{3}$N)=(8$\pm$2)$\times$10$^{13}$ cm$^{-2}$.
In addition, similar to the analysis of SO and SO$_{2}$ emission in Esplugues et al. (2013), we also found a contribution from the component centered at 20.5 km s$^{-1}$, with $N$(HC$_{3}$N)=(3.4$\pm$0.6)$\times$10$^{14}$ cm$^{-2}$. This contribution is negligible in the isotopologues or vibrational modes of HC$_{3}$N, but it is significant for ground state transitions as we can see in the line profiles in Fig. \ref{figure:HC3N_mix_30m}, in particular for $J$$\geq$15.

For the vibrational mode $\nu$$_{7}$ there are no collisional rates available, so we adopted LTE approximation. As for the ground state of HC$_{3}$N, we found the largest column density in the outer hot core, HC1, with a value of (8$\pm$2)$\times$10$^{14}$ cm$^{-2}$. In HC2, the component with the highest temperature, we found a similar column density, (6$\pm$1)$\times$10$^{14}$ cm$^{-2}$.
The plateau component with higher temperature (PL) mainly contributes to the emission of lines at 2 and 3 mm with $N$(HC$_{3}$N)=(4.0$\pm$0.8)$\times$10$^{14}$ cm$^{-2}$. This value is about two times lower than that obtained for the ground state. In the extended and compact ridge, the column densities are similar to those obtained for the $^{13}$C isotopologues of HC$_{3}$N (discussed below).
Figure \ref{figure:HC3N_V7_mix_30m} shows our best LTE model fits for lines observed in the IRAM survey. Fits for lines from HIFI are shown in the Appendix (Fig. \ref{figure:HC3N_V7_mix_hifi}).
From Fig. \ref{figure:HC3N_V7_mix_30m}, we observe that lines at 3 mm present a poor fit at negative velocities. The same behaviour is also observed in Fig. \ref{figure:HC3N_mix_30m} for the ground state of HC$_{3}$N. This could be due to contributions from other blended species, as in the case of $J$=11-10, which is blended with HCOOCH$_{3}$, however, the existence of emission arising from a component centered at $V$$_{\mathrm{LSR}}$$\sim$0 km s$^{-1}$ seems more likely. We are able to fit this part of the line profiles by introducing a new component in the model centered at -1 km s$^{-1}$, with $n$(H$_{2}$)=10$^{5}$ cm$^{-3}$, $T$$_{\mathrm{k}}$=60 K, and a line-width of 10 km s$^{-1}$. Given the low temperature, this component seems to represent a gradient of the extended ridge. However, $\triangle$$V$$_{\mathrm{FWHM}}$=10 km s$^{-1}$ is more than twice the typical line width for the ridge, which could indicate effects from outflows. Finally, we have not to included this component in the final model because of its unknown origin and the low certainty of its real existence, since it could also be due to contamination from other species.

We have assumed LTE excitation due to the lack of collisional rates for the vibrational modes 2$\nu$$_{7}$ and 3$\nu$$_{7}$ of HC$_{3}$N. They present column densities similar to the $\nu$$_{7}$ mode in the inner hot core HC2, in particular the mode 2$\nu$$_{7}$, with $N$=(5$\pm$1)$\times$10$^{14}$ cm$^{-2}$. However, in HC1, the contribution to the emission of 2$\nu$$_{7}$ and 3$\nu$$_{7}$ is up to eight times lower than that of $\nu$$_{7}$ (see Table \ref{table:HC3N_column_densities}). The lowest contributions for both modes come from the extended and the compact ridge. HC$_{3}$N 2$\nu$$_{7}$ is between 1.6 and 8 times more abundant than HC$_{3}$N 3$\nu$$_{7}$.
Figures \ref{figure:HC3N_2V7_mix_30m} and \ref{figure:HC3N_2V7_mix_hifi} show our best fit lines of 2$\nu$$_{7}$ from IRAM and HIFI, respectively, and Fig. \ref{figure:HC3N_3V7_mix_30m} shows the fits to lines of 3$\nu$$_{7}$ from the IRAM line survey.

Figures \ref{figure:HC3N_V5_mix_30m}, \ref{figure:HC3N_V6_mix_30m}, \ref{figure:HC3N_V5_mix_hifi}, and \ref{figure:HC3N_V6_mix_hifi} (see Appendix) show the modeled profiles for the vibrational modes $\nu$$_{5}$ and v$_{6}$ of HC$_{3}$N, observed with the IRAM and Herschel surveys. These fits have also been made considering LTE approximation, due to the lack of collisional rates.
For both vibrational states, the emission comes mainly from the hottest region (HC2), with column densities $\sim$3$\times$10$^{14}$ cm$^{-2}$ (see Table \ref{table:HC3N_column_densities}). This component mainly affects the lines in the HIFI range and the lines at 1.3 mm. In the case of $\nu$$_{6}$, we also find an important contribution from the plateau (PL), with a column density of (1.0$\pm$0.2)$\times$10$^{14}$ cm$^{-2}$. In the case of $\nu$$_{5}$, the line profiles are very narrow and it was not necessary to include any contribution from the plateau.

We have also fitted the lines of the state $\nu$$_{6}$+$\nu$$_{7}$ of HC$_{3}$N (Fig. \ref{figure:HC3N_V6+V7_mix_30m} in the Appendix), with LTE models owing to the absence of collisional rates. We found that the largest contribution arises from the hot core (HC2), with $N$(HC$_{3}$N $\nu$$_{6}$+$\nu$$_{7}$)=(1.5$\pm$0.5)$\times$10$^{13}$ cm$^{-2}$. 
Due to the high energies of all transitions, the contribution from the component HC1 is negligible compared to that from HC2.
We found that the lines in the HIFI data were blended with stronger species, therefore we did not consider any detection of HC$_{3}$N $\nu$$_{6}$+$\nu$$_{7}$ with this instrument.

The emission from the $^{13}$C isotopologues of HC$_{3}$N is dominated by the contribution from the inner hot core with $N$=(0.9-2.4)$\times$10$^{14}$ cm$^{-2}$ for all of them, followed by emission from the HC1, where we find $N$=(6$\pm$1)$\times$10$^{13}$ cm$^{-2}$.
To fit the line profiles at 2 and 3 mm it was necessary to consider a contribution from the plateau PL, where $N$=(3.3$\pm$0.6)$\times$10$^{13}$ cm$^{-2}$ for H$^{13}$CCCN and $N$=(1.0$\pm$0.2)$\times$10$^{13}$ cm$^{-2}$ for the other two $^{13}$C isotopologues. These values are very similar to those obtained for the high velocity plateau as well.
The column densities obtained for the extended ridge are about one order of magnitude lower, $N$=(1-3)$\times$10$^{12}$ cm$^{-2}$ for the $^{13}$C isotopologues.
Figures \ref{figure:H13CCCN_mix_30m}, \ref{figure:HC13CCN_mix_30m}, and \ref{figure:HCC13CN_mix_30m} (see Appendix) show our best fit LVG model for each isotopologue observed with the IRAM-30m, and Figs. \ref{figure:H13CCCN_mix_hifi}, \ref{figure:HC13CCN_mix_hifi}, and \ref{figure:HCC13CN_mix_hifi} show the results for the $^{13}$C isotopologue lines from HIFI. The collisional rates used have been taken from HC$_{3}$N-pH$_{2}$ (Wernli et al. 2007).
All lines from HC$_{3}$$^{15}$N except one, are hidden by other stronger species. Therefore, we consider the column density obtained, $N$(HC$_{3}$$^{15}$N)=4.5x10$^{13}$ cm$^{-2}$, as an upper limit. 
We have also calculated the column densities of the $^{13}$C isotopologues of HC$_{3}$N $\nu$$_{7}$. The main contribution to the emission arises from the hot core (HC1 and HC2), while in the ridge (compact and extended) we found column densities about one order of magnitude lower (see Table \ref{table:HC3N_column_densities} and Figs. \ref{figure:H13CCCN_V7_mix_30m}, \ref{figure:HC13CCN_V7_mix_30m}, and \ref{figure:HCC13CN_V7_mix_30m}).

\begin{figure}
   \centering 
   \includegraphics[angle=-90,width=7cm]{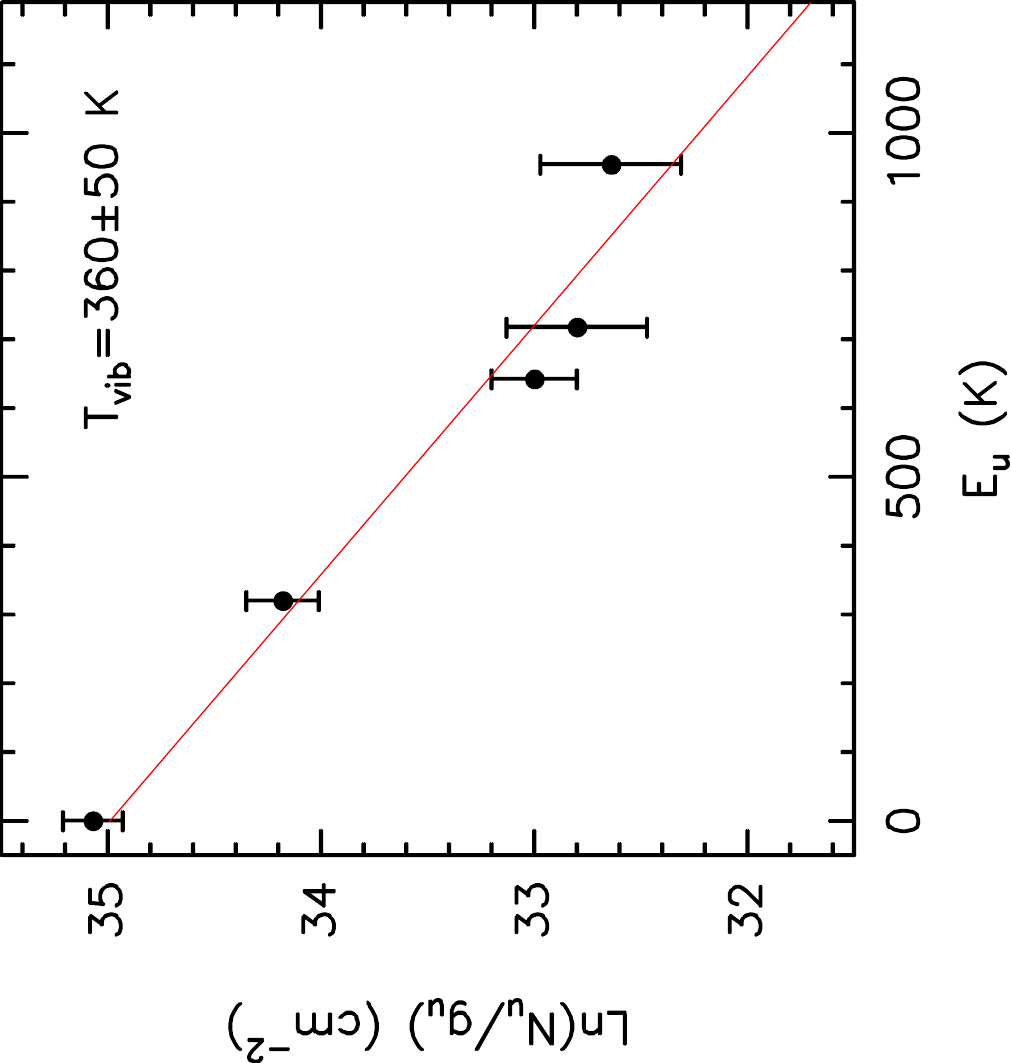}
   \caption{Column densities, Ln($N$$_{u}$/$g$$_{u}$), of HC$_{3}$N $\nu$$_{0}$, $\nu$$_{5}$, $\nu$$_{6}$, $\nu$$_{7}$, and 2$\nu$$_{7}$ vibrational states versus vibrational energy $E$$_{u}$ (K). $g$$_{u}$ is the degeneracy of each mode.}
    \label{figure:temperature_vibrational}
   \end{figure}

We have also calculated the rotation temperature for the vibrational lines (vibrational temperature) in the hot core (HC1+HC2). We have represented in Fig. \ref{figure:temperature_vibrational} the column densities obtained for the ground state and the vibrational modes $\nu$$_{5}$, $\nu$$_{6}$, $\nu$$_{7}$, and 2$\nu$$_{7}$ (see Table \ref{table:HC3N_column_densities}) versus their vibrational energies (0, 954.48, 718.13, 320.45, and 642.67 K, respectively). From a linear fit of the points, we obtain a vibrational temperature of $T$$_{\mathrm{vib}}$=360$\pm$50 K. 
This value is similar to the kinetic temperature we have adopted for the hot core HC2 (310 K). Since vibrational excitation is expected to strongly depend on temperature and density gradients in the region, it is consistent that we find the strongest emission from vibrational modes in the hot core.


\subsection{DC$_{3}$N}

To model the lines of DC$_{3}$N we have assumed LTE conditions for the hot core and LVG calculations for the rest of the components, as we did for HC$_{3}$N, using collisional rates from Wernli et al. (2007) (HC$_{3}$N-p-H$_{2}$).  
Given that this molecule had previously been only detected in dark clouds, we first tried to use only components with rather low temperatures ($T$$_{\mathrm{K}}$$\sim$60 K). However, it was impossible to fit the lines at high frequencies, so we had to consider hotter components in our model. In addition, these hot components should also be narrow (given the observed line widths). 
For DC$_{3}$N we found that only the hot core (HC1) and the compact ridge (CR) are responsible for the emission. The CR affects only the lines at 2 and 3 mm, whereas HC1 contributes to all lines, although mainly to lines at 1.3 mm (Fig. \ref{figure:DC3N}). In this case we have changed the line-width for the hot core (HC1) component with respect to the parameters given in Table \ref{table:components_LVG} in order to better reproduce the line profiles. The adopted value is $\triangle$$V$$_{\mathrm{FWHM}}$=8 km s$^{-1}$. The column densities obtained are (1.5$\pm$0.5)$\times$10$^{13}$ cm$^{-2}$ and (2.7$\pm$0.8)$\times$10$^{12}$cm$^{-2}$ for HC1 and CR, respectively (see Table \ref{table:HC3N_column_densities}).
The value of the column density for the CR is very similar to the values obtained by Howe et al. (1994) in the dark clouds TMC-1, $N$=(5.9$\pm$0.3)$\times$10$^{12}$ cm$^{-2}$, and L1544 $N$=(2.2$\pm$0.5)$\times$10$^{12}$ cm$^{-2}$, although larger than the value obtained by Cordiner et al. (2012) in Cha-MMS1, $N$=9$\times$10$^{11}$ cm$^{-2}$. The column density for HC1, however, is one order of magnitude higher than the values found in these dark clouds.

\begin{table*}
\caption{Sample of molecular cloud cores.}             
\begin{tabular}{l l l l l l l l l }     
\hline\hline       
          &      &    & Core    &      & Infrared                & HC$_{3}$N                 & $N$(HC$_{3}$N)    &  \\ 
Source    & R.A. & Dec. & Mass    & Distance     & Luminosity              & $J$=10-9                  & $\times$10$^{13}$ & References\\ 
          & (J2000)  & (J2000)   &$\times$10$^{3}$(M$_{\odot}$) & (kpc)   & $\times$10$^{3}$(L$_{\odot}$) & $\Delta$$V$ (km s$^{-1}$) & (cm$^{-2}$)       &  \\
\hline 
   W3(OH) & 02:27:03.7       & 61:52:25    & 0.100   & 2.000  & ...    & 3.65    & 0.4        & 2, 3, 4, 5, 6\\  
   Mon R2 & 06:07:46.2       & -06:23:08.3 & 1.000   & 0.950  & 50     & 2.00    & 1.2        & 7, 8, 9, 10, 11\\
   NGC 2264 & 06:41:09.80    & 09:29:32.0  & 1.664   & 0.700  & 0.367  & 2.11    & 1.4        & 6, 12, 13\\
   M17 SW & 18:20:27.6       & -16:12:00.9 &30.000   & 2.400  & ...    & 5.60    & 1.0        & 10, 14, 15, 16\\
   Orion KL & 5:35:14.3      & -5:22:36.7  & 1.800   & 0.414  & 100    & 9.00    & 78$^{a}$  & 1, 17, 18, 19, 21\\
   S140 & 22:19:19.1         & 63:18:50.3  & 0.771   & 0.910  & 20.56  & 2.32    & 2.0        & 6, 10, 12, 13\\
   DR21S & 20:39:01.01       & 42:22:49.9  & 1.941   & 3.000  & 50     & 4.80    & 2.0        & 6, 10, 12, 20\\
   OMC2 & 5:35:27.0          & -05:10:06   & 1.000   & 0.450  & 3.5    & 1.30    & 1.2        & 24, 25, 26\\
   Sgr B2N & 17:47:20.3      & -28:22:19   & 0.800   & 7100   &10$^{4}$& 23.60   & 10$^{5}$   & 30\\
   Sgr B2M & 17:47:20.3      & -28:23:07   & 0.500   & 7100   &2$\times$10$^{4}$ & 19.30  & 7000    & 30\\
   NGC 6334N & 17:20:53.2    & -35:47:17   & 3000    & 1700  & 7      & 10.40   & ...          & 10, 23, 31, 22\\
MCLD 123.5+24.9 & 2:17:20.07 & 87:41:18.3  & 0.003   & 0.150  & ...      & 0.22  & 0.27       & 27, 28, 29\\
   Cha-MMS1 & 11:06:33.0     & -77:23:46.0 & 0.0015  & 0.150  & ...      & 0.65  & 45         & 32, 33\\
   CrA C & 19:03:58          & -37:16:00   & 0.007   & 0.170  & ...      & 0.34  & 0.057      & 32\\
   TMC1 (HCL2-A)& 04:41:45.9 & +25:41:27   & 0.005   & 0.140  & ...      & 0.57  & 0.4        & 34, 35 \\
   B1-b &  03:33:20.8        & +31:07:34   & 0.006   & 0.200  & ...      & 0.89  & 2.8        & 36, 37 \\
   L1544 & 05:04:18.1        & +25:10:48   & 0.0027  & 0.140  & ...      & 0.462 & 2.6        & 37, 38 \\
   L183  & 15:54:08.6        & -02:52:10   & ...     & 0.160  & 0.0013   & 0.478 & 2.6        & 37, 39 \\
\hline 
\label{table:correlations_HC3N}                 
\end{tabular}
\tablefoot{Column 1 indicates the
molecular cloud core, Col.2 the right ascension, Col. 3 gives the
declination, Col. 4 the core mass, 
Col. 5 the distance, Col. 6 the infrared luminosity, Col. 7 the line width of the transition $J$=10-9 of HC$_{3}$N, Col. 8 the HC$_{3}$N column density, and Col. 9 the references.\\
\tablefoottext{a}{result obtained from data shown in Table \ref{table:sHC3N_parameters_gaussian} through a rotational diagram.}
}
\tablebib{
(1) This Paper; (2) Kawamura $\&$ Masson (1998); (3) Polychroni et al. (2012); (4) Hachisuka et al. (2006); (5) Kim et al. (2006); (6) Li et al. (2012); (7) Tafalla et al. (1997); (8) Racine $\&$ van der Bergh (1970); (9) Thronson et al. (1980); (10) Vanden Bout et al. (1983); (11) Rizzo et al. (2005); (12) Alakoz et al. (2003); (13) Ridge et al. (2003); (14) Wang et al. (1993); (15) Lada et al. (1976); (16) Bergin et al. (1997); (17) Morris et al. (1976); (18) Tatematsu et al. (1993); (19) Peng et al. (2012); (20) Chung et al. (1991); (21) Gezari et al. (1998); (22) Gezari (1982); (23) Sandell (2000); (24) Liu et al. (2011); (25) Nielbock et al. (2003); (26) Thronson (1978); (27) Shimoikura et al. (2012); (28) Heithausen et al. (2008); (29) B\"ottner et al. (2003); (30) de Vicente et al. (2000); (31) Tieftrunk et al. (1998); (32) Kontinen et al. (2000); (33) Cordiner et al. (2012); (34) Cernicharo et al. (1984); (35) Howe et al. (1996); (36) Hirano et al. (1999); (37) Thesis Nuria Marcelino (2007); (38) Doty et al. (2005); (39) Sargent et al. (1983).}\\
\end{table*}

\subsection{HC$_{5}$N}

Column densities for HC$_{5}$N have been derived using LVG models with collisional rates from Deguchi \& Uyemura et al. (1984). We only find contribution to the emission from three components: the hot core (HC1), the compact ridge, and the extended ridge. 
The hot core (HC1) presents the largest HC$_{5}$N column density, with $N$(HC$_{5}$N)=(7$\pm$2)$\times$10$^{13}$ cm$^{-2}$, while the column densities for both ridges are an order of magnitude lower (see Table \ref{table:HC3N_column_densities}). Comparing these results with other sources, Broten et al. (1976) found a column density of 1.5$\times$10$^{14}$ cm$^{-2}$ in Sgr B2, whereas in dark clouds such as L1489, L1521E or TMC-2 the column density is $\sim$10$^{12}$ cm$^{-2}$ (Suzuki et al. 1992). In Orion A (KL), Bujarrabal et al. (1981) found $N$(HC$_{5}$N)=(5$\pm$2)$\times$10$^{12}$ cm$^{-2}$ for the ridge component, consistent with our values.

\section{HC$_{3}$N in other sources}
\label{section:HC3N_other_sources} 

In this section we compare our results for HC$_{3}$N with results obtained by other authors in a sample of 18 molecular cloud cores: 11 belong to giant clouds (GMCs) and 7 (MCLD123.5+24.9, Cha-MMS1, CrA C, TMC1, B1-b, L1544, and L183) to small dark clouds (SMCs). Table \ref{table:correlations_HC3N} shows the properties of each source.
In addition, we study possible correlations between the column density of HC$_{3}$N, the observed line width of this species, the cloud core mass, and the distance from the galactic center.

\begin{figure}
   \centering 
   \includegraphics[angle=-360,width=8cm]{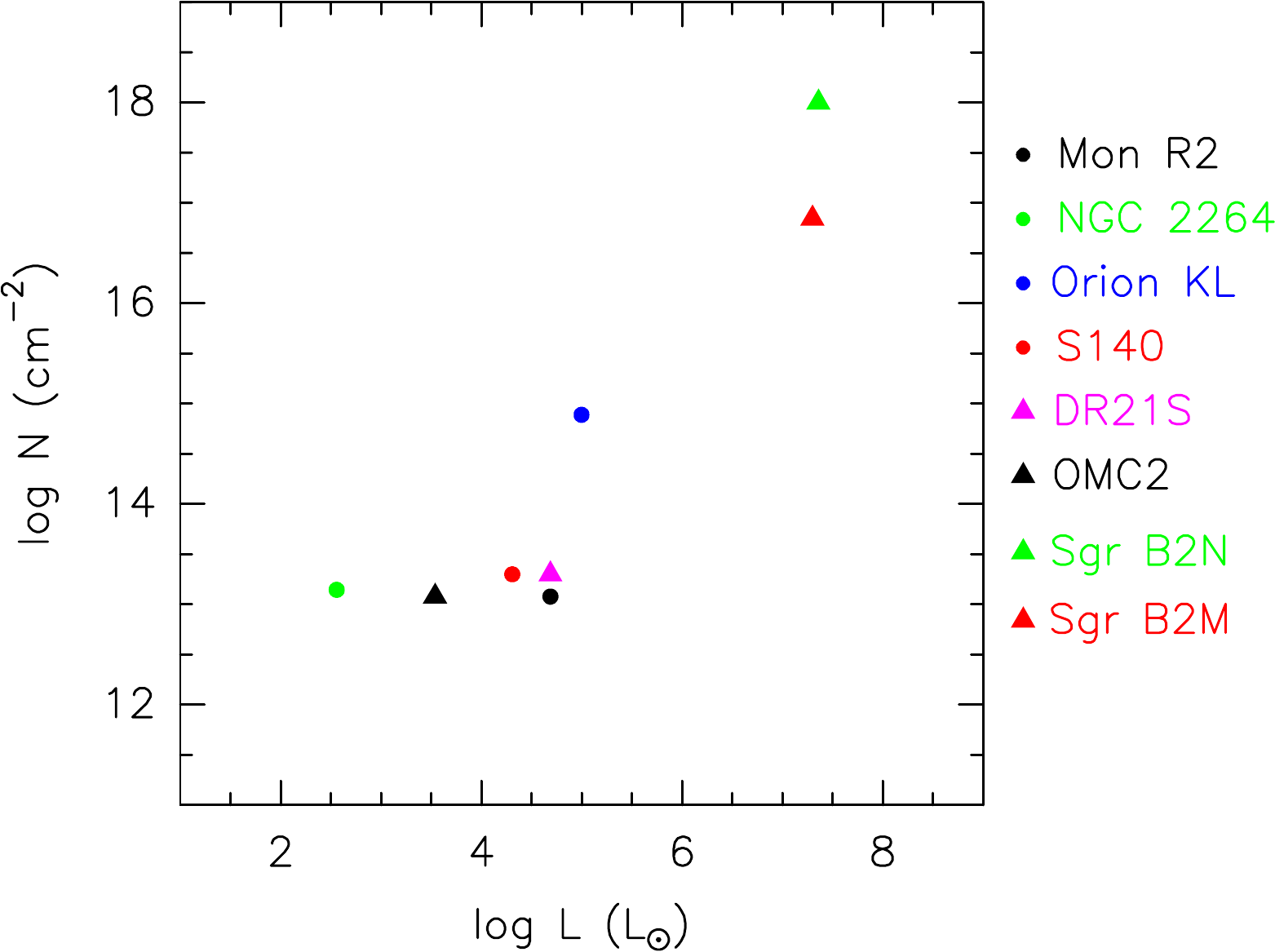}
   \caption{Correlation between the column density of HC$_{3}$N observed in different cloud cores and their infrared luminosity (L$_{\odot}$).}
   \label{figure:L_N}
   \end{figure}

First we have considered the cloud cores of the sample with results for the column density of $N$(HC$_{3}$N). These values (see references in Table \ref{table:correlations_HC3N}) were calculated considering LTE approximation. In the case of Orion KL, to be able to compare with the other sources, we have also calculated the column density of the ground state of HC$_{3}$N for the dominant (hot core) component, assuming that the source fills the beam and that the lines are optically thin. With the data from Table \ref{table:sHC3N_parameters_gaussian} we obtain a rotational diagram by applying the expression:

\begin{equation}
\mathrm{ln}({\gamma_{\mathrm{u}}W/g_{\mathrm{u}}})=\mathrm{ln}(N)-\mathrm{ln}(Z)-(E_{\mathrm{u}}/kT)
\end{equation}

\noindent where $W$ is the integrated line intensity, $g$$_{\mathrm{u}}$ is the statistical weight of each level, $N$ the column density, $Z$ is the partition function at temperature $T$, $E$$_{\mathrm{u}}$ the energy of the upper level, $k$ is the Boltzman constant, $T$ the rotational temperature, and $\gamma$$_{\mathrm{u}}$ a constant which depends on the transition frequency and the Einstein coefficient $A$$_{\mathrm{ul}}$ (see Goldsmith \& Langer 1999 for more details).
Figure \ref{figure:L_N} shows the column density of HC$_{3}$N for each source against its infrared luminosity. 
We observe an increasing trend of the column density of HC$_{3}$N with the luminosity. A linear fit of these points provides a determination coefficient $R$$^{2}$=0.79. 



\begin{figure}
   \centering 
   \includegraphics[angle=-90,width=8.65cm]{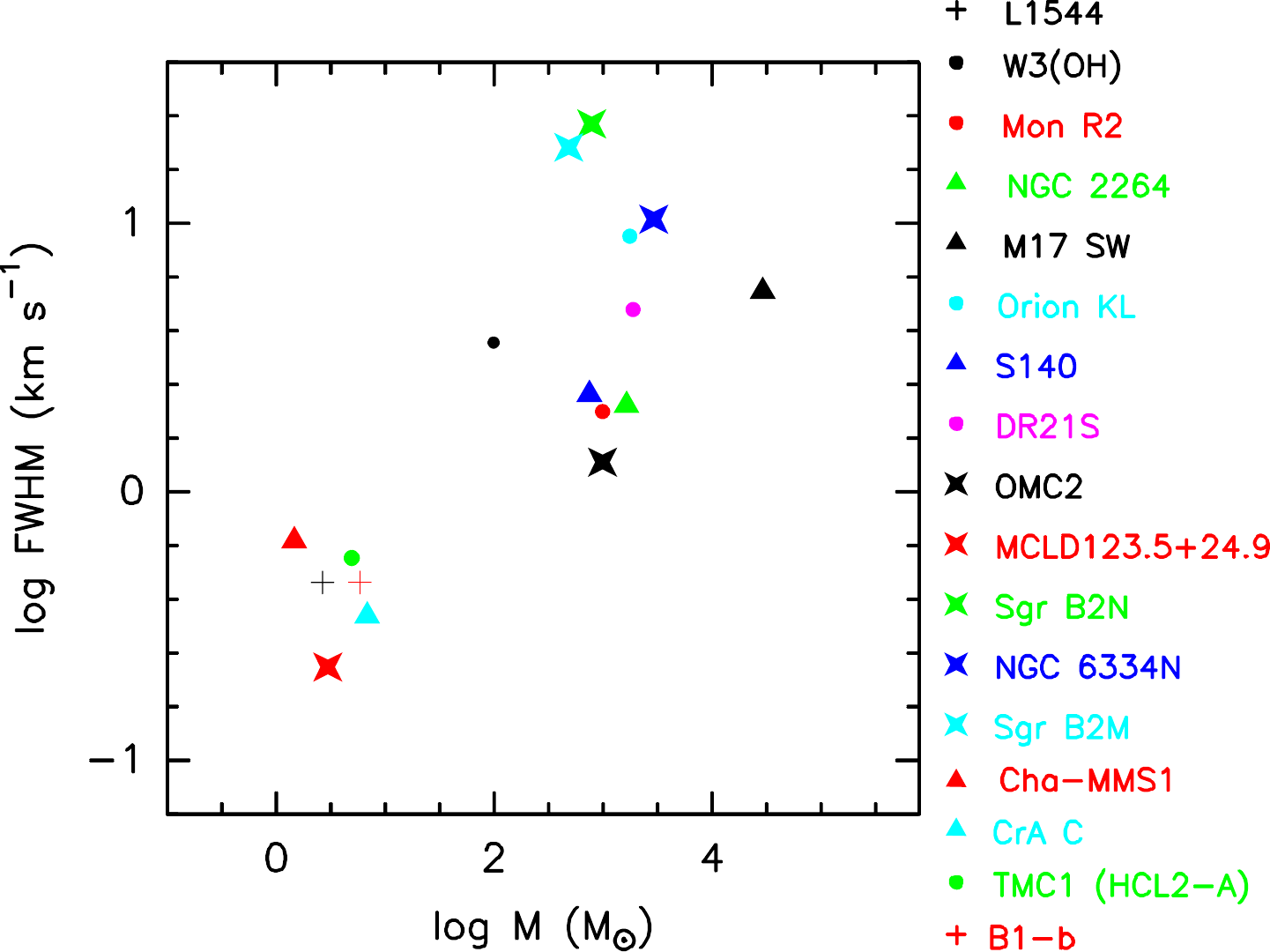}
   \caption{Correlation between the observed line widths of HC$_{3}$N $J$=10-9 in different cloud cores and their mass (M$_{\odot}$).}
   \label{figure:mass_FWHM}
   \end{figure}

Figure \ref{figure:mass_FWHM} shows the observed line widths of HC$_{3}$N ($J$=10-9) in 17 cloud cores according to their masses. Considering only the cores belonging to giant clouds, we do not find any apparent correlation between these parameters. However, if we include the dark clouds, we see a trend ($R$$^{2}$=0.60), also increasing, between the FWHM of the HC$_{3}$N lines and the mass of the sources. A bigger sample (including more cores of dark cloud cores) could confirm the correlation, however we have not included them due to the lack of observations of this transition of HC$_{3}$N in these kind of clouds.



\begin{figure}
   \centering 
   \includegraphics[angle=-90,width=8cm]{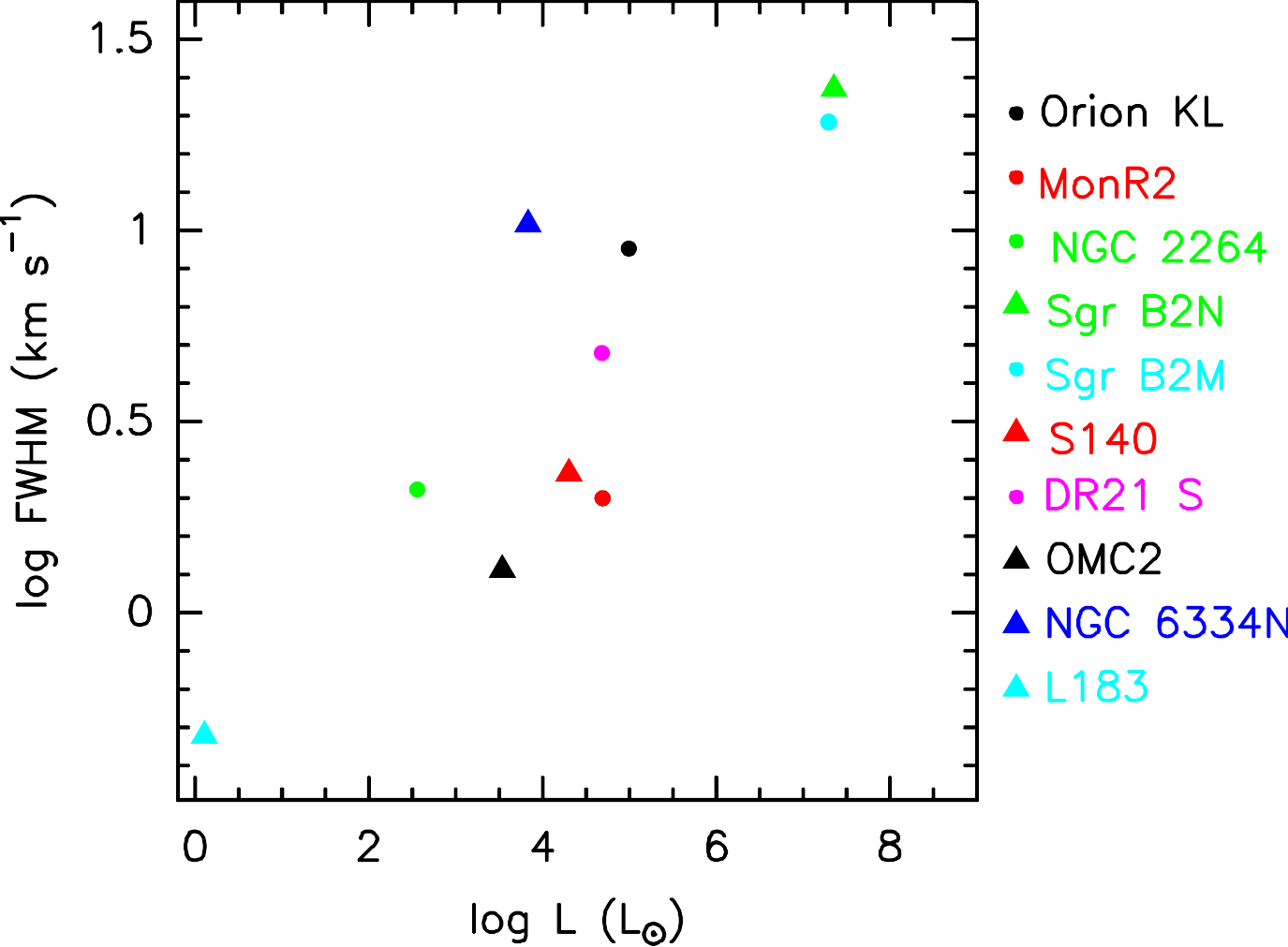}
   \caption{Correlation between observed line width, FWHM, and infrared luminosity (L$_{\odot}$) of different cloud cores.}
   \label{figure:L_FWHM}
   \end{figure}

We have also studied a possible correlation between the luminosity of the cloud cores and the line widths for the transition $J$=10-9 (see Fig. \ref{figure:L_FWHM}). In this case, the FWHM increases as luminosity increases (determination coefficient is $R$$^{2}$=0.76).
We obtain therefore the largest FWHMs, luminosities, and HC$_{3}$N column densities in the cloud cores with higher masses. It suggests that these cores harbor the most massive stars responsible for their high luminosities. However, another posibility should be considered, mainly from the trend of Fig. \ref{figure:L_FWHM}. We see that some of the cores present large FWHMs and since line widths depend on the kinematics of the gas and mainly on the dispersion velocity, these results could suggest the presence of many lower mass sources within these massive cores, yielding larger velocity dispersions and therefore larger FHWMs. Further studies of these cores could provide new hints concerning the high mass star formation and the role of low mass stars in this process.
Observations of HC$_{3}$N in a larger sample of SMCs are required to obtain firmer conclusions about the HC$_{3}$N properties in GMCs and SMCs.

\begin{table*}
\centering
\caption{Isotopic and molecular ratios.}
\begin{tabular}{lllllll}
\hline 
\hline
      & Extended  & Compact  &  High velocity & Plateau &  Hot  &  Hot     \\
Ratio & ridge     & ridge    &  plateau       & (PL)    & core  & core     \\ 
      & (ER)      & (CR)     &  (HP)          &         & (HC1) & (HC2)    \\    
\hline
\hline 
HC$_{3}$N/H$^{13}$CCCN                          & 27$\pm$16 & 27$\pm$14   & 25$\pm$13 & 21$\pm$12   &  17$\pm$11   &  3$\pm$2      \\  
HC$_{3}$N/HC$^{13}$CCN                          & 80$\pm$44 & 20$\pm$11   & 25$\pm$13 & 70$\pm$41   &  14$\pm$8    &  5$\pm$3      \\ 
HC$_{3}$N/HCC$^{13}$CN                          & 80$\pm$44 & 24$\pm$14   & 25$\pm$13 & 70$\pm$41   &  10$\pm$6    &  7$\pm$4      \\ 
\hline
HC$_{3}$N/HC$_{5}$N   & 16$\pm$10     & 67$\pm$37      & ...      & ...    & 14$\pm$8        & ...\\
DC$_{3}$N/HC$_{3}$N   & ...          & 0.014$\pm$0.008  & ...    & ...    & 0.015$\pm$0.009  & ...\\    
\hline
\end{tabular}
\label{table:abundancesI}
\end{table*}

\section{Discussion}
\label{section:discussion}


We have shown that cyanoacetylene is an excellent dense gas tracer (see also Morris et al. 1976, Chung et al. 1991, Bergin 1996). In particular, we find that the column density of HC$_{3}$N in the hot core is up to two orders of magnitude higher than in the other Orion cloud components. The large number of high energy transitions of HC$_{3}$N (especially from the vibrational modes $\nu$$_{5}$, $\nu$$_{6}$, and $\nu$$_{7}$) observed with the Herschel telescope has revealed the existence of a temperature and density gradient in the hot core (the HC2 component), that probably would have not been detected from the IRAM-30m data alone, due to the fact that the emission from this component affects only the highest energy transitions. Moreover, the high vibrational temperature obtained in Sect. \ref{subsection:HC3N}, shows the possible presence of an even hotter inner region than HC2 within the hot core. 


From the 2$\arcmin$$\times$2$\arcmin$ maps (Sect. \ref{subsection:maps}), we find that the emission for the ground state and the $\nu$7 vibrational mode, reaches its peak in the velocity range associated with the hot core and the high velocity plateau (a hot region affected by shocks). These maps also show the kinematics of the gas from two low energy transitions of HC$_{3}$N. The transition $J$=12-11 shows how the emission is elongated in the NE direction toward the extended ridge, with the emission toward the SW being much weaker. This elongation toward the NE could originate from the interaction of the hot spot situated at $\sim$18$\arcsec$ east and $\sim$25$\arcsec$ north of IRc2 (that we observe in the mapped transition $J$=26-25), with the ambient gas in the ridge component. The hot spot could be associated with the CS1 condensation in the extended ridge noted by Mundy et al. (1988).

\subsection{Isotopic and molecular abundances}

Column density ratios have been calculated for the different isotopologues of HC$_{3}$N as well as molecular ratios (see Table \ref{table:abundancesI}). We show only the cases where the column densities were obtained considering LVG approximation in order to avoid large uncertainties in the ratio values.

$^{12}$C/$^{13}$C: in the ground state, we get the same ratio in the high velocity plateau for the three $^{13}$C isotopologues. In the hot core (HC1), we also obtain similar ratios for the three isotopologues, however in the plateau (PL) and in the extended ridge we obtain a larger difference (a factor $\sim$3) between the abundance of H$^{13}$CCCN and the other two $^{13}$C isotopologues with respect to HC$_{3}$N. These last are in agreement with the solar abundance ($^{12}$C/$^{13}$C=90) from Anders \& Grevesse (1989), in the plateau (PL) and extended ridge.
In the rest of the regions, the results are much lower than the solar abundance, in particular in the hot core (HC2). In this region with the lowest size and the highest temperature  of Orion KL, we obtain the largest abundance of the $^{13}$C isotopologue respect to $^{12}$C.
From observations of OCS, C$^{34}$S, and H$_{2}$CS in Orion KL, Tercero et al. (2010) obtained a $^{12}$C/$^{13}$C ratio of 15, 18, and 20, respectively in the hot core, while in the plateau the ratio is 25, and 7-14 in the compact ridge. These values agree with our results (considering the uncertainty), especially those obtained from the C$^{34}$S molecule.
Comparing with other sources, Milam et al. (2004) found $^{12}$C/$^{13}$C=20.5 in the molecular cloud W31 and $^{12}$C/$^{13}$C=47.8 in G49.2, while in the cloud WB391 the value was $^{12}$C/$^{13}$C$\sim$135. 
 

In Table \ref{table:abundancesI}, we also list column density ratios for HC$_{3}$N with respect to its deuterated counterpart (DC$_{3}$N) and HC$_{5}$N. 
For HC$_{3}$N/HC$_{5}$N, we obtain ratios $>$14. Comparing with other authors, Bujarrabal et al. (1981) obtained HC$_{3}$N/HC$_{5}$N=13$\pm$7 in the ridge component of Orion KL, Cernicharo et al. (1984) obtained a ratio HC$_{3}$N/HC$_{5}$N between 3 and 8 in TMC-1, and HC$_{3}$N/HC$_{5}$N=66 in IRC+10216 according to results from Bujarrabal et al. (1981) and Jewell \& Snyder (1984).
But especially interesting is the obtained $D/H$ ratio (0.015), which is comparable to that for dark clouds, like the rato obtained by Langer et al. (1980) in TMC1, $D/H$=0.02-0.08.

Table \ref{table:molecular_abundances} shows the derived molecular abundances of HC$_3$N, HC$_5$N, and DC$_3$N with respect to hydrogen in each component (see Tercero et al. (2011) for the assumed values of $N$$_{\mathrm{H_2}}$). We find that the highest abundance of HC$_3$N is obtained in the high velocity plateau, whereas in the hot core this abundance is twice lower. The extended ridge has the lowest abundance of cyanoacetylene. 
HC$_5$N is up to four times more abundant in the hot core than in the rest of components, and DC$_3$N presents a similar abundance in the hot core and in the compact ridge.

\begin{table}
\caption{Molecular abundances, $X$, with respect to H$_{2}$.}
\begin{center} 
\centering
\begin{tabular}{lllllll}
\hline 
\hline
Region & Species & $X$ (This work)       & $X$ (Other works) \\
       &         & ($\times$10$^{-10}$)  & ($\times$10$^{-10}$)\\
\hline 
Extended       &  HC$_{3}$N & 10   & ... \\
Ridge$^{a}$    &  DC$_{3}$N & ...  & ... \\
               &  HC$_{5}$N & 0.66 & ... \\
\hline
Compact        &  HC$_{3}$N & 27   & 20$^{(1)}$\\
Ridge$^{b}$    &  DC$_{3}$N & 0.36 & ...\\
               &  HC$_{5}$N & 0.40 & ...\\
\hline
               &  HC$_{3}$N & 33  & 24$^{(2)}$\\
Plateau$^{c}$  &  DC$_{3}$N & ... & ..\\
               &  HC$_{5}$N & ... & ..\\
\hline
High           &  HC$_{3}$N & 81  & ... \\
velocity       &  DC$_{3}$N & ... & ...\\
Plateau$^{d}$  &  HC$_{5}$N & ... & ...\\
\hline
Hot            &  HC$_{3}$N & 40   & 18$^{(2)}$\\
core$^{e}$     &  DC$_{3}$N & 0.36 & ... \\
(HC1+HC2)      &  HC$_{5}$N & 1.7  & ... \\
\hline 
\hline
\end{tabular}
\label{table:molecular_abundances}
\end{center} 
\tablefoot{Derived molecular abundances assuming:\\ 
\tablefoottext{a}{$N$$_{\mathrm{H_2}}$=7.5$\times$10$^{22}$ cm$^{-2}$,}
\tablefoottext{b}{$N$$_{\mathrm{H_2}}$=7.5$\times$10$^{22}$ cm$^{-2}$,}
\tablefoottext{c}{$N$$_{\mathrm{H_2}}$=2.1$\times$10$^{23}$ cm$^{-2}$,}
\tablefoottext{d}{$N$$_{\mathrm{H_2}}$=6.2$\times$10$^{22}$ cm$^{-2}$,}
\tablefoottext{e}{$N$$_{\mathrm{H_2}}$=4.2$\times$10$^{23}$ cm$^{-2}$.}
}
\tablebib{
(1) Morris et al. (1976); (2) Sutton et al. (1995).}\\
\end{table}

\subsection{On the origin of the DC$_3$N emission}

We have presented here a tentative detection of DC$_{3}$N in a giant molecular cloud. Prior to this work, it was only observed in dark clouds, where deuterium fractionation is very effectively enhanced above the deuterium cosmic abundance, $D$/$H$$\sim$10$^{-5}$, due to their low temperatures and large CO depletion (see e.g. Roberts et al. 2003). 
We have obtained in the hot core of Orion KL a ratio $D$/$H$=0.015$\pm$0.009. In the same cloud, Walmsley et al. (1987) obtained $D$/$H$=0.003 (from NH$_2$D observations), Mauersberger et al. (1988) deduced $D$/$H$=0.01-0.06 (from CH$_3$OD), Jacq et al. (1993) $D$/$H$=0.01-0.09 (from CH$_{2}$DOH), and Bergin et al. (2010) $D$/$H$=0.02 (from HDO/H$_{2}$O) in broad agreement with our result. Other authors such as Persson et al. (2007) and Neill et al. (2013) obtained, however, lower ratios of $D$/$H$ (0.001 and 0.003, respectively).
Deuterium enrichment is expected to be produced in cold clouds or during the cold phase (prior to the hot core stage) of giant clouds (Fontani et al. 2011). To obtain large ratios of $D$/$H$ (as obtained in this work which is comparable to that obtained in cold clouds), low temperatures and high densities ($\textgreater$10$^{4}$ cm$^{-3}$) are required. In our case, since the temperature of the hot core of Orion is $\textgreater$200 K, one of the most plausible mechanisms to explain the presence of DC$_3$N is to consider that the formation of this molecule occurred prior to the gas phase, i.e., the fractionation occured on grain surfaces when the cloud was much colder, and then the deuterated compounds were released into the gas phase when the icy mantles sublimated. 
However, taking into account that HC$_3$N is not mainly formed on grain surfaces, we should also consider other possible mechanisms for the formation of DC$_3$N, which imply higher temperatures. If we analyse the different ways to form deuterated molecules, we observe that deuterium is mainly locked into HD and transferred to other species by exothermic ion-molecule reactions (Howe et al. 1994):

\begin{equation}
\mathrm{H^{+}_{3} + HD \rightarrow H_{2}D^{+} + H_{2}}
\label{equation:H3}
\end{equation}

\begin{equation}
\mathrm{CH^{+}_{3} + HD \rightarrow CH_{2}D^{+} + H_{2}} 
\label{equation:CH2D}
\end{equation}

\begin{equation}
\mathrm{C_{2}H^{+}_{2} + HD \rightarrow C_{2}HD^{+} + H_{2}} 
\label{equation:C2HD}
\end{equation}

\noindent H$_{2}$D$^{+}$, CH$_{2}$D$^{+}$, and C$_{2}$HD$^{+}$ are molecules that can transfer their deuterium to other species. However, depending on the temeperature of the cloud, some may be more important than others. Equation (\ref{equation:H3}) is only important at low temperatures ($\leq$25 K), because the reverse reaction is exothermic and inefficient, whereas (\ref{equation:CH2D}) and (\ref{equation:C2HD}) (with exothermicities $\sim$390 and $\sim$550 K respectively) are efficient at temperatures up to 100 K.

If we consider the case where HC$_3$N is formed on grains, HC$_3$N can be considered a reservoir from which DC$_3$N could be formed subsequently in the gas phase, as a result of deuterated ions reacting with HC$_3$N (Langer et al. 1980):

\begin{equation}
\mathrm{H_{2}D^{+} + HC_3N \rightarrow HDC_3N^{+} + H_{2}} 
\label{equation:HDC3N+}
\end{equation}

\begin{equation}
\mathrm{HDC_3N^{+} + e \rightarrow DC_3N + H}
\label{equation:DC3N}
\end{equation}

\noindent However, if the DC$_3$N/HC$_3$N ratio is larger than DCO$^{+}$/HCO$^{+}$, the formation of HC$_3$N primarily on grains is ruled out (Langer et al. 1980). The ratio for DCO$^{+}$/HCO$^{+}$ observed in Orion is $\textless$0.009 (Rodr\'iguez-Kuiper et al. 1978). This would suggest that the formation of HC$_3$N did not take place during the cold phase. In this case, some of the mechanisms for the gas-phase production of HC$_3$N are the reactions of H$_2$CN$^{+}$ with C$_2$H$_2$, followed by dissociative recombination, are proposed by Mitchell et al. (1979), or reactions of CH$^{+}_{\mathrm{3}}$ with C$_2$H or C$_2$ to produce C$_3$H$^{+}_{\mathrm{2}}$, followed by a reaction of this ion with nitrogen (Langer et al. 1980). 
The production of DC$_3$N would then be through ion-molecule reactions for example:

\begin{equation}
 \left.
 \begin{array}{r@{$\ +\ $}l}
  \mathrm{HDCN^+} & \mathrm{C_2H_2} \\
  \mathrm{H_2CN^+} & \mathrm{C_2HD} \\
 \end{array}
 \right\}
 \mathrm{HDC_3N^+ + H_2}
\end{equation}

\noindent In fact, other routes of formation may also be important. In order to investigate further the origin of the deuterated cyanoacetylene in Orion KL, we have used a gas-grain time-dependent chemical model (see Sect. \ref{subsubsection:chemcial model}).

\subsubsection{Chemical model}
\label{subsubsection:chemcial model}

We have modeled the hot core of Orion KL using the chemical model UCL$\_$CHEM (Viti et al. 2004), a time-dependent gas-grain model where the chemistry, density, and temperature are calculated at each time-step, producing chemical abundances. It is a two-phase calculation: phase I simulates the free-fall collapse of the core, until densities typical of hot cores ($\sim$10$^{7}$ cm$^{-3}$) are reached. In this phase atoms and molecules are frozen onto grain surfaces. Phase II follows the chemical evolution of the core once a source of radiation is present. We simulate the effect of the presence of an infrared source in the center of the core or in its vicinity increasing the gas and dust temperature up to $T$=300 K. This increase of temperature is based on the luminosity of the protostar by using the observational luminosity function of Molinari et al. (2000).

\begin{figure*}
   \centering 
   \includegraphics[angle=-90,width=18cm]{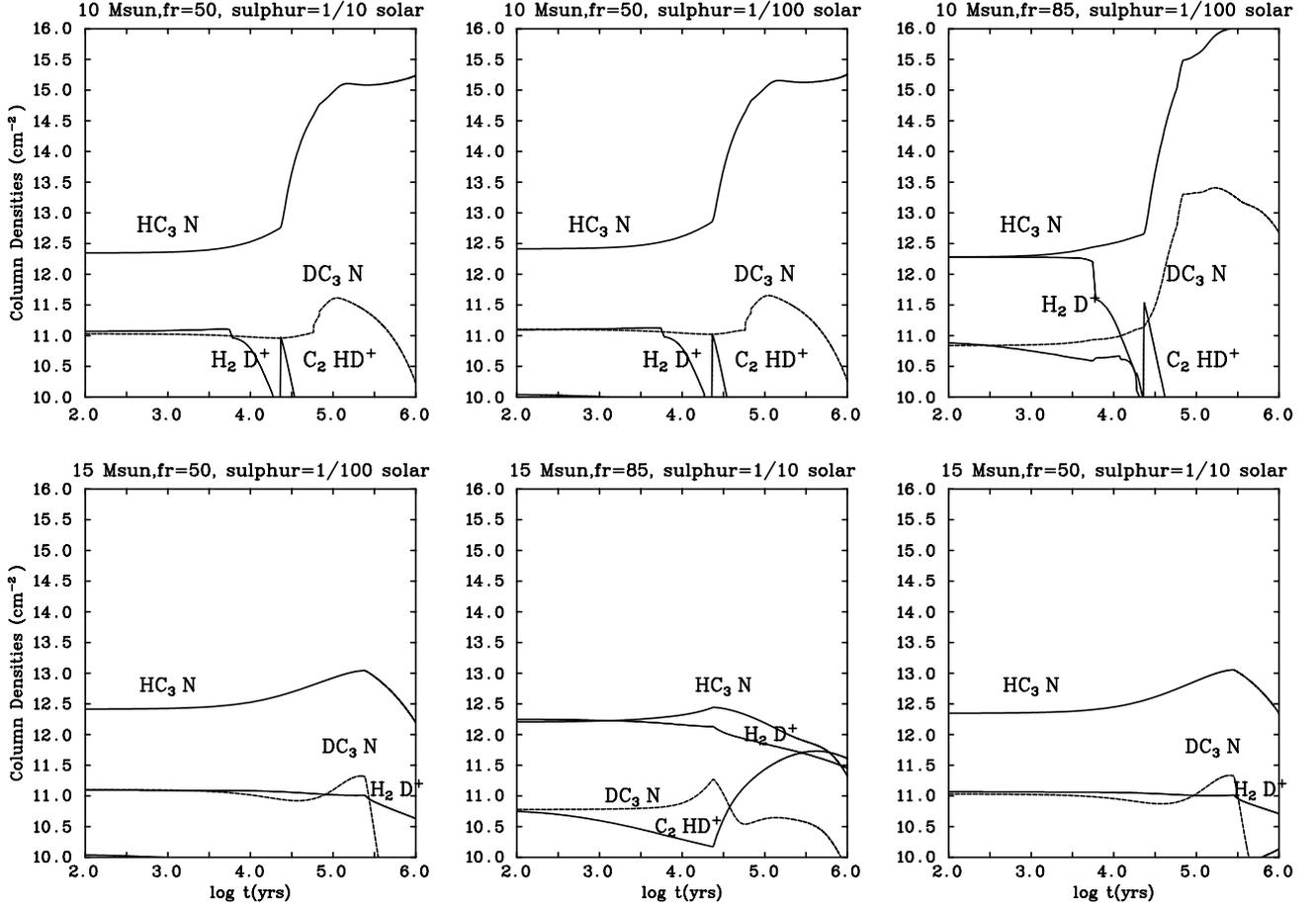}
   \caption{Time evolution of the column densities of HC$_{3}$N, DC$_{3}$N, H$_{2}$D$^{+}$, and CH$_{2}$D$^{+}$ for selected hot cores chemical models. The input parameters for the mass of the formed star, the depletion efficiency ($f$$_{\mathrm{r}}$), and the sulphur solar abundance are indicated on the top of each panel.}
   \label{figure:chemical_models}
   \end{figure*}

We have considered a source diameter of 10 arcsec, a final volume density $n$$_{H}$=2$\times$10$^{7}$ cm$^{-3}$, an efficiency of freeze out $f$$_r$ in phase I of 50\% and 85\%, and a mass for the star of 10M$_{\odot}$ and 15M$_{\odot}$ (see Table \ref{table:hotcore_model}). The initial elemental abundances of the main species (H, He, C, O, N, Mg, S) are also input parameters. We have adopted the solar abundance for all of them, except for sulphur that we have depleted by a factor 10 and 100, as it is well known that it may be depleted onto grains in star-forming regions (Bergin et al. 2001, Pagani et al. 2005). Figure \ref{figure:chemical_models} shows the time evolution of HC$_{3}$N, DC$_{3}$N,  H$_{2}$D$^{+}$, and C$_{2}$HD$^{+}$ for selected hot core models during phase II. 

\begin{table}
\caption{Hot core chemical models and their parameters.}             
\begin{center}      
\centering          
\begin{tabular}{lc c c c }     
\hline\hline       
             
Model & $M$$_{\star}$  & $f$$_{\mathrm{r}}$ & Sulphur \\ 
      &  (M$_{\odot}$) & (\%)               & (S$_{\odot}$) \\
\hline 
   1 &  10 & 50 & 0.1 \\
   2 &  10 & 50 & 0.01 \\  
   3 &  10 & 85 & 0.1\\
   4 &  10 & 85 & 0.01\\
   5 &  15 & 50 & 0.1 \\
   6 &  15 & 50 & 0.01 \\
   7 &  15 & 85 & 0.1 \\
   8 &  15 & 85 & 0.01 \\
\hline
\end{tabular}
\label{table:hotcore_model} 
\end{center}                
\tablefoot{Column 1 indicates the model, Col. 2 the final density of hydrogen at the end of phase I, Col. 3 the mass of the formed star, Col. 4 the percentage of depletion onto grains, and Col. 5 the initial abundance of sulphur in units of the solar abundance.}\\
\end{table}

From Fig. \ref{figure:chemical_models} we find that the initial abundance of sulphur barely affects the time evolution of the column densities shown here. Comparing the theoretical and observed 
(Table \ref{table:HC3N_column_densities}) column densities of HC$_{3}$N and DC$_{3}$N, we find that the models that best reproduce our observations correspond to those for a hot core with a star of mass 10M$_{\odot}$ and with a depletion efficiency of 85$\%$. In this case, the observations are reproduced at $t$$\sim$6$\times$10$^{4}$ years after the central star switches on. 
In the models with $f$$_r$=50\%, we obtain (at $t$$\sim$10$^{5}$ years) values for the column density of HC$_{3}$N similar to those obtained from our line fits; however, the values for DC$_{3}$N are too low. For all models with 15M$_{\odot}$, the column densities obtained for both HC$_{3}$N and DC$_{3}$N are too low. 

It is interesting to note, however, that DC$_3$N reaches the observed value during phase II, i.e, its enhancement occurs during the hot core and not the cold phase, in conjunction with an increase in HC$_3$N. A chemical analysis of these two species after 10$^4$ years (when they start to increase for the best fit models) shows that their main route of formation is via reactions of CN with C$_2$H$_2$ and C$_2$HD; in fact when the core temperature starts increasing above 50 K, C$_2$H$_2$ and C$_2$HD increase by several orders of magnitudes, due to their evaporation from the icy mantles. At later stages however, these species are destroyed, leading to a decrease in HC$_3$N and DC$_3$N as well. 
We note that the time at which the different species sublimate from the mantle is a function of the mass of the star, as well as their binding (on the grains) properties (see Collings et al. 2004 and Viti et al. 2004 for more details). Hence, crucial to the interpretation of our results is our treatment of the time and mass dependence of the increase of the temperature once the protostar is born (Viti \& Williams 1999); we therefore restrain from drawing any conclusion on the age of the core, but note that a too fast increase of temperature (as occurs for a 15M$_{\odot}$ star) leads to a more efficient destruction chemistry, hence \textquotedblleft missing\textquotedblright  the enhancement phase for HC$_3$N and DC$_3$N. Regardless of the details of the individual models, what is clear is that the DC$_3$N that forms during the cold phase is not enough to account for what is observed in the Orion KL hot core and that DC$_3$N must partly form during the hot core phase.

\section{Summary and Conclusions}
\label{section:summary}

We have reported a tentative detection of DC$_3$N in Orion KL with the IRAM 30m line survey. We have also studied the presence of HC$_5$N and HC$_3$N in this region with IRAM-30m and Herschel/HIFI observations, covering a total frequency range from 80 to 1907 GHz. We have detected 35 lines of HC$_5$N and 41 lines of HC$_3$N in its ground state, as well as 68 transitions of its $^{13}$C isotopologues, and a large number of lines (297) from six vibrational states, $\nu$$_5$, $\nu$$_6$, $\nu$$_7$, 2$\nu$$_7$, 3$\nu$$_7$, and $\nu$$_6$+$\nu$$_7$ of HC$_3$N. This large number of observed lines has allowed us to consider a temperature and density gradient in the hot core of Orion KL, that with only observations from ground-based telescopes would probably not be possible. 

For HC$_3$N, we found the emission to arise mainly from the hot core. But unlike other works, we conclude, by applying a LVG code, that the emission comes from the outer part of the hot core, where the temperature is $T$$_{\mathrm{K}}$$\sim$220 K. 
For the vibrational modes $\nu$$_5$, $\nu$$_6$, 2$\nu$$_7$, and 3$\nu$$_7$, we detected emission from the inner most part of the hot core with $T$$_{\mathrm{K}}$$\sim$310 K, this being the component most responsible for the emission of these modes. Their column densities are 1-6$\times$10$^{14}$ cm$^{-2}$. Similar to HC$_3$N, we obtained from LVG models that most of the emission of HC$_5$N is coming from the outer hot core, although its column density in this region is $\sim$15 times lower than that of HC$_3$N. Comparing with other sources, the derived column density of HC$_5$N is a factor 3 larger than the column densities of HC$_3$N found in dark clouds by Cernicharo et al. (1984).
With respect to DC$_3$N, our best model indicates that the emission comes from the hot core and from the compact ridge. Calculating the ratio $R$=DC$_3$N/HC$_3$N from the column densities of both molecules, we obtained $R$=0.015$\pm$0.009 for the hot core. 
In order to study the possible origin of DC$_3$N in this region and to explain the high level of deuteration in the warm gas, we ran several chemical models. We reproduced the observed column densities of HC$_3$N and DC$_3$N in the gas-phase with models of hot core, with a central star of 10M$_{\odot}$ and high depletion efficency (85\%). From these results, we concluded that the likely observed DC$_3$N in Orion KL is mainly formed during the warm gas-phase.

From maps of HC$_3$N, we confirmed that the peak of the emission is found at velocities associated with the hot core, in agreement with the results of our best fit models, and observed at low energies a northest-southwest gradient in the gas excitation, which traces the ridge component. For higher energies ($E$$\sim$150 K), we detected a second peak of emission to the NE of the hot core that may represent the shocked ridge material, similar to the CH$_3$CN study by Bell et al. (submitted). For the vibrational modes $\nu$$_6$ and $\nu$$_7$ we found a compact distribution of the emission centred on the hot core.

Finally, we have compared the results for HC$_3$N in Orion KL with those for other sources. With a sample of 18 cloud cores, we found correlations between their masses, HC$_3$N column densities, the line widths, and their infrared luminosities. We deduced that the most massive cloud cores present the largest column densities and line widths of HC$_3$N.

\begin{acknowledgements}

We thank the Spanish MINECO for funding support through grants AYA2006-14876, CSD2009-00038, AYA2009-07304, and AYA2012-32032. G.B.E. is supported by a CSIC grant JAE PreDoc2009. J.R.G. is supported by a Ram\'on y Cajal research contract. A.P. is supported by a JAE-Doc CSIC fellowship co-funded with the European Social Fund under the program ''Junta para la Ampliaci\'on de Estudios'', by the Spanish MICINN grant AYA2011-30228-C03-02 (co-funded with FEDER funds), and by the AGAUR grant 2009SGR1172 (Catalonia). T. A. B. is supported by a JAE-Doc research contract.
  
\end{acknowledgements}

\pagebreak

\begin{appendix}
\section{Figures and Tables}

\pagebreak

\longtab{1}{
\begin{longtable}{lrlrlrlrl}
\caption{Observed lines of HC$_{3}$N and its isotopologues (ground and vibrational states) \label{table:tab_HC3N_vib}.}\\
\hline\hline                      
Species & Transition & Predicted   & $S$$_{\mathrm{ij}}$ & $E$$_{\mathrm{u}}$ & Observed    & Observed                         & Observed        &  Blended \\
        & $_{J}$-$_{J'}$ & freq.       &                    &  (K)       & freq.              & $V$$_{\mathrm{LSR}}$  & $T$$_{\mathrm{MB}}$  &   with \\
        &                &  (MHz)      &                    &            & (MHz)              &  (km s$^{-1}$)        & (K)                  &        \\
\hline
\endfirsthead
\caption{continued.}\\
\hline\hline
Species & Transition & Predicted   & $S$$_{\mathrm{ij}}$ & $E$$_{\mathrm{u}}$ & Observed    & Observed                         & Observed     &  Blended \\
        & $_{J}$-$_{J'}$ & freq.       &                    &  (K)       & freq.              & $V$$_{\mathrm{LSR}}$  & $T$$_{\mathrm{MB}}$  &   with \\
        &                &  (MHz)      &                    &            & (MHz)              &  (km s$^{-1}$)        & (K)                  &        \\
\hline
\endhead
\hline
\endfoot
HC$_{3}$N &   9-8                           &   81881.462 &   9.0  & 19.6  & 81881.5  & 8.9   & 7.01  &  \\
HC$_{3}$N &  10-9                           &   90978.988 &  10.0  & 24.0  & 90979.5  & 7.3   & 7.90  &  \\
HC$_{3}$N & 11-10                           &  100076.384 &  11.0  & 28.8  & 100076.5 & 8.7   & 9.31  & HCOOCH$_{3}$ \\
HC$_{3}$N & 12-11                           &  109173.637 &  12.0  & 34.1  & 109173.5 & 9.4   & 10.91 &  \\
HC$_{3}$N & 15-14                           &  136464.402 &  15.0  & 52.4  & 136465.1 & 7.5   & 15.40 &  \\
HC$_{3}$N & 16-15                           &  145560.949 &  16.0  & 59.4  & 145561.4 & 8.1   & 18.31 &  \\
HC$_{3}$N & 17-16                           &  154657.287 &  17.0  & 66.8  & 154657.6 & 8.4   & 20.40 &  \\
HC$_{3}$N & 18-17                           &  163753.403 &  18.0  & 74.7  & 163753.9 & 8.1   & 21.11 &  \\
\end{longtable}
\tablefoot{Column 1 indicates the
species, Col.2 the quantum numbers of the line transition, Col. 3 gives the
assumed rest frequencies, Col. 4 the line strength, 
Col. 5 the energy of the upper level, Col. 6 observed frequency
assuming a $V$$_{\mathrm{LSR}}$ of
9.0 km s$^{-1}$, Col. 7 the observed radial velocities, Col. 8 the peak line main beam temperature, and Col. 9 the blended species.}
}

\begin{landscape}
\begin{table}
\caption{HC$_{3}$N parameters from Gaussian fits.}
\begin{center}
\begin{tabular}{lllllllllllllllllll}
\hline 
\hline
Transition & \multicolumn{3}{c}{Plateau} & \multicolumn{3}{c}{High velocity plateau} & \multicolumn{3}{c}{Hot core} & \multicolumn{3}{c}{Compact ridge} & \multicolumn{3}{c}{Extended ridge} & \multicolumn{3}{c}{20.5 km s$^{-1}$ component}\\ 
$J$-$J$$^{\prime}$ & v$_{\mathrm{LSR}}$ & $\Delta$$V$ & $T$$_{\mathrm{MB}}$ &
$V$$_{\mathrm{LSR}}$ & $\Delta$$V$ & $T$$_{\mathrm{MB}}$ &
$V$$_{\mathrm{LSR}}$ & $\Delta$$V$ & $T$$_{\mathrm{MB}}$ & 
$V$$_{\mathrm{LSR}}$ & $\Delta$$V$ & $T$$_{\mathrm{MB}}$ & 
$V$$_{\mathrm{LSR}}$ & $\Delta$$V$ & $T$$_{\mathrm{MB}}$ & 
$V$$_{\mathrm{LSR}}$ & $\Delta$$V$ & $T$$_{\mathrm{MB}}$  \\
\hline
   9-8 & 6.0$\pm$0.9 & 23$\pm$1 & 2.73 & ...      & ...     & ... & 5.0$\pm$0.8 & 8.7$\pm$0.1 & 1.91  & ... & ...          & ... & 9$\pm$1 & 3.66$\pm$0.01 & 2.74 & ... & ... & ...\\
  10-9 & 6.0$\pm$0.9 & 24$\pm$2 & 2.70 & ...      & ...     & ... & 5.0$\pm$0.8 & 9.0$\pm$0.2 & 3.69  & ... & ...          & ... & 9$\pm$1 & 3.64$\pm$0.03 & 2.86 & 21$\pm$4 & 7.2$\pm$0.2 & 0.53\\
 11-10 & 5.8$\pm$0.8 & 22$\pm$3 & 4.69 & ...      & ...     & ... & 3.0$\pm$0.5 &11.2$\pm$0.8 & 3.24  & ... & ...          & ... & 9$\pm$1 & 3.03$\pm$0.09 & 3.51 & 21$\pm$4 & 7.5$\pm$0.3 & 0.33 \\
 12-11 & 5.8$\pm$0.8 & 25$\pm$2 & 1.57 & 10$\pm$1 &30$\pm$9 &1.82 & 4.5$\pm$0.7 &11.3$\pm$0.1 & 5.06  & ... & ...          & ... & 9$\pm$1 & 2.9$\pm$0.5   & 3.75 & 21$\pm$4 & 6.8$\pm$0.6 & 0.15\\
 15-14 & 5.8$\pm$0.8 & 25$\pm$6 & 3.20 & 9$\pm$1 &27$\pm$8 &2.66 & 4.5$\pm$0.7 &11.0$\pm$0.5 & 5.13  & ... & ...          & ... & 9$\pm$1 & 3.7$\pm$0.4   & 5.06 & 21$\pm$4 & 7.6$\pm$0.3 & 0.48\\
 16-15 & 5.8$\pm$0.8 & 24$\pm$4 & 5.27 & 9$\pm$1 &27$\pm$3 &3.23 & 5.5$\pm$0.8 &11.8$\pm$0.2 & 5.66  & ... & ...          & ... & 9$\pm$1 & 2.7$\pm$0.2   & 5.66 & 21$\pm$4 & 7.3$\pm$0.2 & 0.61 \\
\hline 
\end{tabular}
\end{center}
\tablefoot{The fit errors are provided by CLASS. $V$$_{\mathrm{LSR}}$ is the LSR central velocity, $\Delta$$V$ is the line-width, and $T$$_{\mathrm{MB}}$ is the main beam temperature. The units of $V$$_{\mathrm{LSR}}$, $\Delta$$V$, and $T$$_{\mathrm{MB}}$ are (km s$^{-1}$), (km s$^{-1}$), and (K), respectively, in all the cases.}\\
\label{table:sHC3N_parameters_gaussian}
\end{table}
\end{landscape}

\begin{table*}
\caption{Lines of DC$_{3}$N with emission frequencies whithin the 30m-IRAM survey.}   
\centering   
\begin{tabular}{c c c c c c c c c}     
\hline\hline       
Species & Transition     & Predicted   &  $S$$_{\mathrm{ij}}$ & $E$$_{\mathrm{u}}$ & Observed & Observed              & Observed             &  Blended \\
        & $_{J}$-$_{J'}$ & freq.       &                    &  (K)       & freq.              & $V$$_{\mathrm{LSR}}$  & $T$$_{\mathrm{MB}}$  &   with \\
        &                &  (MHz)      &                    &            & (MHz)              &  (km s$^{-1}$)        & (K)                  &        \\
\hline                    
   DC$_{3}$N & 10-9              &  84429.809  &  10.0  & 22.3   &  84430.5 & 6.6   &  0.04  & (CH$_{3}$)$_{2}$CO \\ 
   DC$_{3}$N & 11-10             &  92872.373  &  11.0  & 26.7   &  92872.5 & 8.6   &  0.03  &  \\
   DC$_{3}$N & 12-11             & 101314.817  &  12.0  & 31.6   &  101314.5& 9.9   &  0.04  &   \\ 
   DC$_{3}$N & 13-12             & 109757.131  &  13.0  & 36.9   &  ...     & ...   &  ...   & SO$_{2}$, HCOOCH$_{3}$ $\nu$$_{t}$=1\\
   DC$_{3}$N & 16-15             & 135083.185  &  16.0  & 55.1   &  135083.9& 7.4   &  0.11  &  \\
   DC$_{3}$N & 17-16             & 143524.870  &  17.0  & 62.0   &  ...     & ...   &  ....  & CH$_{3}$CH$_{2}$CN \\
   DC$_{3}$N & 18-17             & 151966.371  &  18.0  & 69.3   &  151966.5& 8.7   &  0.13  &  \\
   DC$_{3}$N & 19-18             & 160407.677  &  19.0  & 77.0   &  ...     & ...   &  ...   & (CH$_{3}$)$_{2}$CO \\
   DC$_{3}$N & 20-19             & 168848.777  &  20.0  & 85.1   &  ...     & ...   &  ...   & H$_{2}$C$^{34}$S, CH$_{3}$CH$_{2}$CN $\nu$$_{13}$/$\nu$$_{21}$\\
   DC$_{3}$N & 21-20             & 177289.660  &  21.0  & 93.6   &  ...     & ...   &  ...   & HCN \\
   DC$_{3}$N & 24-23             & 202610.900  &  24.0  & 121.6  &  202612.5& 6.6   &  0.16  &      \\
   DC$_{3}$N & 25-24             & 211050.808  &  25.0  & 131.7  &  211052.5& 6.6   &  0.18  & \\
   DC$_{3}$N & 26-25             & 219490.444  &  26.0  & 142.2  &  ...     & ...   &  ...   & CH$_{3}$OCH$_{3}$ \\
   DC$_{3}$N & 27-26             & 227929.799  &  27.0  & 153.2  &  ...     & ...   &  ...   & CH$_{2}$CHCN \\
   DC$_{3}$N & 28-27             & 236368.861  &  28.0  & 164.5  &  ...     & ...   &  ...   & HCOOCH$_{3}$ \\
   DC$_{3}$N & 29-28             & 244807.619  &  29.0  & 176.3  &  244809.2& 7.1   &  0.15  & H$^{13}$COOCH$_{3}$ $\nu$$_{t}$=1\\
   DC$_{3}$N & 30-29             & 253246.063  &  30.0  & 188.4  &  ...     & ...   &  ...   & CH$_{3}$OH  \\
   DC$_{3}$N & 31-30             & 261684.183  &  31.0  & 201.0  &  ...     & ...   &  ...   & H$^{13}$COOCH$_{3}$ $\nu$$_{t}$=1 \\
   DC$_{3}$N & 32-31             & 270121.966  &  32.0  & 213.9  &  270123.8& 7.0   &  0.31  & HCOOCH$_{3}$ $\nu$$_{t}$=1\\
   DC$_{3}$N & 33-32             & 278559.402  &  33.0  & 227.3  &  ...     & ...   &  ...   & HCOOCH$_{3}$, CH$_{3}$CH$_{2}$CN \\ 
\hline      
\label{table:DC3N}
\end{tabular}
\tablefoot{Column 1 indicates the
species, Col.2 the quantum numbers of the line transition, Col. 3 gives the
assumed rest frequencies, Col. 4 the line strength, 
Col. 5 the energy of the upper level, Col. 6 observed frequency
assuming a $V$$_{\mathrm{LSR}}$ of
9.0 km s$^{-1}$, Col. 7 the observed radial velocities, Col. 8 the peak line main beam temperature, and Col. 9 the blended species.}
\end{table*}

\longtab{4}{
\begin{longtable}{lrlrlrlrl}
\caption{Observed lines of HC$_{5}$N \label{table:tab_HC5N_vib}.}\\
\hline\hline                      
Species & Transition     & Predicted  & S$_{\mathrm{ij}}$ & E$_{\mathrm{U}}$ & Observed   &  Observed          & Observed                    &  Blended \\
        & $_{J}$-$_{J'}$ & freq.      &                   &(K)               & freq.      & v$_{\mathrm{LSR}}$ &  $T$$_{\mathrm{MB}}$        &   with   \\
        &                & (MHz)      &                   &                  & (MHz)      & (km s$^{-1}$)      &   (K)                       &          \\
\hline
\endfirsthead
\caption{continued.}\\
\hline\hline
Species & Transition         & Predicted   &  S$_{\mathrm{ij}}$ &  E$_{\mathrm{U}}$   & Observed    & Observed                  & Observed                &  Blended \\
        & $_{J}$-$_{J'}$ & freq.      &                   &(K)               & freq.      & $V$$_{\mathrm{LSR}}$ &  $T$$_{\mathrm{MB}}$        &   with   \\
        & $_{J}$-$_{J'}$ & freq.       &                    &  (K)       & freq.              & $V$$_{\mathrm{LSR}}$  & $T$$_{\mathrm{MB}}$  &   with \\
        &                &  (MHz)      &                    &            & (MHz)              &  (km s$^{-1}$)        & (K)                  &        \\

\hline
\endhead
\hline
\endfoot                  
   HC$_{5}$N & 31-30            &  82539.038   & 31.0   & 63.4   &  82539.5  & 7.3  &  0.13  &  \\
   HC$_{5}$N & 32-31            &  85201.345   & 32.0   & 67.5   &  85201.5  & 8.5  &  0.14  &   \\
   HC$_{5}$N & 33-32            &  87863.629   & 33.0   & 71.7   &  87863.5  & 9.4  &  0.09  &\\
   HC$_{5}$N & 34-33            &  90525.889   & 34.0   & 76.0   &  90526.5  & 7.0  &  0.12  &  \\
   HC$_{5}$N & 35-34            &  93188.124   & 35.0   & 80.5   &  93188.5  & 7.8  &  0.11  &  \\
   HC$_{5}$N & 36-35            &  95850.334   & 36.0   & 85.1   &  95850.5  & 8.5  &  0.12  &  \\
   HC$_{5}$N & 37-36            &  98512.519   & 37.0   & 89.8   &  98512.5  & 9.1  &  0.11  &  \\
   HC$_{5}$N & 38-37            & 101174.676   & 38.0   & 94.7   & 101174.5  & 9.5  &  0.12  & \\
   HC$_{5}$N & 39-38            & 103836.806   & 39.0   & 99.7   & 103836.5  & 9.9  &  0.12  &  \\
   HC$_{5}$N & 40-39            & 106498.908   & 40.0   & 104.8  & 106499.5  & 7.3  &  0.11  &  \\
   HC$_{5}$N & 41-40            & 109160.980   & 41.0   & 110.0  & 109160.5  &10.3  &  0.14  & HC$_{3}$N \\
   HC$_{5}$N & 42-41            & 111823.024   & 42.0   & 115.4  & 111823.5  & 7.7  &  0.12  & CH$_{3}$COOCH$_{3}$ \\
   HC$_{5}$N & 43-42            & 114485.036   & 43.0   & 120.9  & 114485.5  & 7.8  &  0.13  &  \\
   HC$_{5}$N & 49-48            & 130456.436   & 49.0   & 156.5  & 130456.4  & 9.1  &  0.15  & \\
   HC$_{5}$N & 50-49            & 133118.217   & 50.0   & 162.9  & 133117.6  &10.4  &  0.14  & \\
   HC$_{5}$N & 51-50            & 135779.961   & 51.0   & 169.4  &  ...      & ...  &  ...   & $^{34}$SO \\
   HC$_{5}$N & 52-51            & 138441.668   & 52.0   & 176.1  & 138443.9  & 4.2  &  0.21  &  \\
   HC$_{5}$N & 53-52            & 141103.338   & 53.0   & 182.9  & 141106.4  & 2.5  &  0.18  & $^{13}$CH$_{3}$OH \\
   HC$_{5}$N & 54-53            & 143764.970   & 54.0   & 189.8  &  ...      & ...  &  ...   & CH$_{2}$CHCN \\
   HC$_{5}$N & 55-54            & 146426.562   & 55.0   & 196.8  & 146427.6  & 6.9  &  0.12  &  \\
   HC$_{5}$N & 56-55            & 149088.115   & 56.0   & 203.9  & 149088.9  & 7.4  &  0.15  &  \\
   HC$_{5}$N & 57-56            & 151749.628   & 57.0   & 211.2  &  ...      & ...  &  ...   & $^{13}$CH$_{3}$CH$_{2}$CN, CH$_{3}$SH \\
   HC$_{5}$N & 58-57            & 154411.099   & 58.0   & 218.6  &  ...      & ...  &  ...   & HNCO \\
   HC$_{5}$N & 59-58            & 157072.528   & 59.0   & 226.2  & 157072.6  & 8.9  &  0.19  &  \\
   HC$_{5}$N & 60-59            & 159733.915   & 60.0   & 233.8  &  ...      & ...  &  ...   & HCOOCH$_{3}$ \\
   HC$_{5}$N & 61-60            & 162395.259   & 61.0   & 241.6  & 162397.6  & 4.7  &  0.13  &  \\
   HC$_{5}$N & 62-61            & 165056.558   & 62.0   & 249.6  &  ...      & ...  &  ...   & CH$_{3}$OH \\
   HC$_{5}$N & 63-62            & 167717.813   & 63.0   & 257.6  & 167718.9  & 7.1  &  0.22  &  \\
   HC$_{5}$N & 64-63            & 170379.022   & 64.0   & 265.8  &  ...      & ...  &  ...   & CH$_{3}$CH$_{2}$CN \\
   HC$_{5}$N & 65-64            & 173040.185   & 65.0   & 274.1  & 173040.1  & 9.1  &  0.19  & \\
   HC$_{5}$N & 66-65            & 175701.302   & 66.0   & 282.5  &  ...      & ...  &  ...   & $^{13}$CH$_{2}$CHCN, CH$_{3}$CH$_{2}$OH \\
   HC$_{5}$N & 74-73            & 196988.455   & 74.0   & 354.6  & 196988.6  & 8.7  &  0.09  & $^{13}$CH$_{2}$CHCN \\
   HC$_{5}$N & 75-74            & 199649.117   & 75.0   & 364.2  &  ...      & ...  &  ...   & CH$_{3}$CH$_{2}$CN \\
   HC$_{5}$N & 76-75            & 202309.724   & 76.0   & 373.9  &  ...      & ...  &  ...   & CH$_{3}$CN \\
   HC$_{5}$N & 77-76            & 204970.277   & 77.0   & 383.7  &  ...      & ...  &  ...   & HCOOCH$_{3}$ $\nu$$_{t}$=2 \\ 
   HC$_{5}$N & 78-77            & 207630.774   & 78.0   & 393.7  &  ...      & ...  &  ...   & HNCO \\ 
   HC$_{5}$N & 79-78            & 210291.214   & 79.0   & 403.8  & 210292.3  & 7.5  &  0.39  &  \\
   HC$_{5}$N & 81-80            & 215611.924   & 81.0   & 424.3  & 215613.6  & 6.7  &  0.40  & t-CH$_{3}$CH$_{2}$OH \\
             &                  &              &        &        &           &      &        & HCOOCH$_{3}$ $\nu$$_{t}$=1\\
   HC$_{5}$N & 82-81            & 218272.192   & 82.0   & 434.8  & 218272.3  & 8.9  &  0.23  & CH$_{3}$CH$_{3}$CHO \\
   HC$_{5}$N & 83-82            & 220932.400   & 83.0   & 445.4  & 220936.1  & 4.0  &  0.49  & HCOOCH$_{3}$\\
   HC$_{5}$N & 84-83            & 223592.549   & 84.0   & 456.1  &  ...      & ...  &  ...   & CH$_{3}$$^{13}$CH$_{2}$CN \\ 
   HC$_{5}$N & 85-84            & 226252.637   & 85.0   & 467.0  &  ...      & ...  &  ...   & CH$_{2}$CHCN \\ 
   HC$_{5}$N & 86-85            & 228912.664   & 86.0   & 478.0  & 228911.1  &11.0  &  0.36  & HC$_{3}$N $\nu$$_{7}$=2, DNC \\
   HC$_{5}$N & 87-86            & 231572.628   & 87.0   & 489.1  & 231573.6  & 7.7  &  0.13  &  \\
   HC$_{5}$N & 88-87            & 234232.530   & 88.0   & 500.3  & 234234.6  & 6.4  &  0.45  & CH$_{3}$CH$_{2}$CN \\
   HC$_{5}$N & 89-88            & 236892.369   & 89.0   & 511.7  & 236896.5  & 3.7  &  0.35  & CH$_{2}$$^{13}$CHCN \\ 
   HC$_{5}$N & 90-89            & 239552.143   & 90.0   & 523.2  &  ...      & ...  &  ...   & g-CH$_{3}$CH$_{2}$OH \\ 
   HC$_{5}$N & 91-90            & 242211.853   & 91.0   & 534.8  &  ...      & ...  &  ...   & CH$_{3}$CH$_{2}$CN \\ 
   HC$_{5}$N & 92-91            & 244871.496   & 92.0   & 546.6  & 244872.3  & 8.0  &  0.49  & HCOOCH$_{3}$ \\
   HC$_{5}$N & 93-92            & 247531.074   & 93.0   & 558.5  & 247532.4  & 7.4  &  0.27  &  $^{13}$CH$_{2}$CHCN \\
   HC$_{5}$N & 94-93            & 250190.584   & 94.0   & 570.5  & 250191.3  & 8.1  &  0.21  & t-HCOOD \\
   HC$_{5}$N & 95-94            & 252850.027   & 95.0   & 582.6  &  ...      & ...  &  ...   & CH$_{3}$SH, HCOO$^{13}$CH$_{3}$ \\ 
   HC$_{5}$N & 96-95            & 255509.401   & 96.0   & 594.9  & 255511.3  & 6.8  &  0.34  & \\
   HC$_{5}$N & 97-96            & 258168.706   & 97.0   & 607.3  &  ...      & ...  &  ...   & CH$_{3}$CN $\nu$$_{8}$=1, $^{13}$CH$_{3}$OH \\ 
   HC$_{5}$N & 98-97            & 260827.941   & 98.0   & 619.8  &  ...      & ...  &  ...   & CH$_{3}$CH$_{2}$CN \\ 
   HC$_{5}$N & 100-99           & 266146.199   &100.0   & 645.2  &  ...      & ... &  ...   & CH$_{3}$CH$_{2}$CN $\nu$$_{13}$/$\nu$$_{21}$  \\ 
   HC$_{5}$N & 101-100          & 268805.220   &101.0   & 658.1  &  ...      & ... &  ...   & CH$_{3}$CH$_{2}$CN   \\ 
   HC$_{5}$N & 102-101          & 271464.168   &102.0   & 671.1  &  ...      & ... &  ...   & $^{13}$CH$_{3}$CH$_{2}$CN, (CH$_{3}$)$_{2}$CO  \\ 
   HC$_{5}$N & 103-102          & 274123.043   &103.0   & 684.3  & 274123.9  & 8.1 &  0.08  & HCOO$^{13}$CH$_{3}$ \\
   HC$_{5}$N & 104-103          & 276781.843   &104.0   & 697.6  & 276786.5  & 3.9 &  0.37  & $^{13}$CH$_{3}$OH  \\ 
   HC$_{5}$N & 105-104          & 279440.569   &105.0   & 711.0  &  ...      & ... &  ...   & $^{33}$SO$_{2}$, $^{34}$SO$_{2}$ \\ 
\hline 
\end{longtable}
\tablefoot{Column 1 indicates the
species, Col.2 the quantum numbers of the line transition, Col. 3 gives the
assumed rest frequencies, Col. 4 the line strength, 
Col. 5 the energy of the upper level, Col. 6 observed frequency
assuming a $V$$_{\mathrm{LSR}}$ of
9.0 km s$^{-1}$, Col. 7 the observed radial velocities, Col. 8 the peak line main beam temperature, and Col. 9 the blended species.}
}

\pagebreak

\begin{figure*}[ht!]
   \centering 
   \includegraphics[angle=0,width=18cm]{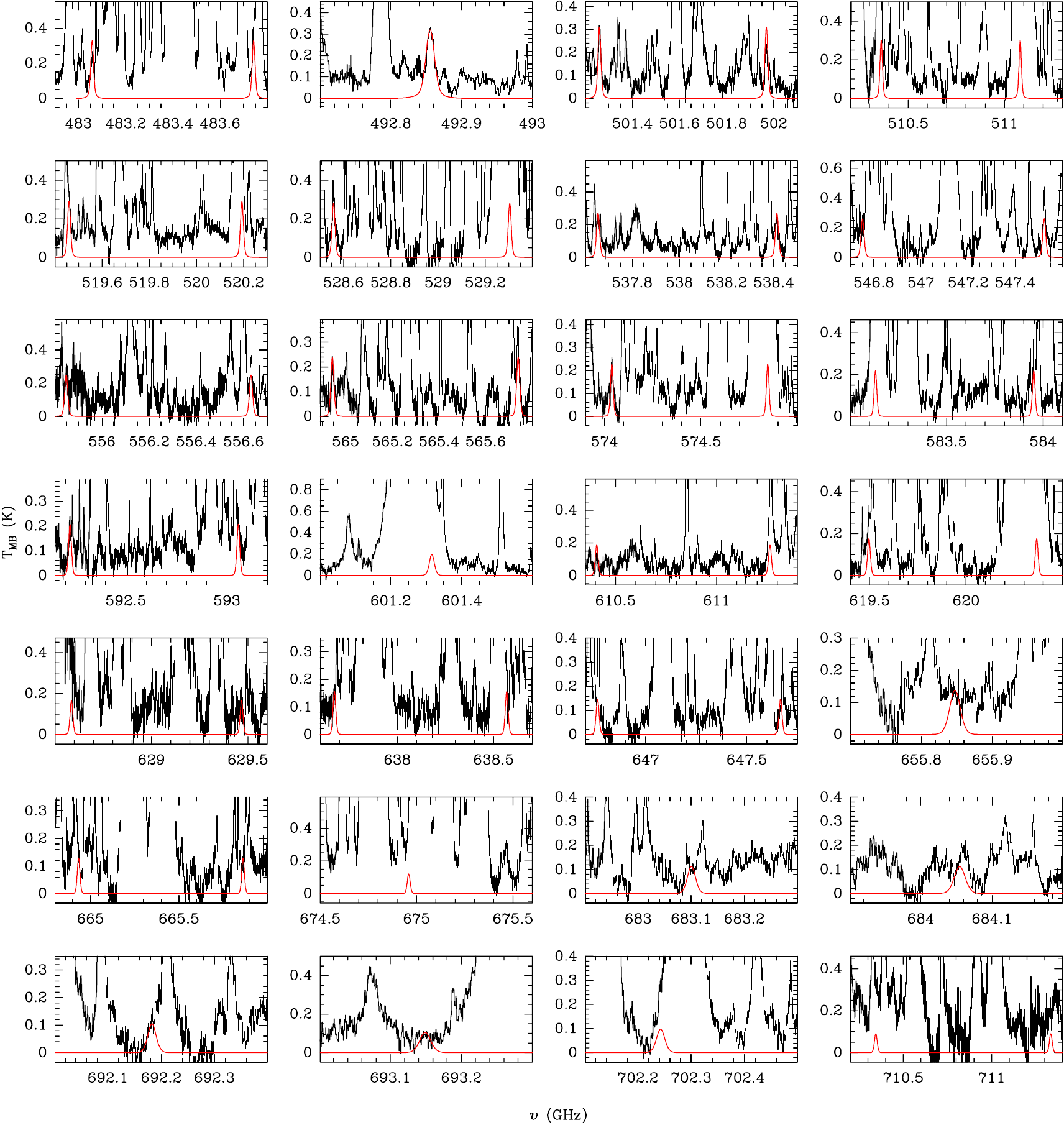}
   \caption{Observed spectra (black histogram) in the HIFI survey. Best fit LTE model results for HC$_{3}$N $\nu$$_{7}$ are shown in red.}
   \label{figure:HC3N_V7_mix_hifi}
   \end{figure*}

\begin{figure*}
   \centering 
   \includegraphics[angle=0,width=16cm]{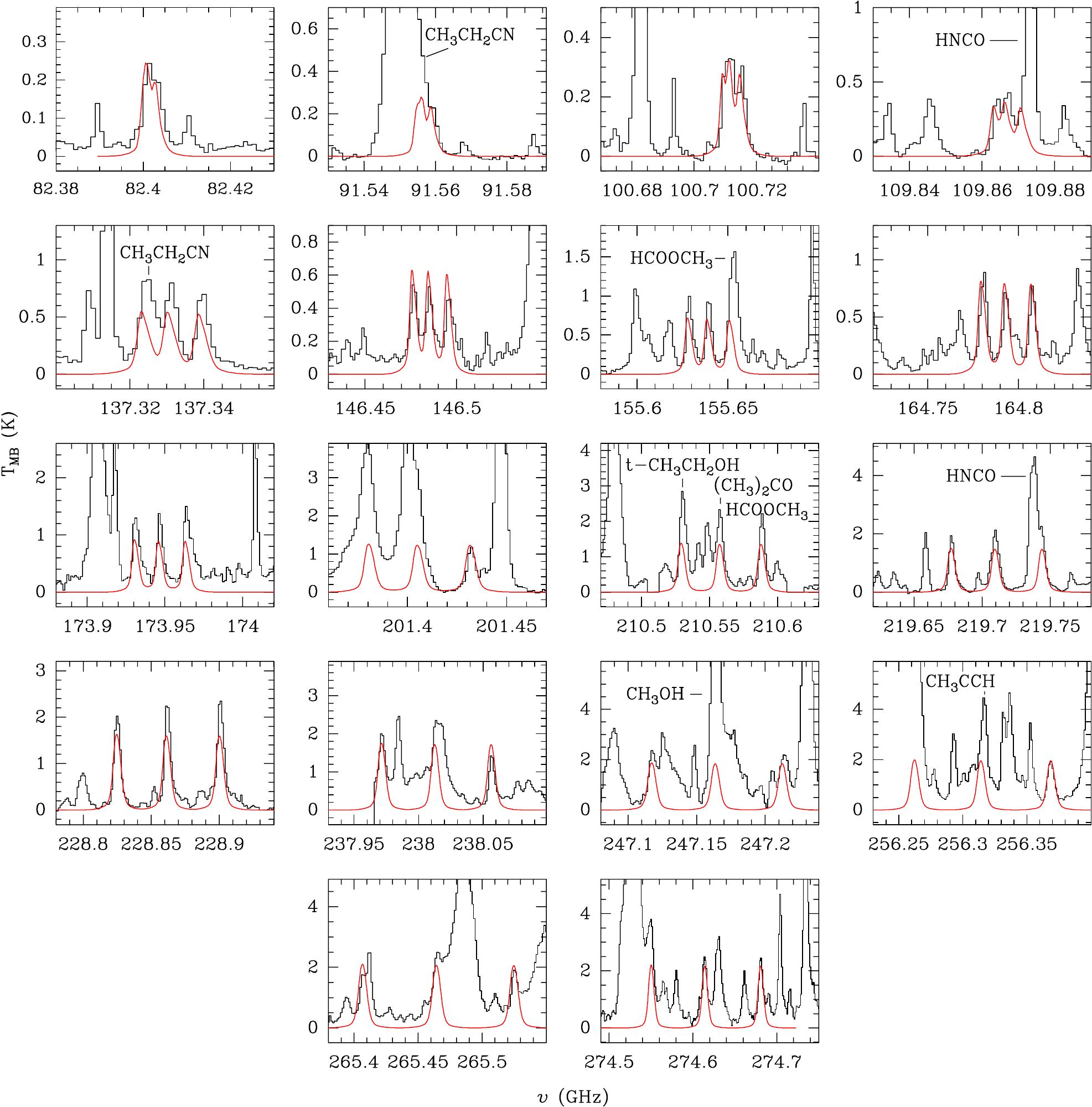}
   \caption{Observed lines of HC$_{3}$N 2$\nu$$_{7}$ (black histogram) in the IRAM survey. Best fit LTE model results are shown in red.}
    \label{figure:HC3N_2V7_mix_30m}
   \end{figure*}

\begin{figure*}
   \centering 
   \includegraphics[angle=0,width=18cm]{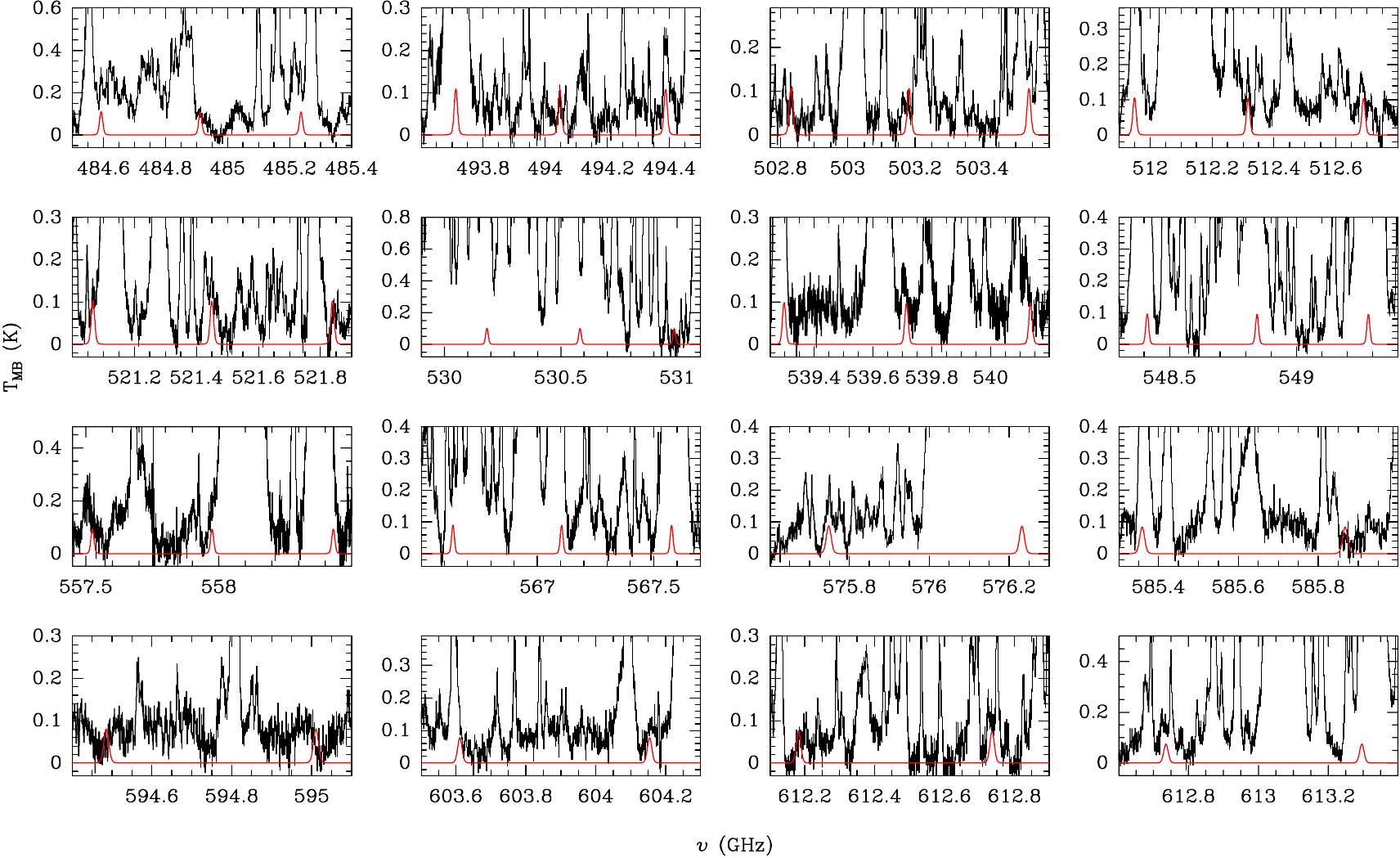}
   \caption{Observed spectra (black histogram) in the HIFI survey. Best fit LTE model results for HC$_{3}$N 2$\nu$$_{7}$ are shown in red.}
   \label{figure:HC3N_2V7_mix_hifi}
   \end{figure*}

\pagebreak

\begin{figure*}
   \centering 
   \includegraphics[angle=0,width=16cm]{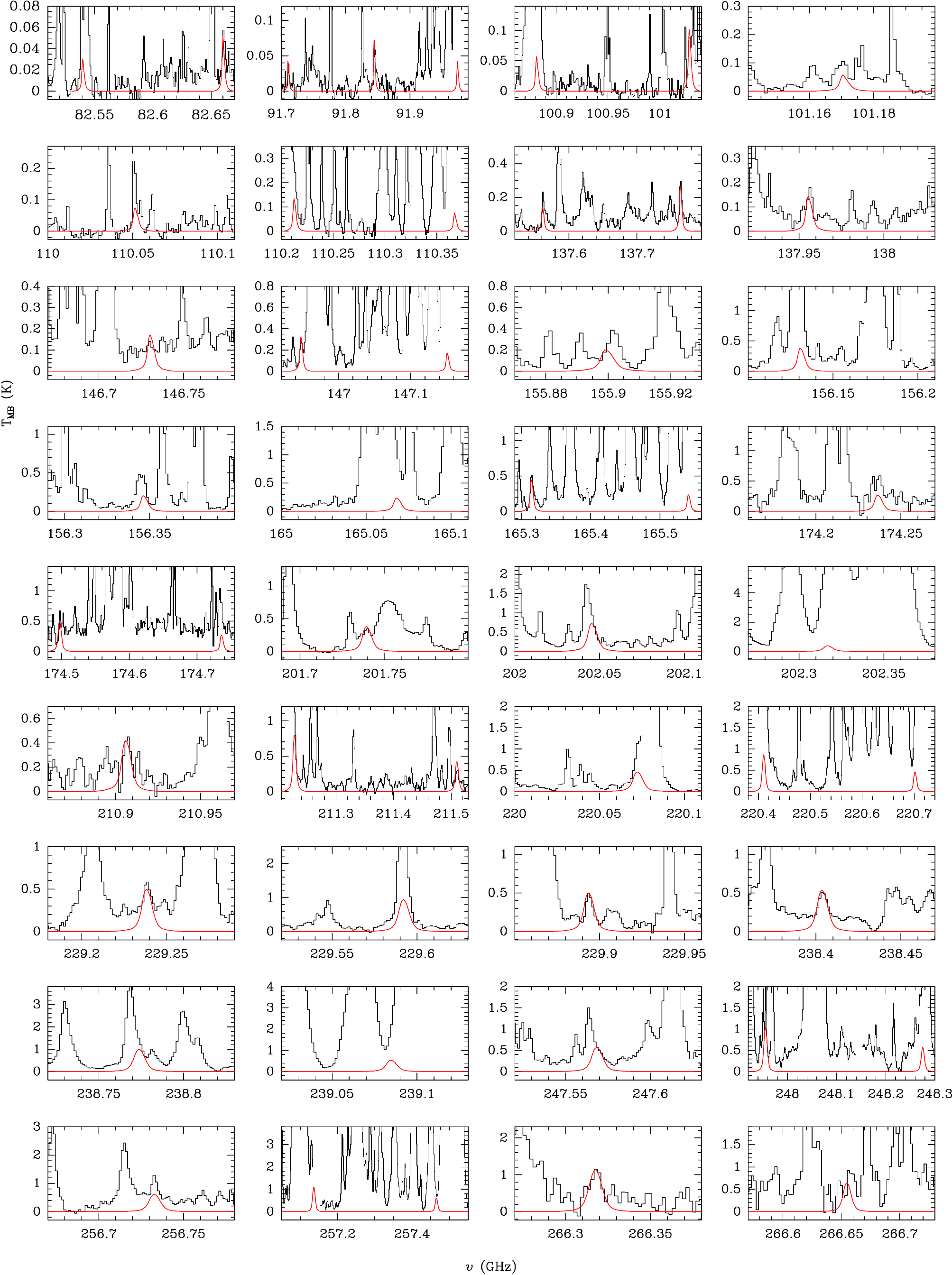}
   \caption{Observed lines of HC$_{3}$N 3$\nu$$_{7}$ (black histogram) in the IRAM survey. Best fit LTE model results are shown in red.}
    \label{figure:HC3N_3V7_mix_30m}
   \end{figure*}

\begin{figure*}
   \centering 
   \includegraphics[angle=0,width=16cm]{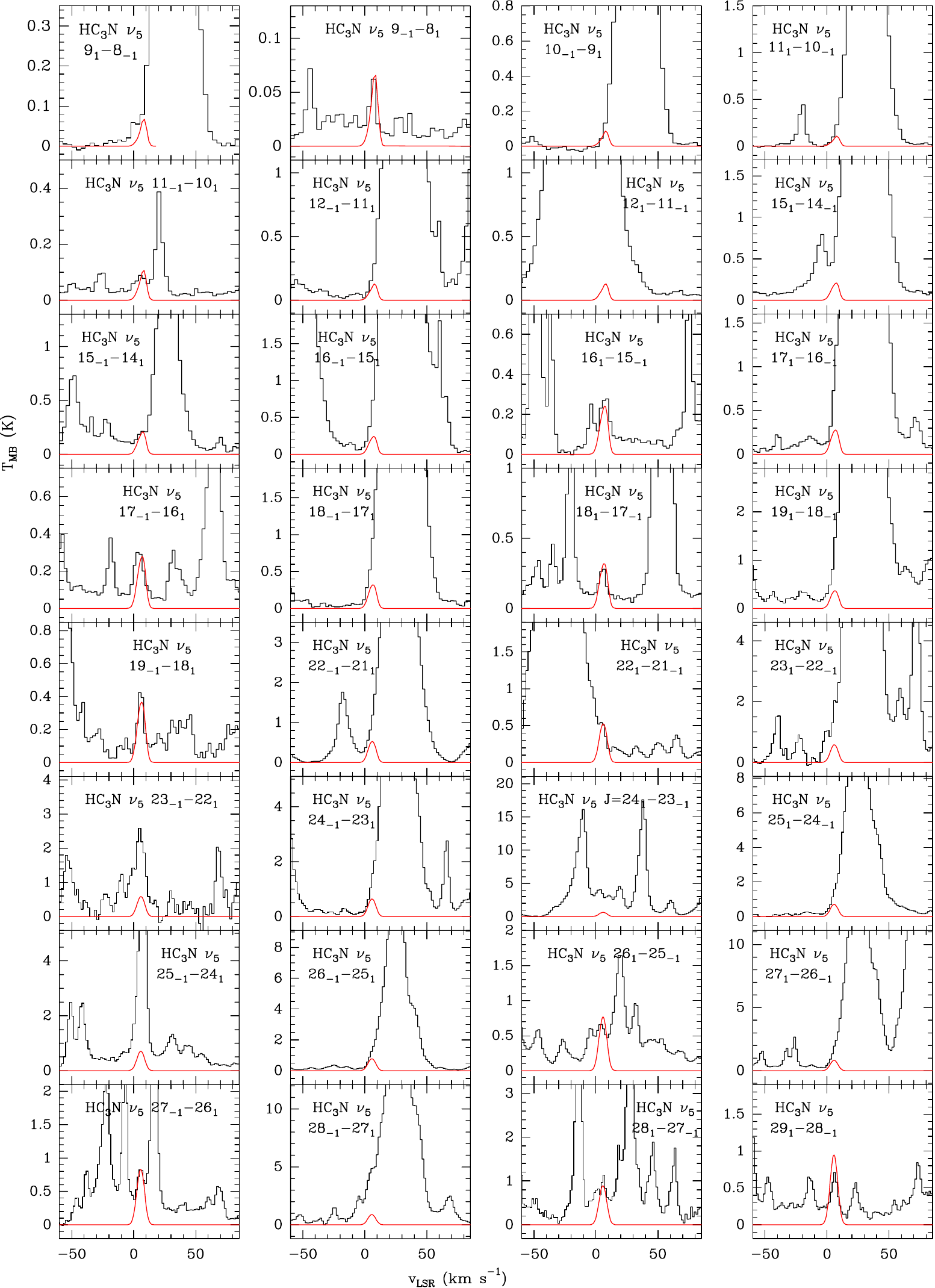}
   \caption{Detected lines of HC$_{3}$N $\nu$$_{5}$ (black histogram) in the IRAM survey. Best fit LTE model results are shown in red.}
    \label{figure:HC3N_V5_mix_30m}
   \end{figure*}

\begin{figure*}
   \centering 
   \includegraphics[angle=0,width=16cm]{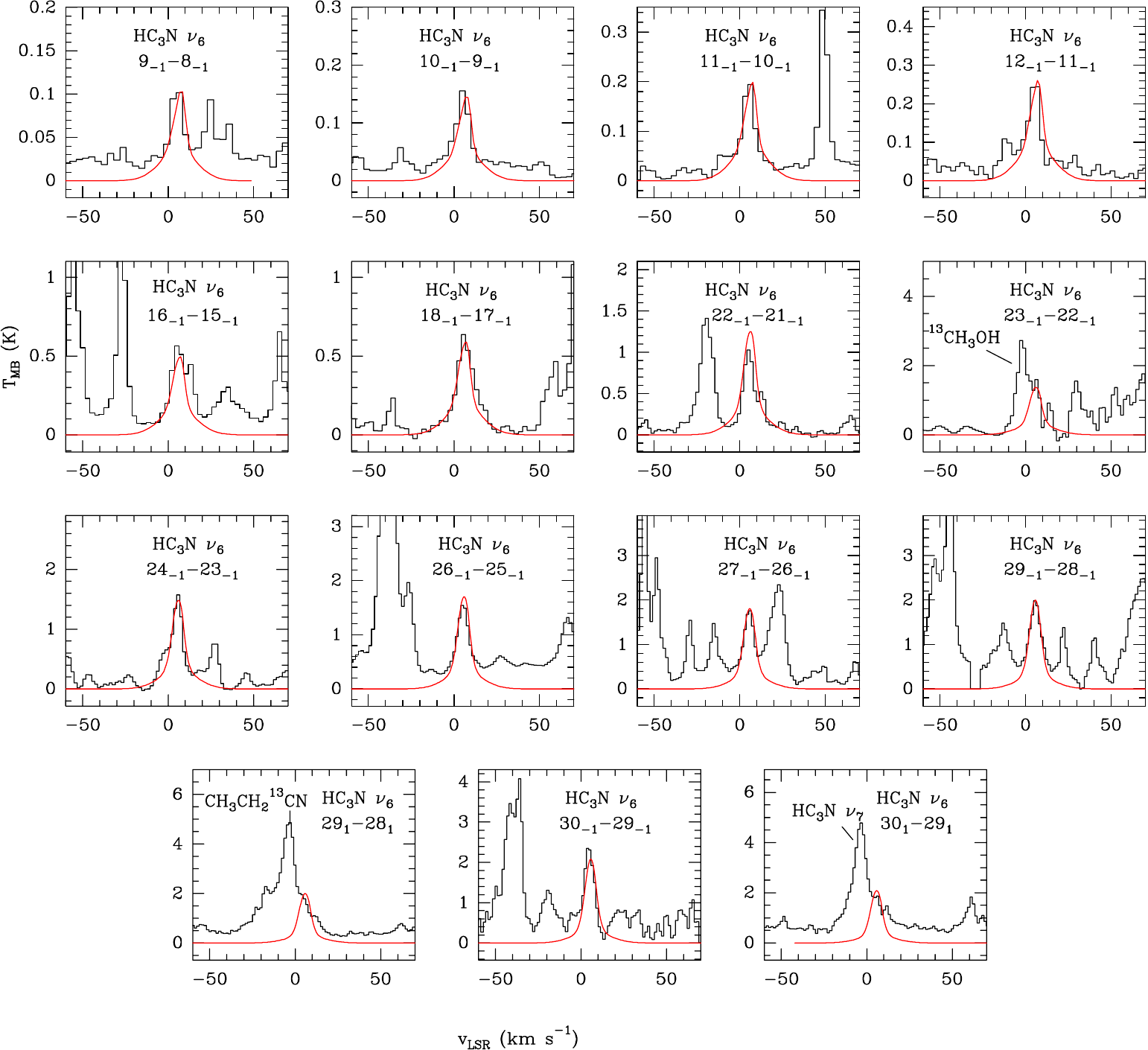}
   \caption{Observed lines of HC$_{3}$N $\nu$$_{6}$ (black histogram) in the IRAM survey. Best fit LTE model results are shown in red.}
    \label{figure:HC3N_V6_mix_30m}
   \end{figure*}

\begin{figure*}
   \centering 
   \includegraphics[angle=0,width=16cm]{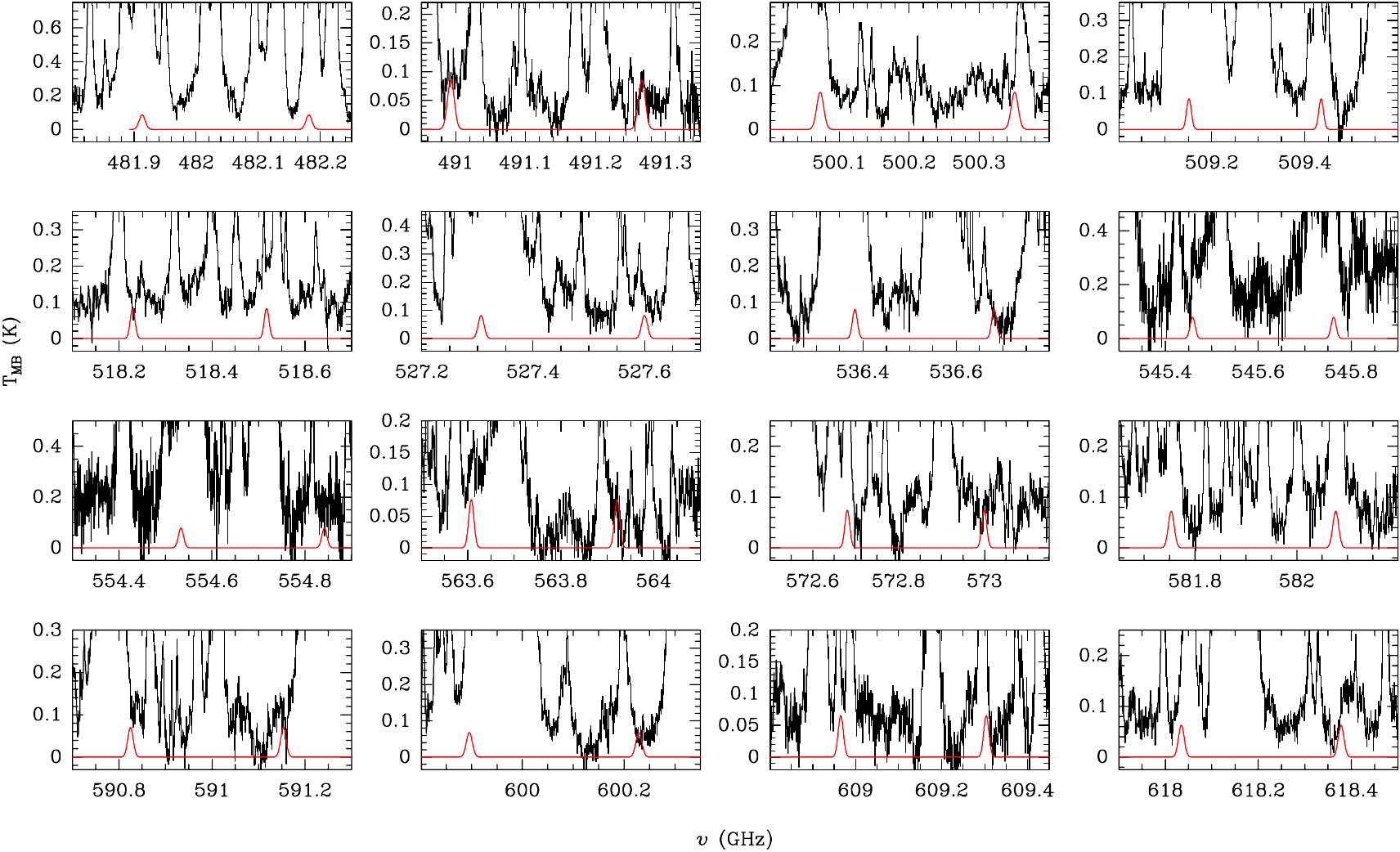}
   \caption{Observed spectra (black histogram) in the HIFI survey. Best fit LTE model results for HC$_{3}$N $\nu$$_{5}$ are shown in red.}
    \label{figure:HC3N_V5_mix_hifi}
   \end{figure*}

\begin{figure*}
   \centering 
   \includegraphics[angle=0,width=16cm]{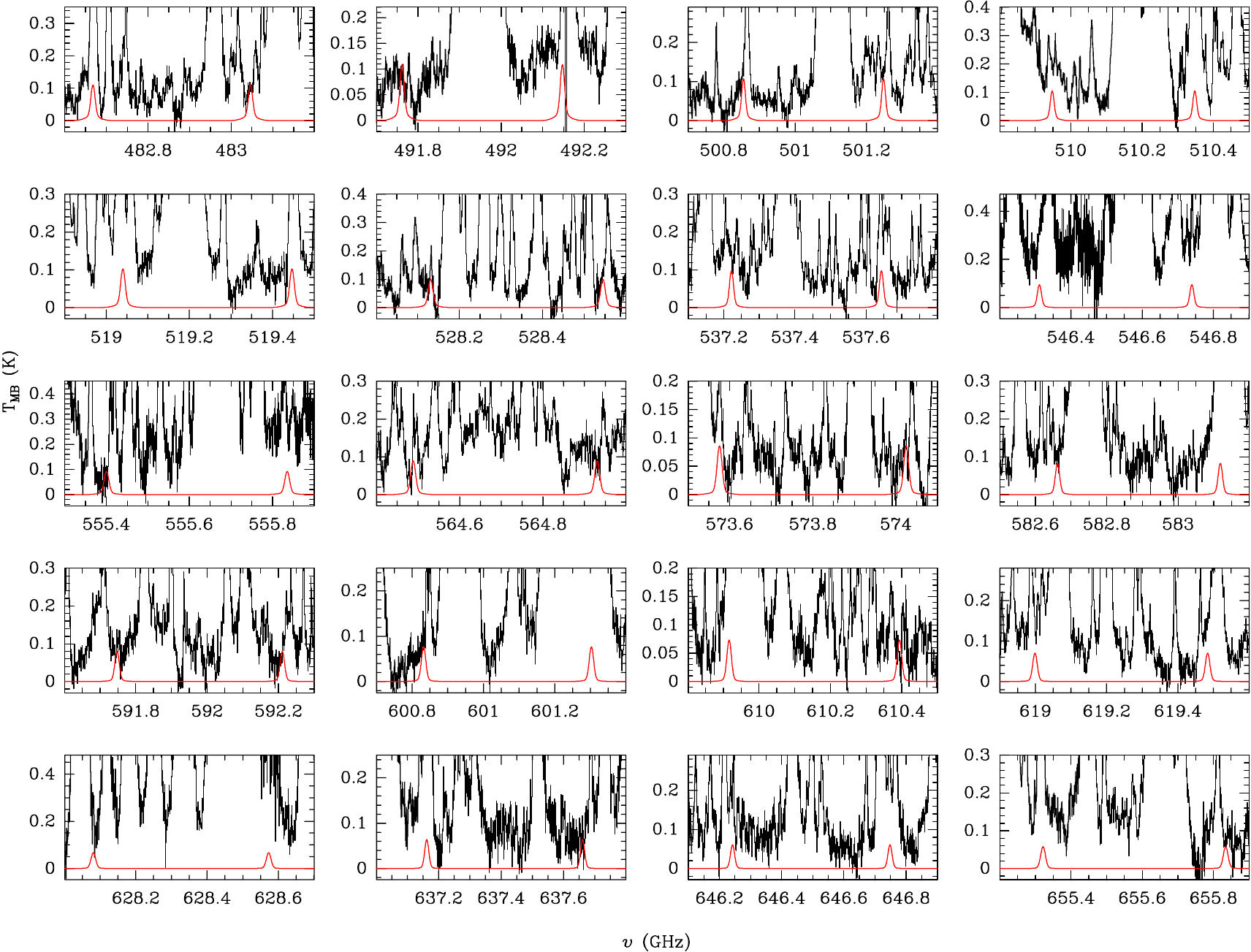}
   \caption{Observed spectra (black histogram) in the HIFI survey. Best fit LTE model results for HC$_{3}$N $\nu$$_{6}$ are shown in red.}
    \label{figure:HC3N_V6_mix_hifi}
   \end{figure*}

\begin{figure*}
   \centering 
   \includegraphics[angle=0,width=16cm]{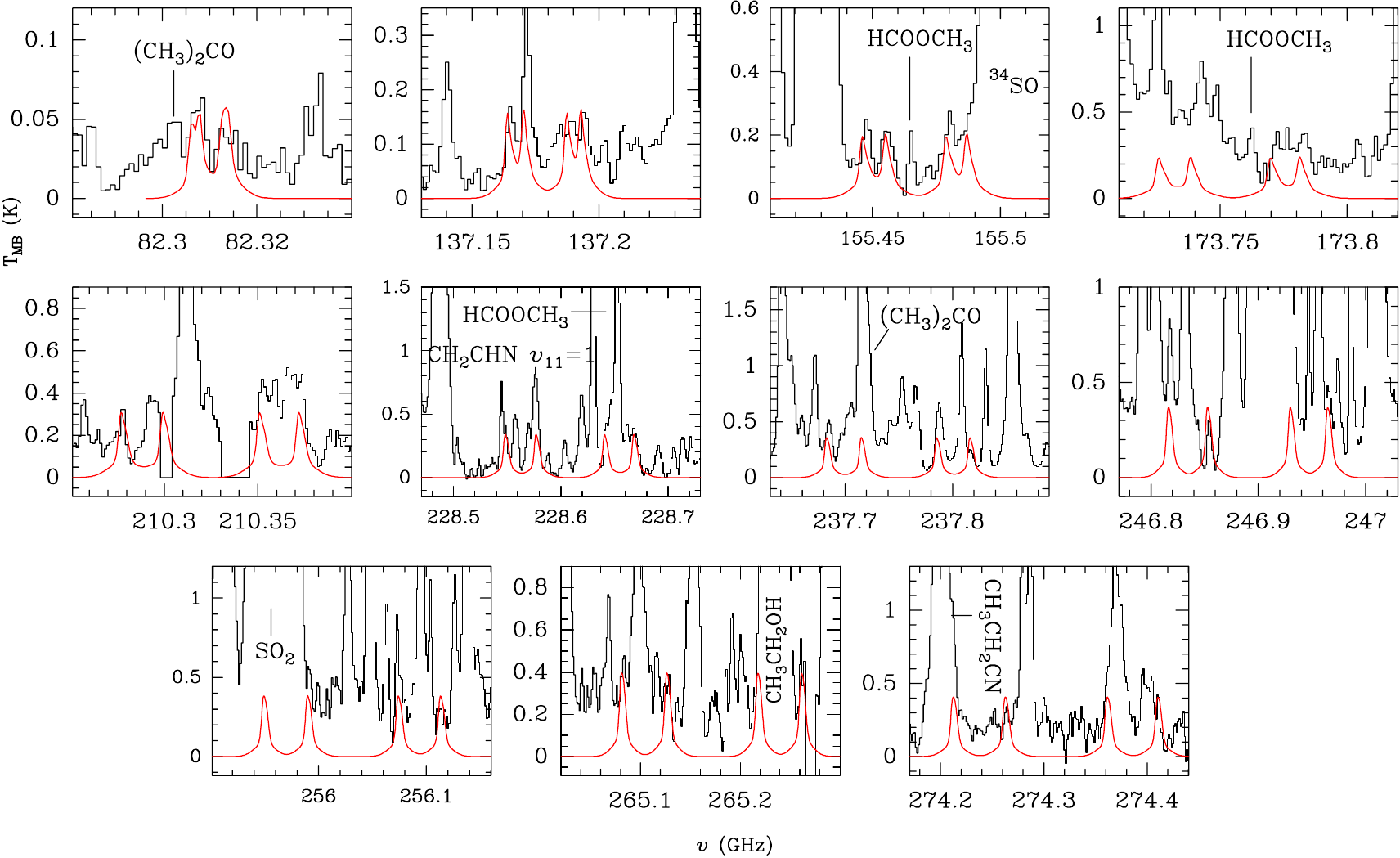}
   \caption{Observed lines of HC$_{3}$N $\nu$$_{6}$+$\nu$$_{7}$ (black histogram) in the IRAM survey. Best fit LTE model results are shown in red.}
    \label{figure:HC3N_V6+V7_mix_30m}
   \end{figure*}

\pagebreak

\begin{figure*}
   \centering 
   \includegraphics[angle=0,width=16cm]{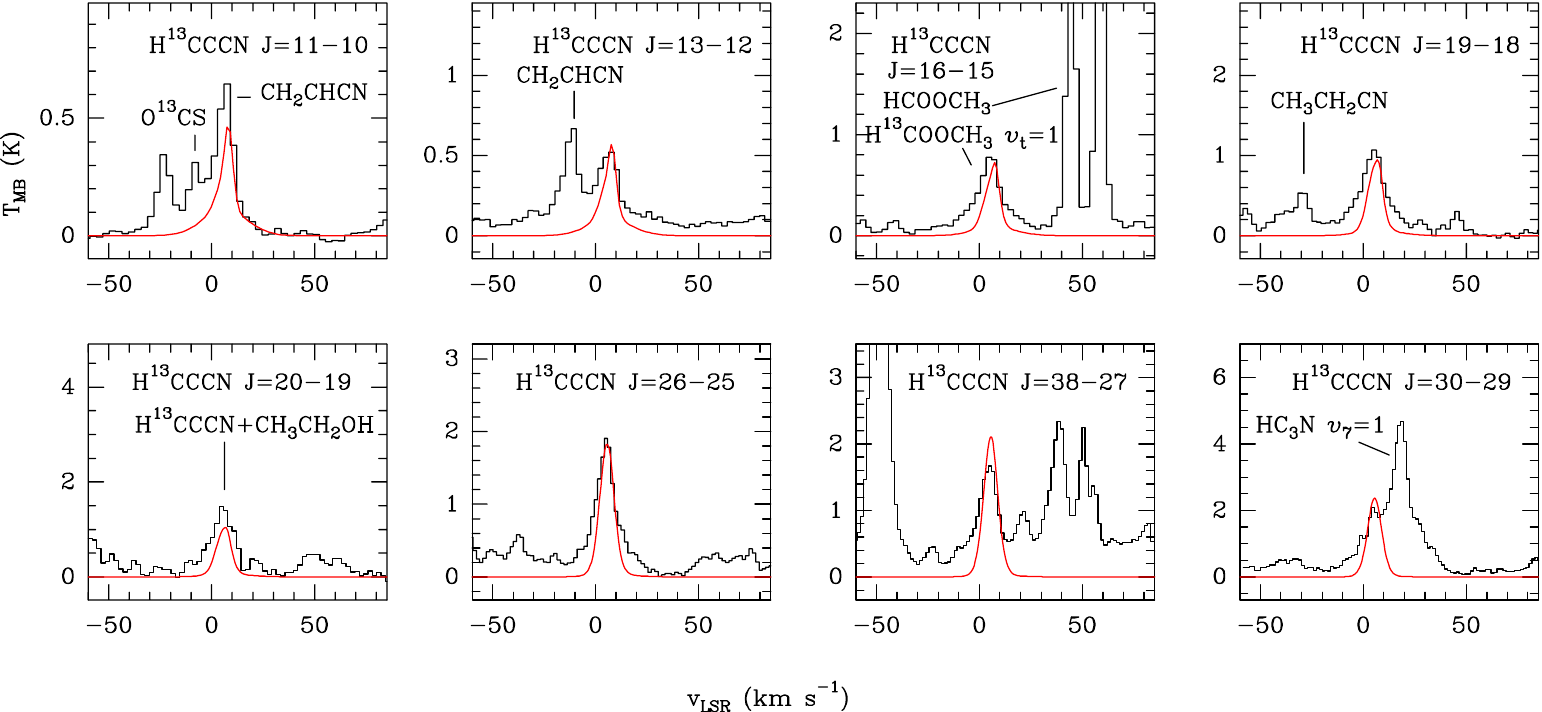}
   \caption{Observed lines of H$^{13}$CCCN (black histogram) in the IRAM survey. Best fit LVG model results are shown in red.}
    \label{figure:H13CCCN_mix_30m}
   \end{figure*}

\begin{figure*}
   \centering 
   \includegraphics[angle=0,width=16cm]{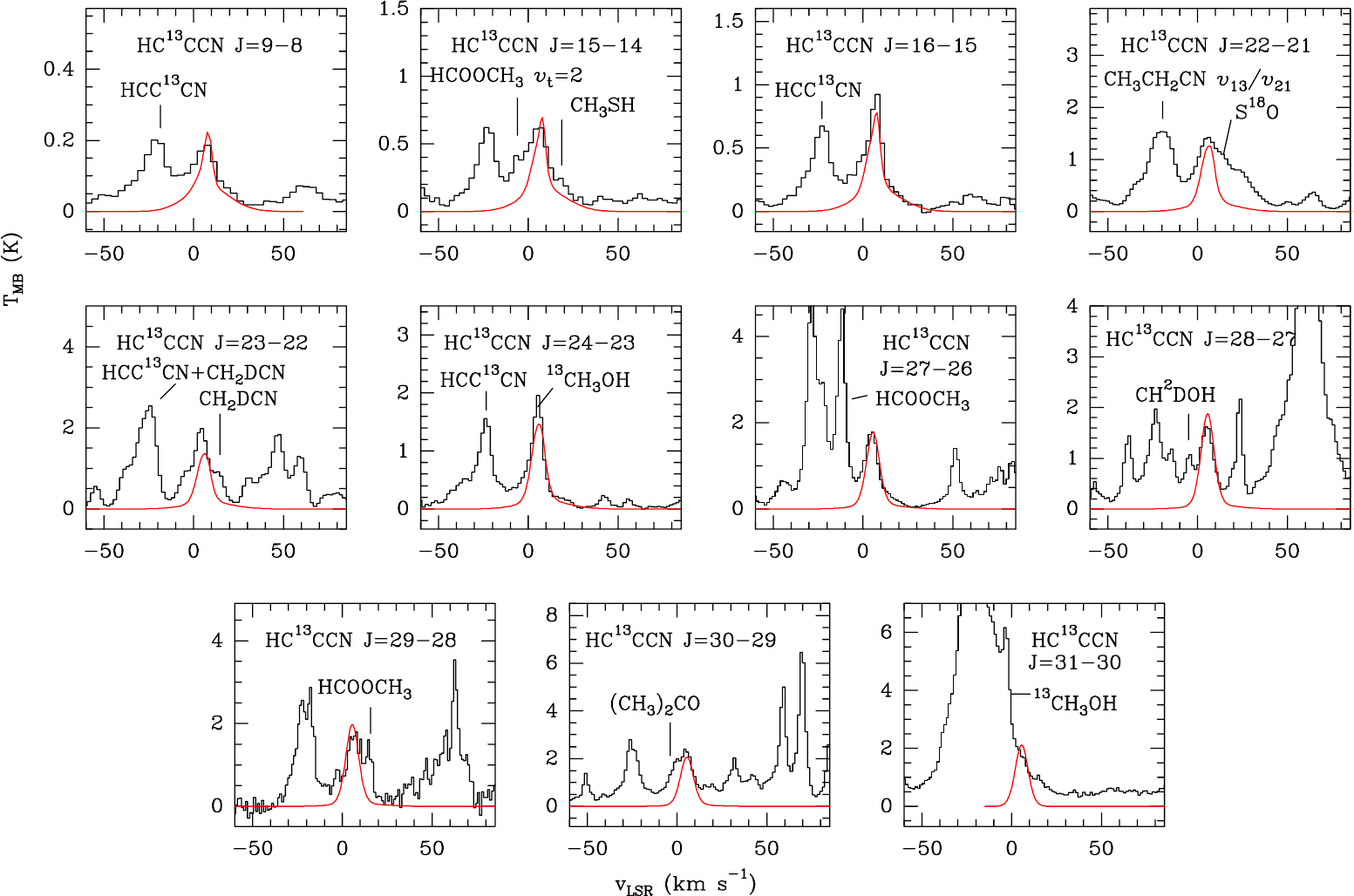}
   \caption{Observed lines of HC$^{13}$CCN (black histogram) in the IRAM survey. Best fit LVG model results are shown in red.}
    \label{figure:HC13CCN_mix_30m}
   \end{figure*}

\begin{figure*}
   \centering 
   \includegraphics[angle=0,width=16cm]{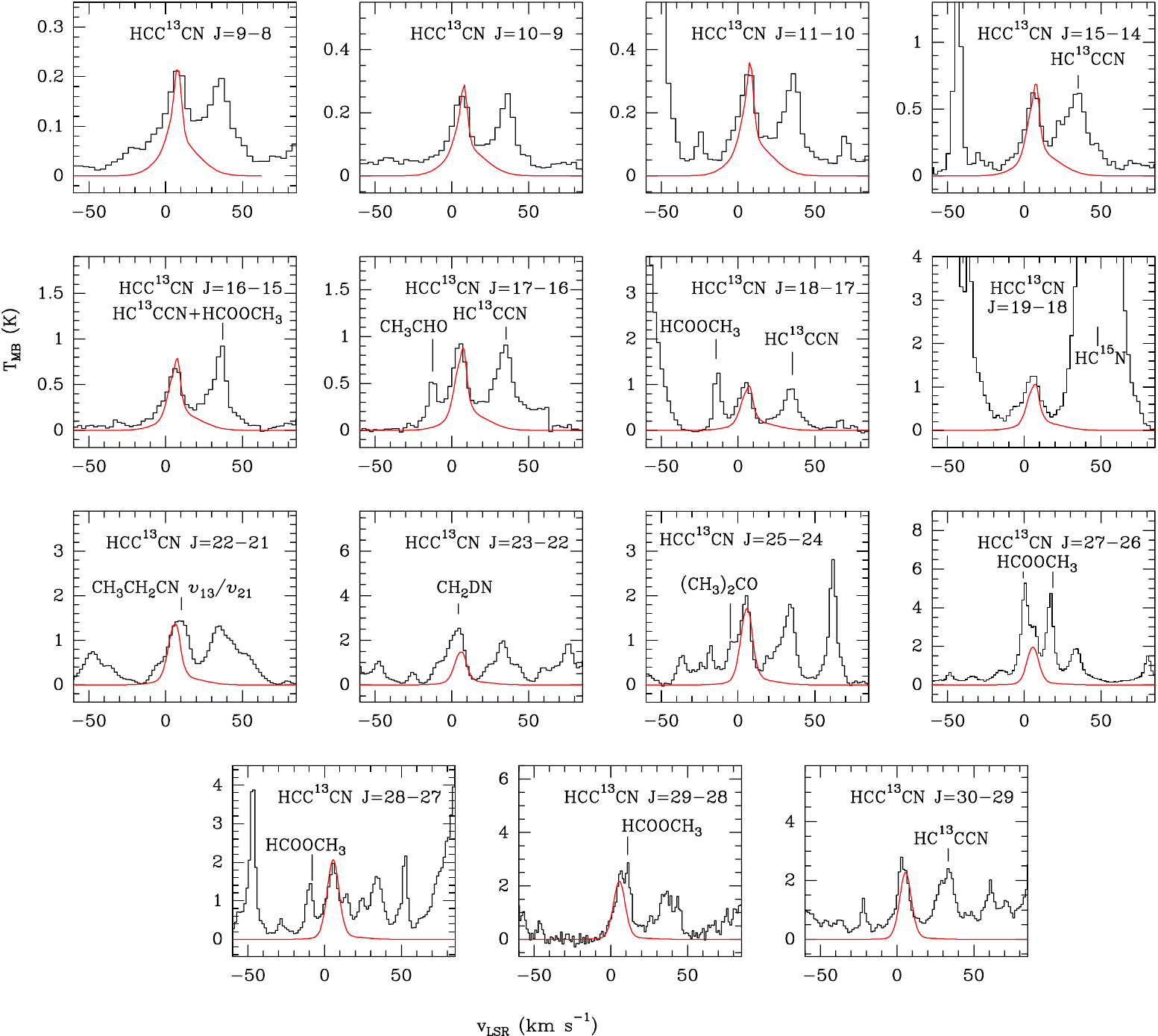}
   \caption{Observed lines of HCC$^{13}$CN (black histogram) in the IRAM survey. Best fit LVG model results are shown in red.}
    \label{figure:HCC13CN_mix_30m}
   \end{figure*}

\begin{figure*}
   \centering 
   \includegraphics[angle=0,width=17cm]{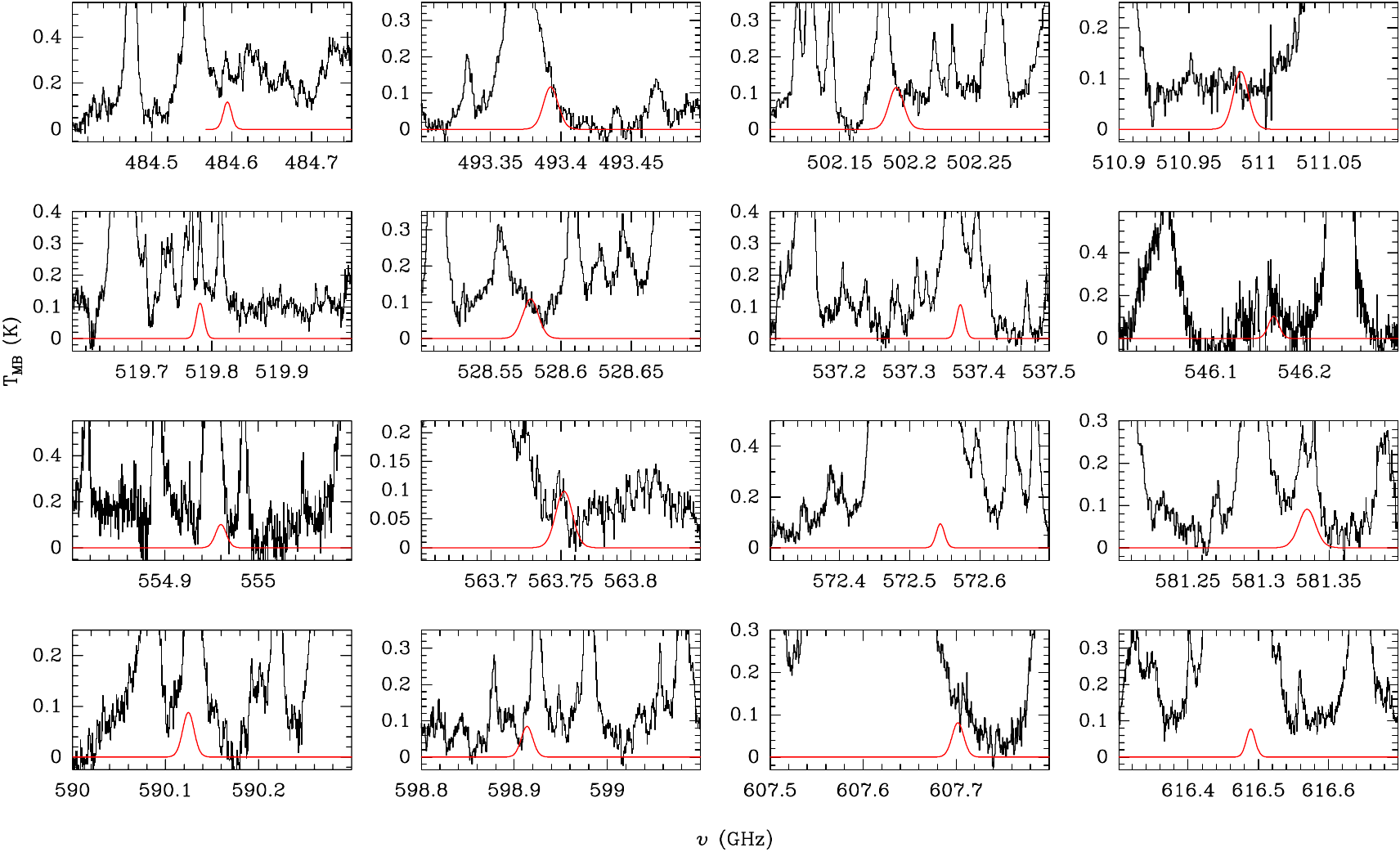}
   \caption{Observed spectra (black histogram) in the HIFI survey. Best fit LVG model results for H$^{13}$CCCN are shown in red.}
    \label{figure:H13CCCN_mix_hifi}
   \end{figure*}

\pagebreak

\begin{figure*}
   \centering 
   \includegraphics[angle=0,width=17cm]{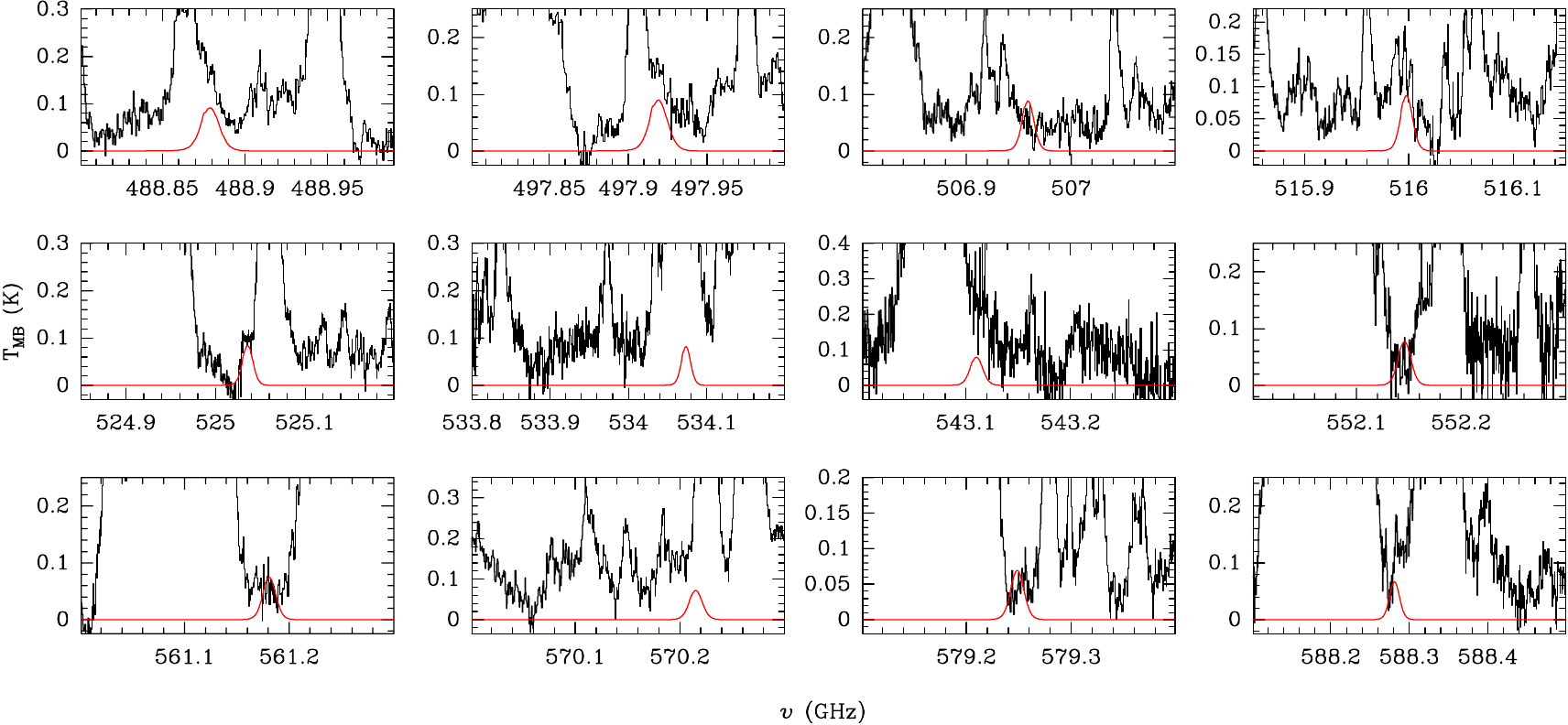}
   \caption{Observed spectra (black histogram) in the HIFI survey. Best fit LVG model results for HC$^{13}$CCN are shown in red.}
    \label{figure:HC13CCN_mix_hifi}
   \end{figure*}

\begin{figure*}
   \centering 
   \includegraphics[angle=0,width=17cm]{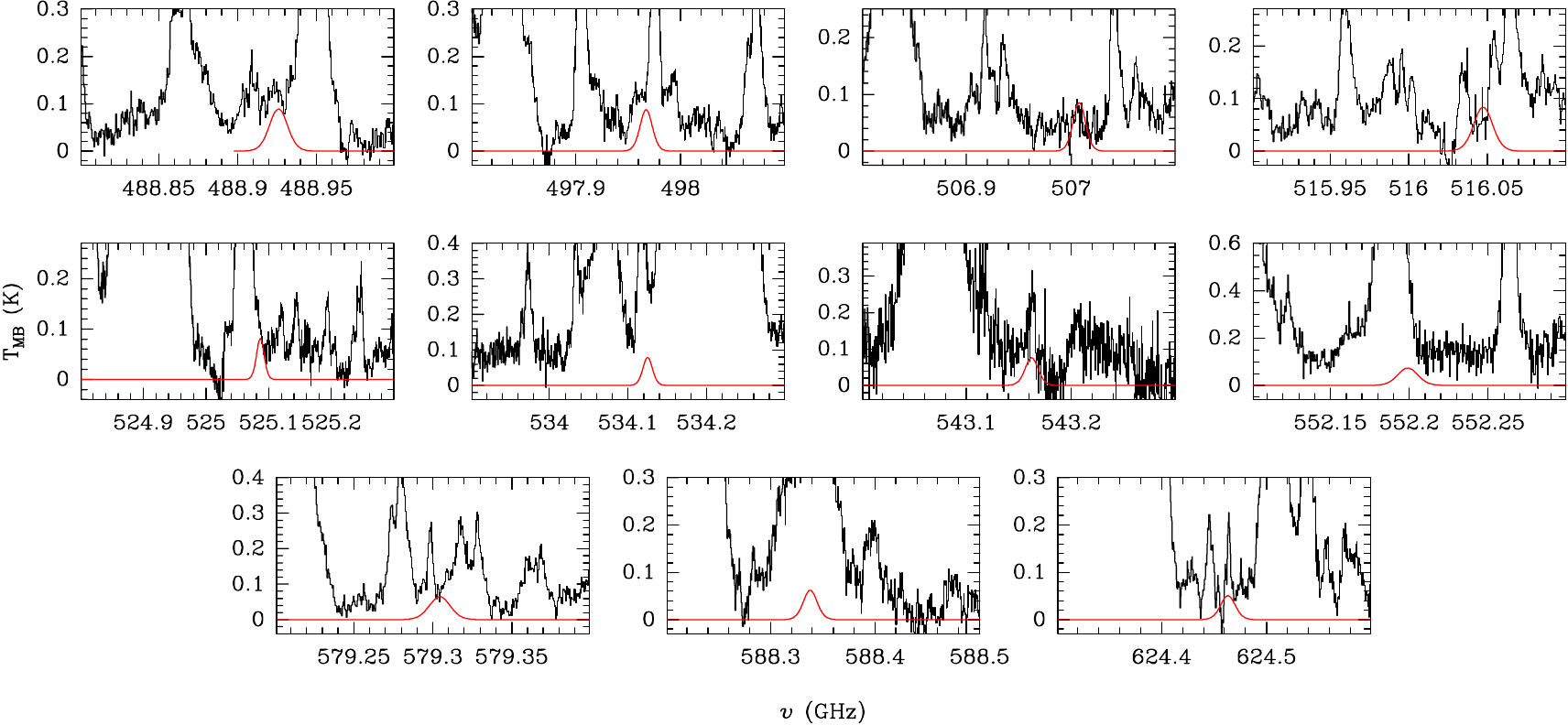}
   \caption{Observed spectra (black histogram) in the HIFI survey. Best fit LVG model results for HCC$^{13}$CN are shown in red.}
    \label{figure:HCC13CN_mix_hifi}
   \end{figure*}

\begin{figure*}
   \centering 
   \includegraphics[angle=0,width=17cm]{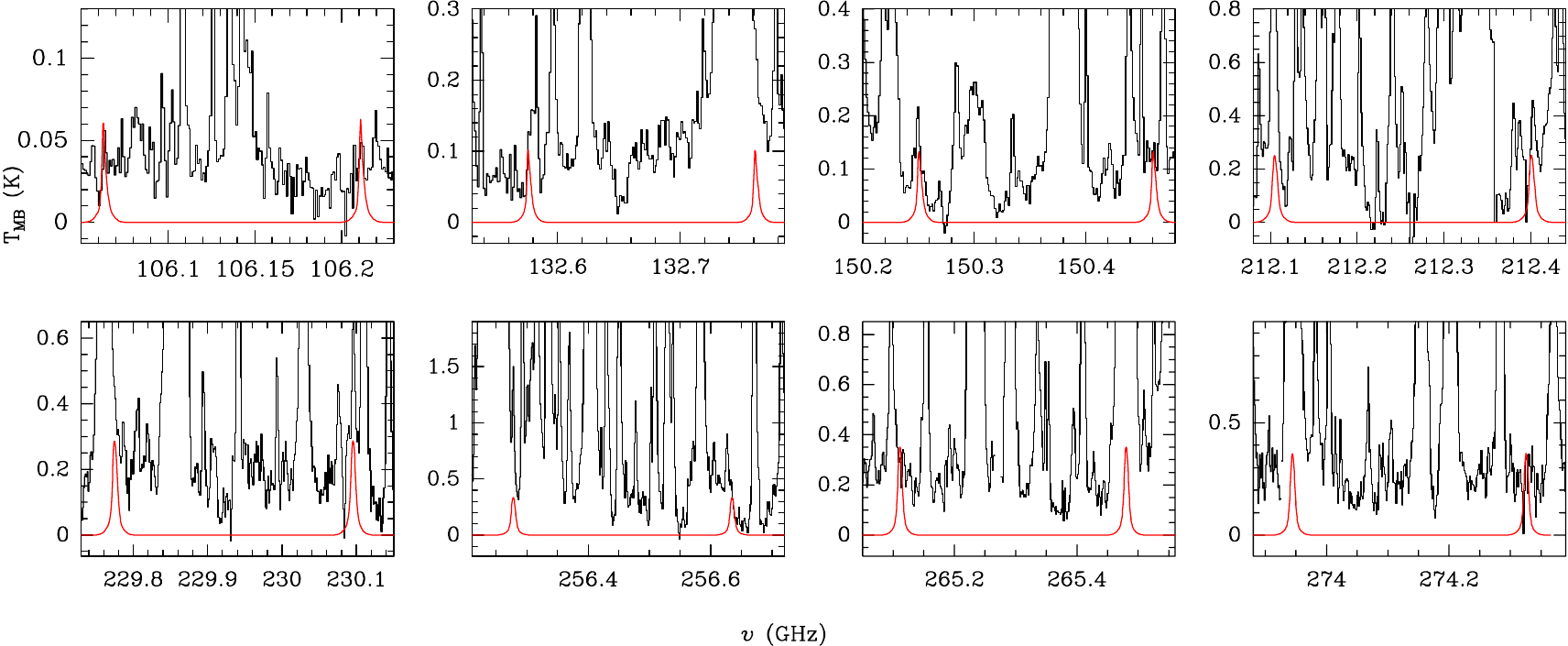}
   \caption{Observed lines of H$^{13}$CCCN $\nu$$_{7}$ (black histogram) in the IRAM survey. Best fit LTE model results are shown in red.}
    \label{figure:H13CCCN_V7_mix_30m}
   \end{figure*}

\begin{figure*}
   \centering 
   \includegraphics[angle=0,width=17cm]{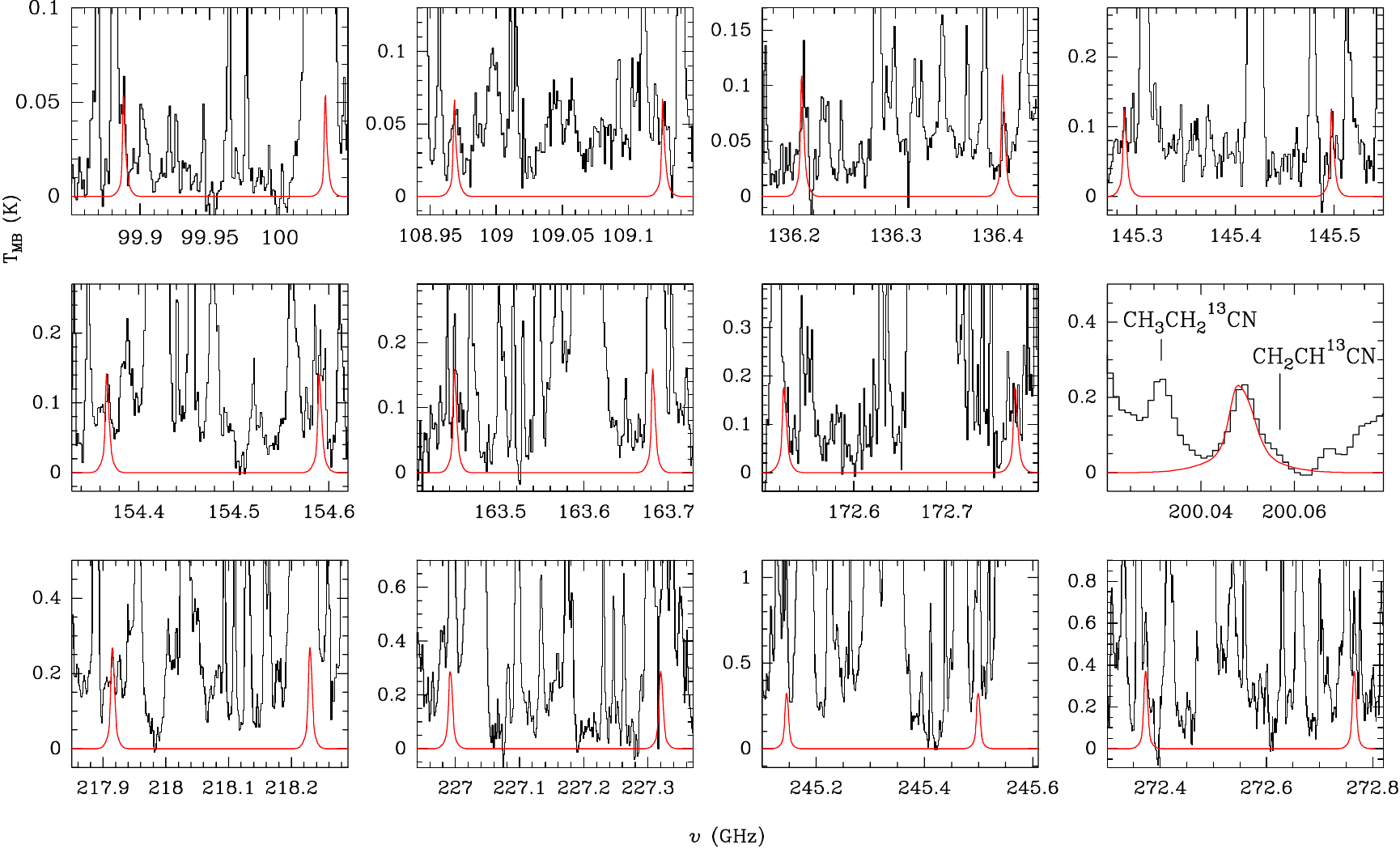}
   \caption{Observed lines of HC$^{13}$CCN $\nu$$_{7}$ (black histogram) in the IRAM survey. Best fit LTE model results are shown in red.}
    \label{figure:HC13CCN_V7_mix_30m}
   \end{figure*}

\pagebreak

\begin{figure*}
   \centering 
   \includegraphics[angle=0,width=17cm]{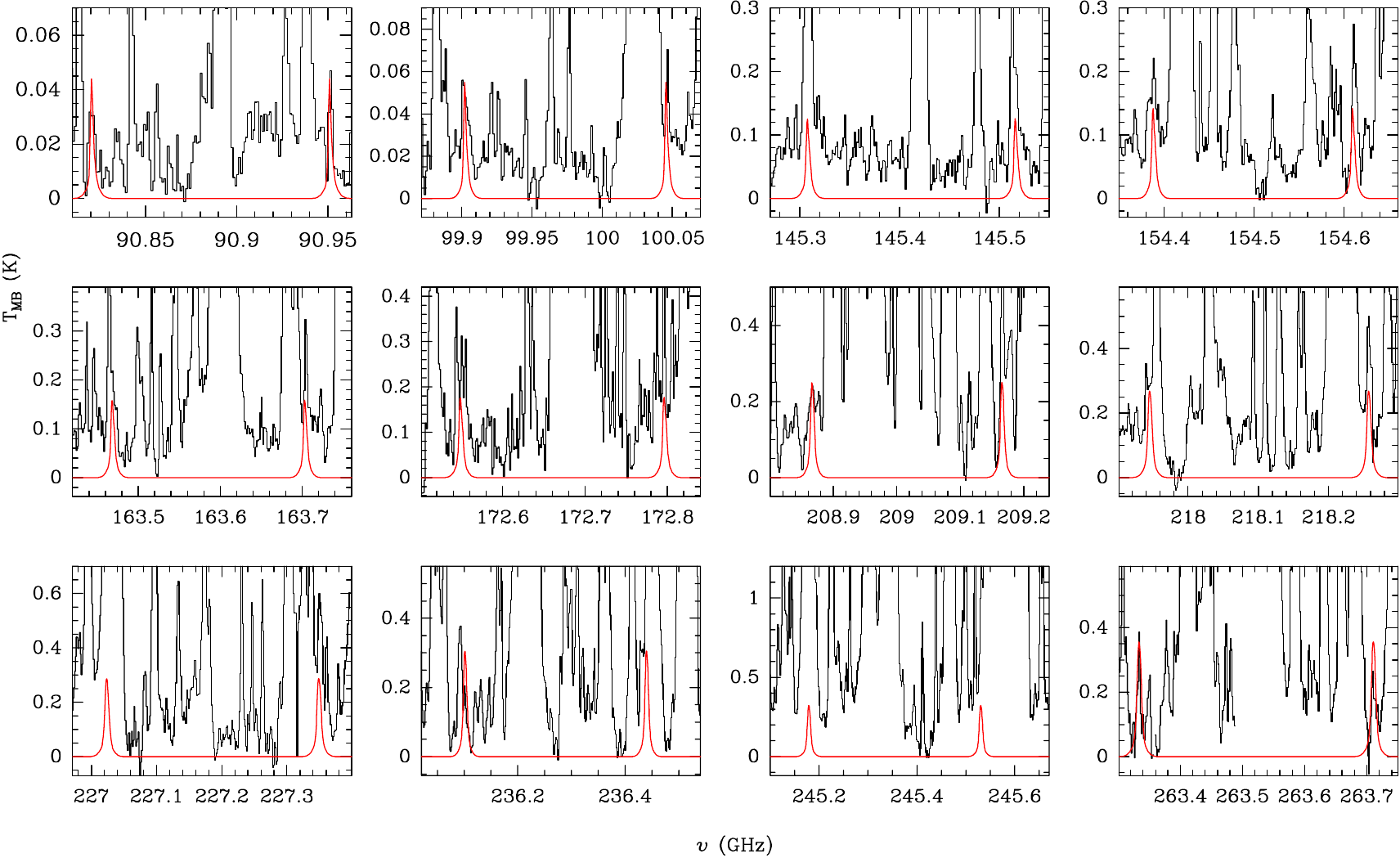}
   \caption{Observed lines of HCC$^{13}$CN $\nu$$_{7}$ (black histogram) in the IRAM survey. Best fit LTE model results are shown in red.}
    \label{figure:HCC13CN_V7_mix_30m}
   \end{figure*}


\pagebreak



\pagebreak
\end{appendix}

\end{document}